\documentclass[a4paper,fleqn,usenatbib]{mnras}

\usepackage{newtxtext,newtxmath}

\usepackage[T1]{fontenc}
\usepackage{ae,aecompl}

\usepackage{graphicx}	
\usepackage{amsmath}	
\usepackage{amssymb}	
\usepackage{gensymb}
\usepackage{tabulary}
\usepackage{subfigure}
\usepackage{epstopdf}
\usepackage{natbib}
\usepackage{afterpage}
\usepackage{float}
\usepackage{booktabs}
\usepackage{caption}

\usepackage{xr}		


\newcommand{\ratio}{$\sigma_{\rm{PSF}}/R_e^{\rm{maj}}$}

\title[SDSS-IV MaNGA: Stellar angular momentum]{SDSS-IV MaNGA: Stellar angular momentum of about 2300 galaxies: unveiling the bimodality of massive galaxy properties} 

\author[M. T. Graham et al.]{Mark T. Graham$^1$\thanks{E-mail: mark.graham@physics.ox.ac.uk},
Michele Cappellari$^1$,
Hongyu Li$^{2,3}$,
Shude Mao$^{4,2,5}$,\newauthor
Matthew Bershady$^6$,
Dmitry Bizyaev$^{7,8}$,
Jonathan Brinkmann$^8$,\newauthor
Joel R. Brownstein$^9$,
Kevin Bundy$^{10}$,
Niv Drory$^{11}$,
David R. Law$^{12}$,\newauthor
Kaike Pan$^{7}$,
Daniel Thomas$^{13}$,
David A. Wake$^{14,15}$,
Anne-Marie Weijmans$^{16}$,\newauthor
Kyle B. Westfall$^{10}$,
Renbin Yan$^{17}$
\\
$^1$Sub-department of Astrophysics, Department of Physics, University of Oxford, Denys Wilkinson Building, Keble Road, Oxford, OX1 3RH, UK\\
 $^2$National Astronomical Observatories, Chinese Academy of Sciences, 20A Datun Road, Chaoyang District, Beijing 100012, China\\
$^3$University of Chinese Academy of Sciences, Beijing 100049, China\\
$^4$Physics Department and Tsinghua Centre for Astrophysics, Tsinghua University, Beijing 100084, China\\
$^5$Jodrell Bank Centre for Astrophysics, School of Physics and Astronomy, The University of Manchester, Oxford Road, Manchester, M13 9PL, UK\\
$^6$Department of Astronomy, University of Wisconsin-Madison, 475N. Charter St., Madison WI 53703, USA\\
$^7$Sternberg Astronomical Institute, Moscow State University, Moscow, Russia\\
$^8$Apache Point Observatory and New Mexico State University, P.O. Box 59, Sunspot, NM, 88349-0059, USA\\
$^9$Department of Physics and Astronomy, University of Utah, 115 S. 1400 E., Salt Lake City, UT 84112, USA\\
$^{10}$UCO/Lick Observatory, University of California, Santa Cruz, 1156 High St. Santa Cruz, CA 95064, USA\\
$^{11}$McDonald Observatory, The University of Texas at Austin, 1 University Station, Austin, TX 78712, USA\\
$^{12}$Space Telescope Science Institute, 3700 San Martin Drive, Baltimore, MD 21218, USA\\
$^{13}$Institute of Cosmology \& Gravitation, University of Portsmouth, Dennis Sciama Building, Portsmouth, PO1 3FX, UK\\
$^{14}$Department of Physics, University of North Carolina Asheville, One University Heights, Asheville, NC 28804, USA\\
$^{15}$School of Physical Sciences, The Open University, Milton Keynes, MK7 6AA, UK\\
$^{16}$School of Physics and Astronomy, University of St Andrews, North Haugh, St Andrews, Fife, KY16 9SS, UK\\
$^{17}$Department of Physics and Astronomy, University of Kentucky, 505 Rose Street, Lexington, KY 40506, USA\\
}

\date{Accepted 2018 February 21. Received 2018 February 21; in original form 2017 November 3}

\pubyear{2018}

\begin{document}
\label{firstpage}
\pagerange{\pageref{firstpage}--\pageref{lastpage}}
\maketitle

\begin{abstract}

We measure $\lambda_{R_e}$, a proxy for galaxy specific stellar angular momentum within one effective radius, and the ellipticity, $\epsilon$, for about 2300 galaxies of all morphological types observed with integral field spectroscopy as part of the MaNGA survey, the largest such sample to date. We use the $(\lambda_{R_e}, \epsilon)$ diagram to separate early-type galaxies into fast and slow rotators. We also visually classify each galaxy according to its optical morphology and two-dimensional stellar velocity field. Comparing these classifications to quantitative $\lambda_{R_e}$ measurements reveals tight relationships between angular momentum and galaxy structure. In order to account for atmospheric seeing, we use realistic models of galaxy kinematics to derive a general approximate analytic correction for $\lambda_{R_e}$. Thanks to the size of the sample and the large number of massive galaxies, we unambiguously detect a clear bimodality in the $(\lambda_{R_e}, \epsilon)$ diagram which may result from fundamental differences in galaxy assembly history. There is a sharp secondary density peak inside the region of the diagram with low $\lambda_{R_e}$ and $\epsilon < 0.4$, previously suggested as the definition for slow rotators. Most of these galaxies are visually classified as non-regular rotators and have high velocity dispersion. The intrinsic bimodality must be stronger, as it tends to be smoothed by noise and inclination. The large sample of slow rotators allows us for the first time to unveil a secondary peak at $\pm90\degree$ in their distribution of the misalignments between the photometric and kinematic position angles. We confirm that genuine slow rotators start appearing above a stellar mass of $2\times10^{11} M_{\odot}$ where a significant number of high-mass fast rotators also exist.
\end{abstract}




%

\section{Introduction}
Among the most fundamental of galaxy properties is the angular momentum $J_{\star}$. Set at early times by perturbations due to the misalignment from nearby protogalaxies (Tidal Torque Theory; \citealp{hoyle1951origin, peebles1969origin, doroshkevich1970spatial, white1984angular}), the specific angular momentum $j_{\star} \equiv J_{\star}/M_{\star}$ is assumed to be conserved during the collapse of the initial gas cloud \citep{thacker2001star, romanowsky2012angular}, itself contained within dark matter haloes assumed to have the same angular momentum \citep{fall1980formation, fall1983galaxy, mo1998formation, zavala2008bulges}. Provided the gas is allowed to cool and sink to the centre of the dark matter halo undisturbed, it will form a stable rotating disk which over time will evolve into a spiral (late-type) galaxy \citep{white1978core}. The angular momentum grows through accretion of cold gas via filaments where the gas is cold enough to sink to the centre of the halo while retaining a high specific angular momentum \citep{kerevs2005galaxies, stewart2013angular}\par
Within the hierarchical framework, the most massive galaxies are built up from smaller progenitors (e.g. \citealp{white1991galaxy}). In the largest haloes which later host galaxy groups and clusters, the gas quickly collapses at the centre of the halo to form a massive galaxy where the majority of the stellar component forms at very early times \citep{de2007formation}. However, as the gas is heated by feedback from the supermassive black hole, the gas is expelled via outflows and the star formation is quenched \citep{silk1998quasars, di2005energy, springel2005simulations, maiolino2012feedback}. When nearby haloes merge, the central galaxies are also able to merge due to their low relative velocities and large mass \citep{aragon1998brightest}. They merge through dissipationless dry mergers where the gas is either absent or dynamically unimportant (\citealp{naab2014atlas3d} and references therein). The angular momentum is redistributed into the merged halo and the resulting central galaxy is dispersion-dominated.\par
Early attempts to measure the specific angular momentum of galaxies found that ellipticals have about an order of magnitude less angular momentum than spirals. \cite{bertola1975dynamics} calculated analytically the angular momentum of a bright elliptical galaxy (NGC 4679) from the rotation curve and found that spirals have about 5-30 times the angular momentum (depending on the assumed mass-to-light ratio). \cite{fall1983galaxy} introduced the $j_{\star}-M_{\star}$ relation and claimed that elliptical and spiral galaxies follow parallel sequences with a slope close to the theoretical prediction ($j_{\star} \propto M_{\star}^{2/3}$; e.g. \citealp{white1984angular, mo1998formation}) and spirals having a factor of $\approx6$ larger $j_{\star}$ than ellipticals for a given stellar mass. The difference was attributed to the bulge fraction with intermediate morphologies such as S0s occupying the space between spirals and ellipticals (see \citealp{romanowsky2012angular}). An alternative measure of the stellar angular momentum comes from the ratio of ordered to random motion of galaxy's stellar component, quantified as the ratio between the maximum rotation velocity $V_{\rm{max}}$ along the major axis and the maximum velocity dispersion $\sigma_0$ at the centre of the galaxy \citep{illingworth1977rotation, binney1978rotation}. The classification of whether a galaxy rotates fast or slow then depends on the ratio between the location of a galaxy on the $(V/\sigma, \epsilon)$ diagram where $\epsilon$ is the ellipticity (flattening), with respect to the prediction for a galaxy with an isotropic velocity ellipsoid \citep{kormendy1982barred, kormendy1982disk}. \cite{davies1983kinematic} used this method to provide the first evidence that high-luminosity ETGs have lower angular momentum than low/normal luminosity ETGs.\par
In the case of spiral galaxies, it is necessary that they form via accretion of cold gas so that a fresh supply of accreted gas can cool and contract to form new stars \citep{bernardi2007brightest}. However, for ETGs, it is insufficient to rely on visual morphology alone to assess whether a galaxy formed via wet, gas-rich processes or via dry, gas-poor mergers. A more accurate and robust classification is the fast/slow rotator classification, where the terms ``fast" and ``slow" correspond to whether a galaxy's rotation is regular (i.e. circular velocity) or non-regular dominated by dispersion (i.e. random motion) \citep{emsellem2007sauron, cappellari2007sauron}. For example, as noted by \cite{cappellari2011atlas3db}, as many as two thirds of disk-like fast rotator ETGs (i.e. S0s or disky ellipticals when seen edge-on) are likely to be misclassified as spheroidal elliptical galaxies, particularly when face-on (see also \citealp{krajnovic2011atlas3d, emsellem2011atlas3d}). In this way, the fast/slow rotator classification is more robust than the traditional Hubble classification of galaxies \citep{hubble1926extragalactic, hubble1936realm, sandage1961hubble} as it is based on motions from stellar kinematics and so is less affected by projection effects. Massive, core slow rotators (SRs) are most commonly found to occupy the centres of clusters and form via the dry merging channel, while fast rotators (FRs) form via the wet merging channel and occupy the outskirts of clusters as well as populate the field \citep{cappellari2011atlas3db, cappellari2013effect}. The two classes form distinct galaxy populations which can be distinguished by measuring a parameter related to the galaxy spin (e.g. \citealp{emsellem2007sauron, emsellem2011atlas3d, d2013fast, fogarty2014sami, scott2014distribution, falcon2015angular, fogarty2015sami, querejeta2015formation, cortese2016sami, veale2016massive, brough2017kinematic, van2017sami, greene2017kinematic, greene2018angular, smethurst2018quenching}). For a complete review, see \cite{cappellari2016structure}, hereafter C16.\par
However, a global measurement of the stellar angular momentum requires a two-dimensional view of the line-of-sight velocity distribution (LOSVD). This has been possible since the advent of integral field spectroscopy (IFS), an observational technique whereby optical fibres or lenslets are placed across the primary mirror, allowing spectra to be measured across the field of view. IFS has made possible the study of stellar kinematics as a new way to directly observe the kinematic properties of galaxies. A proxy for the stellar angular momentum measured within one half-light (effective) radius, $\lambda_{R_e}$, was proposed by \cite{emsellem2007sauron}. $\lambda_{R_e}$ takes into account the spatial structure in the kinematic maps and takes full advantage of the capabilities of IFS. With this approach, a more accurate measure of a galaxy's angular momentum can be obtained that cleanly separates physical properties of galaxies and is nearly insensitive to inclination \citep{cappellari2007sauron}.\par
The first major effort to make a census of $\lambda_{R_e}$ for nearby ETGs was conducted as part of the ATLAS$^{\rm{3D}}$ survey \citep{cappellari2011atlas3d}, a follow-up of the initial SAURON survey \citep{de2002sauron}. ATLAS$^{\rm{3D}}$ utilised the dedicated SAURON spectrograph \citep{bacon2001sauron} on the 4.2-metre William Herschel Telescope to provide gas and stellar kinematics for a volume-limited ($D<42$ Mpc) survey of 260 ETGs, extracted from a complete sample of 871 galaxies. As this was the largest sample to date (SAURON surveyed a sample of 48 ETGs and 24 spirals), further constraints could be placed on the boundary between SRs and FRs, including a dependence on ellipticity. Analysis of the kinematic maps revealed that in almost all cases, FRs have regular velocity fields (hourglass shape, like the rotation of inclined disks) whereas SRs are non-regular rotators showing irregular or complex velocity maps or little overall rotation \citep{krajnovic2011atlas3d}. The SR/FR boundary was defined to be a best-fit in order to separate the two classes with minimal overlap \citep{emsellem2011atlas3d}.\par
More recently, the CALIFA survey \citep{sanchez2012califa} observed $\sim$600 galaxies of all morphological types at low redshift ($z<0.03$) with a single integral field unit (IFU) to provide kinematic maps of velocity and velocity dispersion as well as properties of the stellar population and ionised gas. CALIFA provided $(\lambda_{R_e}, \epsilon)$ values for 300 galaxies comprising the largest homogeneous census of $\lambda_{R_e}$ to date \citep{falcon2015angular}. The tight connection between spiral galaxies and FRs was observed, but a dependence on bulge size was also observed, with Sa and Sb galaxies exhibiting high $\lambda_{R_e}$, and Sc and Sd galaxies showing lower values. A combined $(\lambda_{R_e}, \epsilon)$ diagram from the ATLAS$^{\rm{3D}}$, CALIFA and SAMI pilot surveys \citep{fogarty2015sami} was presented in section 3.6.3 of C16. While ATLAS$^{\rm{3D}}$ and CALIFA provided a large homogeneous sample of nearby galaxies, its main limitation was the availability of only one single IFU, meaning that only one galaxy could be observed at once, making the possibility of larger surveys unlikely.\par
At the present time, two large scale optical IFS surveys are in operation with the capability of observing more than one galaxy simultaneously. The Sydney-AAO Multi-object Integral field spectrograph survey (SAMI; see \citealp{croom2012sydney} for details about the spectrograph and \citealp{fogarty2015sami} for details about the survey) utilises 13 IFUs identical in size and shape (hexabundles containing 61 fibres each) with the aim of mapping 3400 nearby galaxies ($z<0.095$; primary sample; \citealp{bryant2015sami}). Using 488 galaxies from the SAMI survey, \cite{cortese2016sami} found that, rather than galaxies forming two distinct channels on the $j_{\star}-M_{\star}$ plane as found by \cite{fall1983galaxy} and \cite{romanowsky2012angular}, galaxies form a continuous sequence whereby a galaxy's position on the $j_{\star}-M_{\star}$ plane is correlated with morphology and Sérsic index \citep{Sersic1963influence, Sersic1968atlas}. The same result has been found in hydrodynamical simulations \citep{lagos2017angular} using a number of proxies for galaxy morphology. These results may suggest that the formation of spheroids and disks is a continuous process rather than being fundamentally different in nature.\par
The other ongoing large-scale IFS survey is the Mapping Nearby Galaxies at Apache Point Observatory (MaNGA) survey \citep{bundy2015overview}, currently operating as part of the Sloan Digital Sky Survey (SDSS) IV \citep{blanton2017sloan} programme which started in 2014. Over six years, MaNGA will observe approximately 10,000 galaxies with IFS, making MaNGA the largest homogeneous IFS survey to date. MaNGA makes use of the dedicated 2.5-metre diameter Sloan Foundation Telescope \citep{gunn20062} at Apache Point Observatory (APO), New Mexico. The full specification for the survey design is outlined in \cite{yan2016sdss}.\par
As the distances probed by large-scale, ground-based IFU surveys increase, it is becoming necessary to account for atmospheric seeing when measuring a quantity that is derived from the LOSVD, such as $\lambda_{R_e}$. At larger distances, both the apparent sizes of galaxies and the spatial resolution decrease, and therefore atmospheric seeing effects become more important. \cite{d2013fast} investigated the effect of reduced spatial resolution on $\lambda_{R_e}$. By ``reobserving" kinematic models of SAURON data using \texttt{KINEMETRY} \citep{krajnovic2006kinemetry} at the distance of the galaxy cluster A1689 ($\sim$ 50 times the average distance of the SAURON survey), they found they could approximate $\lambda_{R_e}$ by using all available spaxels to measure $\lambda_R(\rm{IFU})$, before applying a correction based on $R_e$. \cite{van2017sami} simulated the effect of seeing on ATLAS$^{\rm{3D}}$ kinematics, which, being very nearby, can be assumed to be unaffected by seeing, in order to provide an estimate for the error on $\lambda_{R_e}$ measurements. They found that the impact is strongest for small galaxies ($R_e \sim \rm{PSF}$ where PSF is the point source function) with $\lambda_{R_e}>0.2$, with a median decrease of 0.08 when the PSF is 3\arcsec at FWHM. Finally, \cite{greene2018angular} took a subset of 50 galaxies in MaNGA that have at least four beams within $R_e$ with radial coverage out to $1.5R_e$, and degraded the resolution so that the PSF was a constant fraction of $R_e$ that matched the typical resolution of the whole MaNGA sample. They found that $\lambda_{R_e}$ decreased by up to 0.075 for $\lambda_{R_e}<0.2$, and up to 0.125 for $\lambda_{R_e}>0.2$. While these tests provide a useful indication of the errors due to atmospheric effects, it remains difficult to correct $\lambda_{R_e}$ on an galaxy-by-galaxy basis without an analytic correction.\par
In this work, we derive such a correction by simulating the effect of seeing on the kinematics of galaxy models using the Jeans Anisotropic Modelling method (JAM; \citealp{cappellari2008measuring}). We then apply our correction to $\lambda_{R_e}$ measurements of a \textit{clean sample} of 2286 galaxies observed thus far by the MaNGA survey, the largest sample observed with IFS to date. We classify galaxies by visual morphology and kinematic structure in order to achieve the most complete description of the $(\lambda_{R_e}, \epsilon)$ diagram possible. We also run simulations to understand how well the true value of $\lambda_{R_e}$ can be recovered when the velocity dispersion, $\sigma$, is low after correcting for the instrumental resolution. Finally, we present the mass-size relation as well as the kinematic misalignment as a function of ellipticity.\par
This paper is structured as follows: In \autoref{sec:samples}, we discuss the data samples used. In \autoref{sec:methods}, we present our methods including the selection criteria used to define our \textit{clean sample} which our final results are based on. We also present our simple analytic correction to apply to observed values of $\lambda_{R_e}$ to account for atmospheric effects due to the PSF, the derivation of which is presented in Appendix \ref{app:beamsmearing}. We present our results and discussion in \autoref{sec:results} and our final conclusions in \autoref{sec:conclusions}. Throughout this work, we adopt standard values for the cosmological parameters, close to the latest measured values \citep{planck2015results}. We take the value of the Hubble constant, $H_0$, to be $70$ km s$^{-1}$ Mpc$^{-1}$, and we assume a flat cosmology where $\Omega_M$, $\Omega_k$ and $\Omega_\Lambda$ are 0.3, 0 and 0.7 respectively.\par
\section{Data Samples}
\label{sec:samples}
\subsection{MaNGA IFU observations}
\label{sec:manga} 
The data used in this study are taken from the fifth MaNGA Product Launch (MPL-5; July 2016), an incremental internal data release. The spectra for MPL-5 were released to the public as part of the SDSS Data Release 14 (DR14; \citealp{abolfathi2017fourteenth}). MPL-5 contains stellar and gas kinematics for 2722 galaxies. A small number of galaxies were observed with more than one IFU bringing the total number of data cubes to 2774. MaNGA utilises 17 IFUs on a single plate varying in size from the smallest, containing 19 fibres (diameter 12\arcsec), to the largest, containing 127 fibres (diameter 32\arcsec; see \citealp{drory2015manga} for a complete description of the IFUs). A set of 12 bundles containing seven fibres each are used for flux calibration \citep{yan2016spectrophotometric} and 92 single fibres are used for sky subtraction.\par
Each IFU is shaped as a hexagon in order to optimise the available space on the detector \citep{law2015observing}. The spatial resolution is set by the 2\arcsec diameter of each fibre, subtending a physical distance of $\sim$2 kpc at $z\sim0.05$. The IFUs are housed on the BOSS spectrographs \citep{smee2013multi} which have a median spectral dispersion of $\sigma_{\rm{inst}}\sim$ 72 km s$^{-1}$ \citep{law2016data}. The spectral range covers the entire visible spectrum from about 3600 - 10300 \AA, resulting in a typical resolving power $R \sim 2000$ \citep{smee2013multi, law2015observing}. The targets have been selected at low redshift ($0.01 < z < 0.15$) to follow a flat distribution across the full stellar mass range ($10^9\textrm{ M}_{\odot}$ - $10^{12}\textrm{ M}_{\odot}$), using the absolute magnitude in SDSS i-band, M$_i$, as a proxy for stellar mass to remove any bias from stellar population models \citep{bundy2015overview}.\par
The full sample that we use from MPL-5 consists of three distinct subsets: the Primary, Secondary and Colour-Enhanced sample. One of the survey targets for MaNGA is to observe all galaxies in the Primary sample out to $\sim 1.5 R_e$ and all galaxies in the Secondary sample out to $\sim 2.5 R_e$. Therefore, the Secondary sample contains galaxies that are systematically at higher redshifts for a given M$_i$ than the Primary sample. The Colour-Enhanced sample is designed to fill in gaps in the colour-magnitude plane \citep{bell2004red, baldry2004bimodal}, covering low-mass red galaxies and high-mass blue galaxies for example. No cuts are made on colour, morphology or environment such that the galaxies observed by MaNGA are fully representative of the local galaxy population. For a complete description of the target selection, see \cite{wake2017sdss}.\par

\subsection{Cross referencing with the 2MASS XSC}
\label{sec:2mass} 
We cross-reference the MaNGA sample with the Two Micron All-Sky Survey Extended Source Catalog (2MASS XSC\footnote{http://irsa.ipac.caltech.edu/}; \citealp{skrutskie2006two}). Our focus is the near-infrared J and K$_{\rm{S}}$ bands which are most sensitive to old stellar populations that form the dominant baryonic mass component of galaxies. The near-infrared is also unobscured by dust extinction allowing the distribution of stellar populations to be revealed. Of the 2722 MaNGA galaxies in MPL-5, 2270 galaxies ($\sim83\%$) were matched by ra and dec to within 5\arcsec of a near-infrared source. The remaining $\sim17\%$ are too faint in the near-infrared to be included in the XSC. They are found in the 2MASS Point Source Catalog but lack any 2D photometric parameters such as $\epsilon$.\par
For each matched target, we record the axis ratio (\texttt{sup\_ba}), the apparent magnitude in $\rm{K}_{\rm{S}}$ band (\texttt{k\_m\_ext}) and the major axis of the isophote enclosing half the total galaxy light in J band (\texttt{j\_r\_eff}). Half-light (or effective) radii are a poorly defined empirical quantity for low signal-to-noise (S/N) images, as they formally require the knowledge of galaxy fluxes out to infinite radii. For this reason, we use an empirical relation to calibrate the semi-major axis with respect to \texttt{j\_r\_eff}. This relation is defined by \cite{cappellari2013effect} (see also \citealp{krajnovic2017channels}) to be $R_{e}^{\rm{maj}} = 1.61 \times \rm{\texttt{j\_r\_eff}}$ where 1.61 is a best-fit factor used to match the 2MASS effective radii to the RC3 catalogue \citep{de1991third}. (A previous relation used in \citet{cappellari2011atlas3d} took $R_{e}^{\rm{maj}}=\rm{MEDIAN}(\texttt{j\_r\_eff}, \texttt{h\_r\_eff}, \texttt{k\_r\_eff})$ where \texttt{h\_r\_eff} is the same quantity in H band etc. However, this definition of $R_{e}^{\rm{maj}}$ required a factor of 1.7 to match RC3.) As noted by \cite{cappellari2013effect}, J band is preferable over other bands as it has a higher signal-to-noise ratio. The circularised 2MASS effective radius is then $R_{e} = R_{e}^{\rm{maj}} \times \sqrt{\rm{\texttt{sup\_ba}}}$.

\subsection{NSA optical data}
\label{sec:NSA}
The target selection for MaNGA was taken from the NASA-Sloan Atlas\footnote{\url{http://www.nsatlas.org}} (NSA) which is based on SDSS imaging \citep{blanton2011improved}. The version of the catalogue used for MaNGA is v1\_0\_1, a summary of which was released as part of DR14\footnote{\url{http://www.sdss.org/dr14/manga/manga-target-selection/nsa/}}. Many relevant quantities are stored in a companion catalogue to each MPL; for MPL-5, we use \texttt{drpall-v2\_1\_2}. Along with the full NSA catalogue, companion FITS files are also available containing images in the SDSS \textit{ugriz} bands. For each galaxy, we use the following quantities from the NSA catalog: the redshift (\texttt{z\_dist}) estimated using the peculiar velocity model of \cite{willick1997homogeneous}, the stellar mass derived from the K-correction (\texttt{mass}, \citealp{blanton2007k}) as well as parameters derived from a 2D Sérsic fit \citep{blanton2011improved}. These are the 50$\%$ light (effective) radius along the major axis (\texttt{sersic\_th50}), the angle (East of North) of the major axis (\texttt{sersic\_phi}), the axis ratio (\texttt{sersic\_ba}) and the Sérsic index $n$ (\texttt{sersic\_n}).\par
As the 2MASS XSC is incomplete for the MaNGA sample, we need to ensure that the NSA $R_e$ measurements are at the same scale as the XSC $R_e$, which are in turn scaled to match RC3. The 2MASS catalogue is accurate over all radii above $\sim 5$\arcsec, below which the effects due to the 2.5\arcsec  PSF (FWHM) are dominant. At small radii, the NSA is more accurate than 2MASS due to the small 1.3\arcsec  PSF (FWHM) \citep{stoughton2002sloan}. We find that there is a systematic offset between the NSA and XSC values of $R_e$, due to differences in depth of photometry, as well as a deviation from a one-to-one correlation at low radii, due to the different PSFs of the two catalogues.\par
In order to bring the NSA $R_e$ to the same RC3 scale as our 2MASS-derived $R_e^{\rm{2MASS}}$ values, we find the smallest $R_e^{\rm{2MASS}}$ above which $R_e^{\rm{2MASS}} \propto R_e^{\rm{NSA}}$ i.e. the slope is equal to one within the errors. We find that for $R_e^{\rm{2MASS}}>7.5\arcsec$, $R_e^{\rm{2MASS}} = 1.17 R_e^{\rm{NSA}}$. We apply this scale factor to all $R_e^{\rm{NSA}}$ including those at low radii where there is not a one-to-one correlation with 2MASS. These scaling ensures that all our $R_e$ are consistent with the sizes adopted in both the RC3 and ATLAS$^{\rm{3D}}$ catalogues, so they can be compared in an absolute sense.\par

\section{Method and data analysis}
\label{sec:methods}
In this section, we describe our methods for extracting the stellar kinematics from the MaNGA data, as well as our process of determining the parameters of the half-light ellipse. We introduce the spin parameter $\lambda_{R_e}$ which acts as a proxy for the stellar angular momentum within the half-light ellipse and is sensitive to the kinematic morphology which we classify. We also derive stellar mass estimates from apparent $\rm{K}_{\rm{S}}$ band magnitudes from the 2MASS XSC using distance estimates derived from the NSA redshift. In order to produce a \textit{clean sample} of galaxies on which to base our final results, we perform a number of quality control steps. As part of this, we run a number of simulations to allow us to assess how an intrinsic velocity dispersion of 0 km s$^{-1}$ affects $\lambda_{R_e}$ after necessarily correcting for the instrumental dispersion. We discuss our method for returning from the MaNGA selection function to a volume-limited sample. Finally, we introduce our approximate analytic correction that can be used to correct $\lambda_{R_e}$ for atmospheric seeing.\par 

\subsection{Extraction of stellar kinematics}
All science-ready data products are produced by the Data Reduction Pipeline (DRP; \citealp{law2016data}), a semi-automated procedure that performs the reduction, calibration and sky-subtraction for MaNGA observations. To allow the user to quickly access the data that is relevant to their particular science goals, as well as minimise the storage space required, specific data products are produced by the Data Analysis Pipeline (DAP; Westfall et al. in prep). We use MAPS files, which are the primary output of the DAP. MAPS files are two-dimensional images of DAP measured properties, such as stellar kinematics. To derive the stellar kinematics, the DAP uses the Penalised Pixel-Fitting (pPXF) method \citep{cappellari2004parametric, cappellari2017improving} to extract the LOSVD by fitting a set of 49 clusters of stellar spectra from the MILES stellar library \citep{sanchez2006library, falcon2011updated}, known internally as MILES-THIN, to the absorption-line spectra. The clusters are determined by a hierarchical clustering method and each cluster contains a number of stellar spectra with similar properties. Each datacube contains 4563 spectral slices covering a wavelength range 3600 - 10300 \AA. Before the extraction of the mean stellar velocity $V$ and velocity dispersion $\sigma$, the spectra are spatially Voronoi binned\footnote{Voronoi binning employs a tessellation to achieve the best possible spatial resolution given a minimum signal-to-noise threshold.} \citep{cappellari2003adaptive} to achieve a minimum signal-to-noise ratio of $\sim$10 per spectral bin of width 70 km s$^{-1}$. We extract the mean velocity and velocity dispersion for each Voronoi bin, as well as the bin coordinates (in ra and dec). At each pixel, we obtain the median signal (corresponding to flux) so that the surface brightness is measured across the field of view at a constant resolution.\par

\subsection{Determination of the half-light ellipse}
\label{sec:half-light ellipse}
For each galaxy, we require robust and accurate parameters of the half-light ellipse in order to calculate $\lambda_{R_e}$. The half-light ellipse is defined as an ellipse which covers the same area as the half-light circle, i.e. a circle containing half of the visible light of a galaxy with a radius equal to the effective radius. Even though the NSA and 2MASS XSC catalogues provide parameters for the half-light ellipse, neither are individually suitable for covering the wide range of galaxy sizes in the MaNGA sample. Rather than seeking a combination of the two, we measure our own values for effective radius, ellipticity and position angle.\par

\subsubsection{Multi-Gaussian Expansion method}
\label{sec:mgemethod}
We start by fitting the NSA \textit{r} band photometry for each galaxy using the Multi-Gaussian Expansion (MGE) method \citep{emsellem1994multi, cappellari2002efficient}. MGEs are an efficient way of describing the surface brightness and morphology of galaxies. They consist of a sum of two dimensional Gaussians each described by three parameters: the dispersion $\sigma$, the axial ratio $q$ and the luminosity $L$ (see Equation 9 in \citealp{cappellari2013atlas3db}). For all MaNGA galaxies, we use 12 Gaussians which we find is more than sufficient to successfully describe the photometry. We use the Python version of the \texttt{mge\_fit\_sectors} package\footnote{\label{Capwebpage}\url{http://purl.org/cappellari/software}} described in \cite{cappellari2002efficient}.\par
Before starting the fitting procedure, we subtract the sky background measured using \texttt{measure\_sky}\footnote{\url{https://gist.github.com/jiffyclub/1310947\#file-msky-py}}. Simply, the routine fits a second degree polynomial $Ax^2+Bx+C$ to a logarithmic histogram of the flux in pixels. The sky level is calculated as $-0.5B/A$ where $A$ and $B$ are the coefficients of the fitted polynomial. This is equivalent to fitting a Gaussian to the original fluxes and finding the peak, but the pixels are weighted differently. Once the sky background is subtracted, we fit an ellipse with an area (number of pixels) equal to $\pi {R_e^{\rm{NSA}}}^2$ to the largest connected group of pixels in the NSA \textit{r} band photometry using \texttt{find\_galaxy}\footnote{We use the keyword fraction where fraction = $\pi R_e^2/N$ where $N$ is the total number of pixels in the image.} in the \texttt{mge\_fit\_sectors} package. To ensure we are fitting to the correct target in the NSA cutout (size $\sim 2$'/$\sim10R_e$ FOV), we restrict our search field to 2$R_{e}^{\rm{maj}}$ (from 2MASS or NSA) or the size of the IFU, whichever is larger.\par
The photometry is divided into sectors linearly spaced in eccentric anomaly using \texttt{sectors\_photometry}, and the MGE fit is performed on the sectors using \texttt{mge\_fit\_sectors}. Once we have the parameters of the MGE, we convert the total counts (TotalCounts) of each Gaussian into peak surface brightness $C_0$ using Equation 1 from \cite{cappellari2002efficient},
\textbf{
\begin{equation}
C_0=\frac{\rm{TotalCounts}}{2 \pi \sigma^2 q_{\rm{obs}}},
\end{equation}}
where $q_{\rm{obs}}$ is the axial ratio of each Gaussian. To make the fit independent of distance, we convert $\sigma$ into units of arcsec.\par
Finally, we determine the semi-major axis of the isophotal contour containing half the MGE luminosity using the routine \texttt{mge\_half\_light\_isophote}\textsuperscript{\ref{Capwebpage}} which implements steps (i) to (iv) found before Equation 12 in \cite{cappellari2013atlas3db}. In short, the routine constructs a synthetic galaxy image from the MGE and finds the surface brightness enclosed by a number of different isophotes each with different radii. To save computing time, the routine only considers one quadrant as 2D Gaussians naturally have symmetry about both axes. By using linear interpolation, the routine finds the isophote that encloses half the surface brightness. The semi-major axis is then the $x$-coordinate of that isophote. The ellipticity of the isophote is also determined by the same routine using Equation 12 of \cite{cappellari2013atlas3db}. We also obtain the photometric position angle $\Psi_{\rm{phot}}$ and the central coordinates of the half-light ellipse for each target galaxy.\par

\subsubsection{Effective radius}
\label{sec:effectiveradius}
Given that we use the same SDSS photometry, in the same \textit{r} band, we follow \cite{cappellari2013atlas3db} by scaling the (circular) $R_e^{\rm{MGE}}$ by an empirical factor of 1.35 in order to make our $R_e^{\rm{MGE}}$ measurements comparable to ATLAS$^{\rm{3D}}$ (see Figure 7 of \citealp{cappellari2013atlas3db}). In \autoref{fig:mgere}, we plot the correlation between our $R_e^{\rm{MGE}}$ measurements against $R_e^{\rm{2MASS}}$ calculated in \autoref{sec:2mass} for galaxies which are found in the XSC. We find the best-fit correlation using the \texttt{lts\_linefit}\textsuperscript{\ref{Capwebpage}} program described in Section 3.2 of \cite{cappellari2013atlas3db}. The procedure combines the robust Least Trimmed Squares (LTS) technique of \cite{rousseeuw2006computing} into a least-squares fitting algorithm. The routine starts by assuming an intrinsic scatter of zero. It then selects a subset of data points that when fit with a linear relation minimises the $\chi^2$ (see Equation 6 of \citealp{cappellari2013atlas3db}). The whole process is then repeated assuming a different intrinsic scatter. Once all the variables settle, the best fit linear relation is found. In order to reduce covariance between the slope and intercept, and also to reduce uncertainty in the intercept, we perform the fit about a pivot $x_0$ which is equal to the median value of $x$ (also see Equation 6 of \citealp{cappellari2013atlas3db}). We use this method in all similar plots in this paper.\par
By applying the 1.35 factor, \cite{cappellari2013atlas3db} found that the MGE $R_e$ measurements agree remarkably well with $R_e$ measurements from 2MASS and RC3 combined, considering that the two sources are independent. In \autoref{fig:mgere}, we find a similar correlation with a slope of one within the errors. We note the higher level of scatter, likely due to the combination of a larger sample and the larger distances involved. However, we reiterate that both the best-fitting slope and intercept are very close to one, confirming the tight agreement between $R_e^{\rm{MGE}}$, which in our case is measured from the \textit{r} band photometry, and $R_e^{\rm{2MASS}}$, which is taken from fits to the surface brightness in J band.\par
We note that for a small number ($\sim4\%$) of galaxies, nearby foreground stars and/or low surface brightness can result in inaccurate fits from the MGE. For these cases, instead of using $R_e^{\rm{MGE}}$, we use $R_e^{\rm{2MASS}}$ if this value is greater than 7.5\arcsec (i.e. the cutoff found in \autoref{sec:NSA}), and $R_e^{\rm{NSA}}$ if $R_e^{\rm{2MASS}}$ is less than 7.5\arcsec.

\subsubsection{Ellipticity}
\label{sec:ellipticity}
\begin{figure*}
\centering
\begin{minipage}[t]{.49\textwidth}
\includegraphics[width=\textwidth]{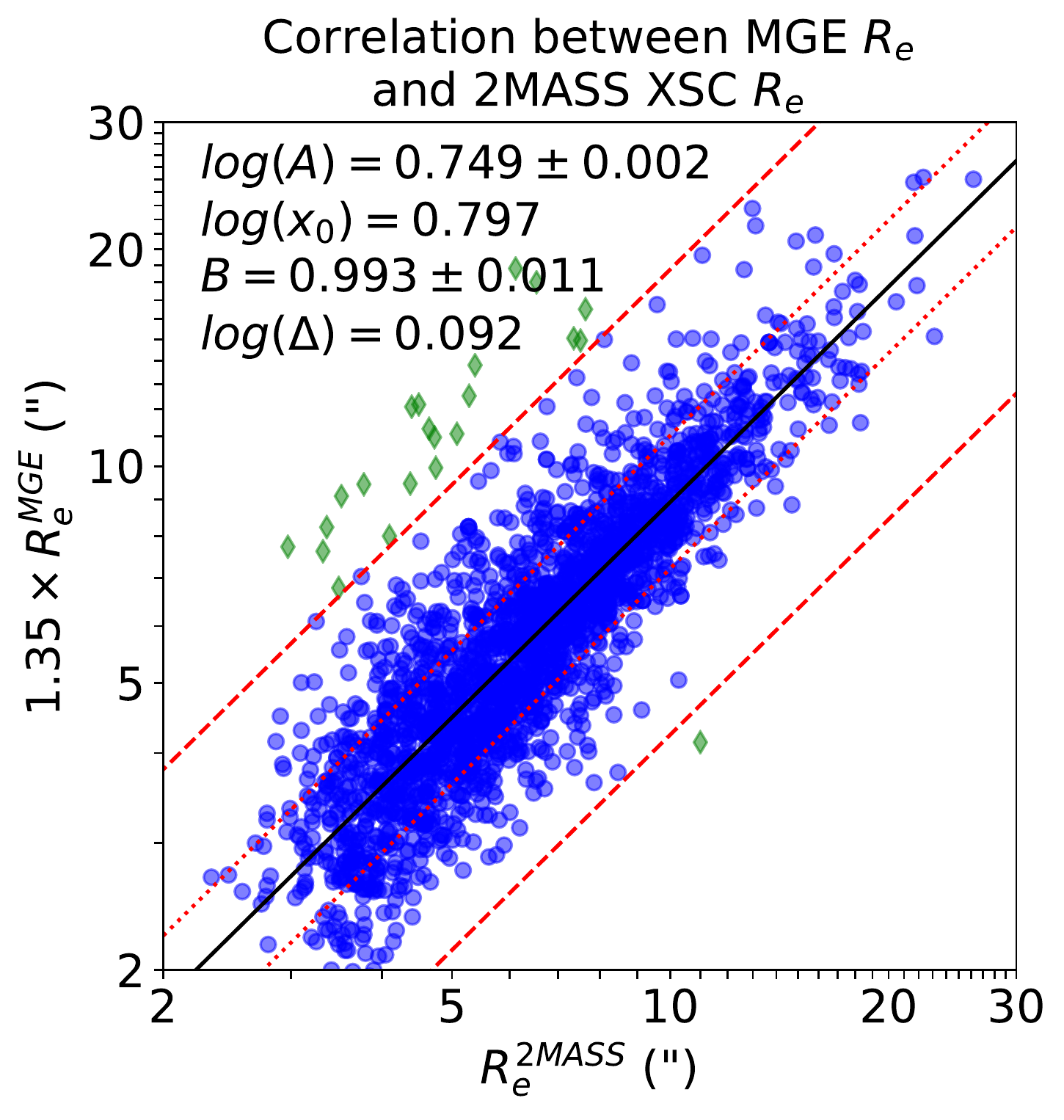}
\caption{Comparison between effective radii measured from MGE fits to the \textit{r} band photometry and the same quantity derived from fits to the J band luminosity from 2MASS XSC. The parameters of the best fitting line are shown. The red dotted and dashed lines are $\Delta$ and 3.5$\Delta$ from the best fit line, where $\Delta$ is the intrinsic scatter. Galaxies which are not included in the fit are shown as green diamonds.}
\label{fig:mgere}
\end{minipage}\hfill
\begin{minipage}[t]{.49\textwidth}
\includegraphics[width=\textwidth]{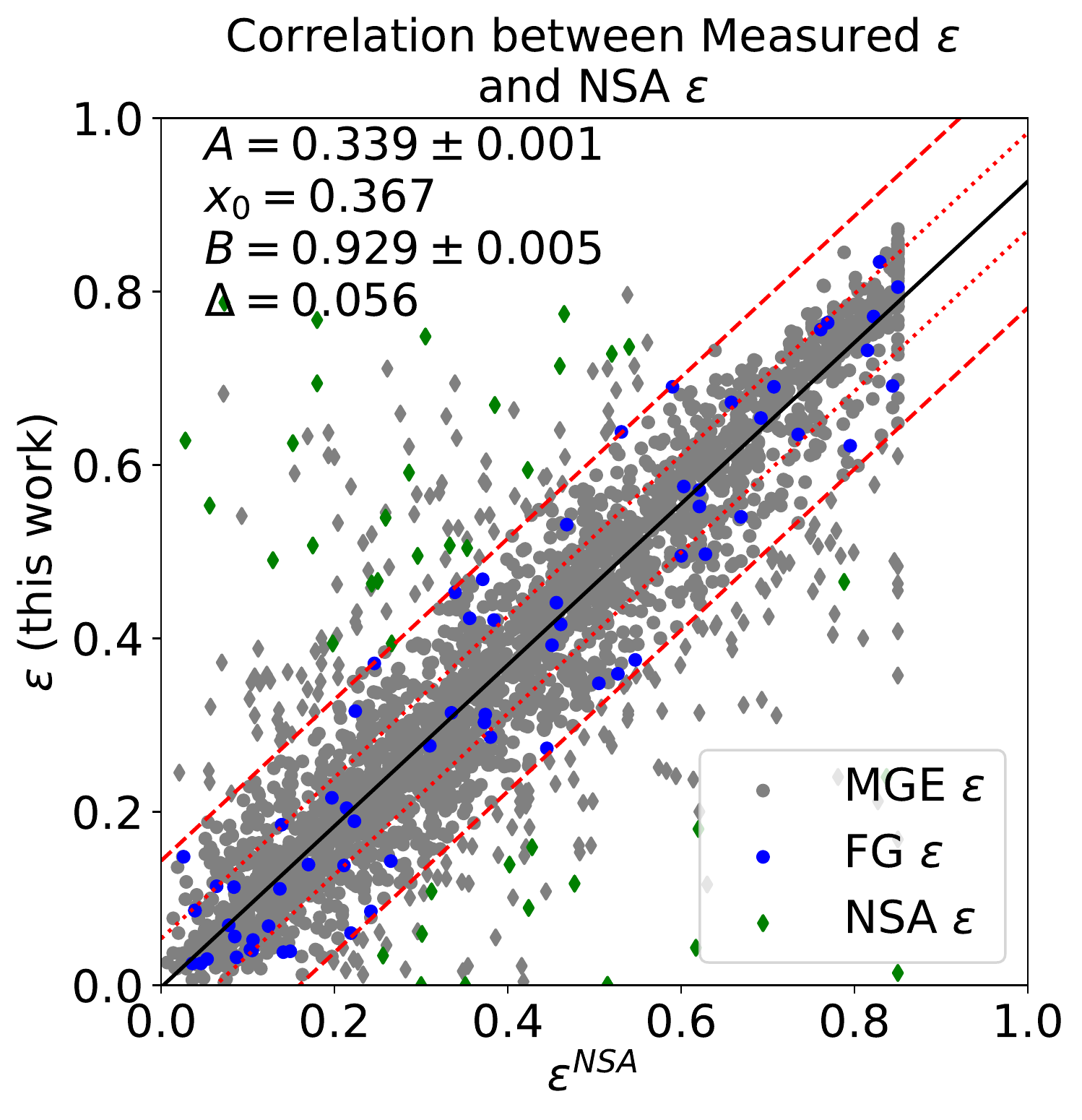}
\caption{Correlation between measured half-light ellipticity from the SDSS NSA \textit{r} band photometry and the half-light ellipticity recorded in the NSA catalogue. For all galaxies, we aim to use the ellipticity measured from the MGE (grey). If for a particular galaxy this measurement is not possible (likely due to nearby foreground stars or low surface brightness), we use either $\epsilon$ measured using \texttt{find\_galaxy} (blue) or the NSA catalog value (lime green) depending on how far the point is from the best-fit line. We make this cut at $2.6\Delta$ from the best fit line, corresponding to 99\% of points for a Gaussian and indicated by the outer red-dashed lines. Points which are not included in the fit are shown as diamonds.}
\label{fig:epscorr}
\end{minipage}
\end{figure*}

We compare our values for the ellipticity determined using the MGE method with those taken from the NSA in \autoref{fig:epscorr}. We find that although the slope is less than one, the best-fit line passes very close to the origin. The likely reason for the slope to be less than one is that our ellipticity is measured within the 1 $R_e$ isophotes, while NSA is globally fitted. As galaxies tend to be rounder near the centre, this likely explains the small systematic difference. Our aim is to use the MGE ellipticity for all galaxies. For the $\sim4\%$ of galaxies for which the MGE fits are inaccurate, we measure $\epsilon$ using \texttt{find\_galaxy}, which fits directly to the photometry, and compare with $\epsilon$ taken from the NSA. If the galaxy falls within $2.6\Delta$ of the best-fit line in \autoref{fig:epscorr} where $\Delta$ is the intrinsic scatter, then the \texttt{find\_galaxy} value is taken. Otherwise, the NSA ellipticity calculated from the axis ratio (see \autoref{sec:NSA}) is taken.\par

\subsection{Proxy for specific angular momentum and effective velocity dispersion}
We measure the luminosity-weighted stellar angular momentum $\lambda_R$ using Equation 1 from \cite{emsellem2007sauron}, reproduced here,
\begin{equation} \label{eq:lambda_R}
\lambda_{R} \equiv \frac{\langle R |V| \rangle}{\langle R \sqrt{V^2 + \sigma^2} \rangle} = \frac{\Sigma ^{N}_{n=1} F_n R_n |V_n|}{\Sigma ^{N}_{n=1} F_n R_n \sqrt{V_n^2 + \sigma_n^2}}
\end{equation}
where the summation is performed over $N$ pixels within the radius $R$. $F_n$, $V_n$ and $\sigma_n$ are the flux, projected velocity and velocity dispersion of the $n$th pixel respectively. As $V$ and $\sigma$ are binned, the binned values are replicated for each pixel belonging to each bin. The quantity $\sqrt{V_n^2 + \sigma_n^2}$ is proportional to mass and ensures that $\lambda_R$ is normalised. The fact that $R$ is present on both the numerator and denominator means that $\lambda_R$ is dimensionless.\par
We also estimate the effective velocity dispersion, $\sigma_e$ within 1 $R_e$ from the projected second velocity moment using Equation 29 from \cite{cappellari2013atlas3db},
\begin{equation} \label{eq:sigma}
\langle v_{\rm{rms}}^2 \rangle_e = \langle v^2 + \sigma^2 \rangle_e \equiv \frac{\Sigma ^{N}_{n=1} F_n (V_n^2 + \sigma_n^2)}{\Sigma ^{N}_{n=1} F_n},
\end{equation}
where the summation is performed analogously to the calculation of $\lambda_{R_e}$. In the ATLAS$^{\rm{3D}}$ works, $\sigma_e$ is measured from the integrated spectra within the half-light ellipse. However, here we calculate $\sigma_e$ as $\langle v_{\rm{rms}}^2 \rangle_e ^{\frac{1}{2}}$ in order to fully take advantage of the IFS data available to us. $\sigma_e$ is found to be approximately equal to $\langle v_{\rm{rms}}^2 \rangle_e ^{\frac{1}{2}}$ within the random errors \citep{cappellari2013atlas3db}. Since $\sigma_e$ is defined as an observed (projected) quantity, we do not attempt to deproject the velocities.

\subsection{Kinematic properties}
\label{sec:kinprop}
We measure the kinematic position angle $\Psi_{\rm{kin}}$ using \texttt{fit\_kinematic\_pa}\textsuperscript{\ref{Capwebpage}}. The method is described in Appendix C of \cite{krajnovic2006kinemetry}. Briefly, the routine generates a number of symmetrised models of the data by averaging values over four quadrants such that $V'(x,y) = [V(x,y) + V(x,-y) - V(-x, y) - V(-x, -y)]/4$ where $V'$ denotes the model velocity field and $V$ denotes the observed velocity field. The best-fit position angle $\Psi_{\rm{kin}}$ is the angle that minimises the $\chi^2$ difference between the model and the data. The kinematic misalignment is then $\Psi_{\rm{mis}} = \Psi_{\rm{kin}} - \Psi_{\rm{phot}}$.\par
Kinematic classifications were judged from the maps by eye by MTG following the five kinematic groups introduced in Section 3.2.3 of \cite{krajnovic2011atlas3d}. Here, however, we perform a purely qualitative classification from the map following the illustration in Figure 4 of C16, rather than using \texttt{KINEMETRY} \citep{krajnovic2006kinemetry}. Each galaxy is either a regular or non-regular rotator. A regular rotator exhibits an hourglass-shaped velocity map. Of the non-regular rotators, there are four sub-categories: non-rotators (NRs; showing no overall rotation), complex rotators (CRs; showing non-regular rotation), kinematically decoupled cores (KDCs; showing a rotating core that is small compared to the galaxy which itself is non-rotating), and counter-rotators (``2$\sigma$" galaxies; showing an inner region that is rotating in an opposite direction to the global rotation). The alternate motion in counter-rotating galaxies results in two peaks in the velocity dispersion, hence the name ``2$\sigma$". We flag galaxies that are either ongoing mergers, or close pairs making the galaxy boundaries indeterminate. We also flag galaxies with bad kinematic data or galaxies that do not fit into the categories mentioned (see \autoref{sec:qualcon}).\par
The two most likely border cases for classifying non-regular rotators are between non-rotators/complex rotators and KDCs/2$\sigma$ galaxies. For the first case (i.e. NR/CR), both classes typically have disordered rotation. We make the distinction that the absolute maximum velocity within the half-light ellipse is $|V_{\rm{max}}| \lesssim 30 \textrm{ km s}^{-1}$ for NRs, whereas for CRs, $|V_{\rm{max}}| \gtrsim 30 \textrm{ km s}^{-1}$. The arbitrary cutoff of $30 \textrm{ km s}^{-1}$ is not a physical distinction but is purely driven by data quality and thus serves as an approximate boundary for distinguishing between the two classes.  For the second case (i.e. KDC/2$\sigma$), we require that 2$\sigma$ galaxies show either two strong peaks in $\sigma$ along the major axis, or clear counter-rotation and high $\sigma$ at the  counter-rotation boundary. All other cases are KDCs. Although the definition of KDC given above specifies that there should be no large scale rotation, we allow for some small ($\sim 30 \textrm{ km s}^{-1}$) regular rotation at larger scales. Most border cases, including those not described here, were verified by MC.

\subsection{Stellar mass derived from 2MASS absolute K$_{\rm{S}}$ band magnitude}
Rather than take values for the stellar mass from the NSA, which are derived from the K-correction \citep{blanton2007k} and so are dependent on stellar population models, we estimate the stellar mass using an empirical relation which is calibrated to ATLAS$^{\rm{3D}}$ dynamical models and is based on the absolute $\rm{K}_{\rm{S}}$ band magnitude, $\textrm{M}_{\rm{K}_{\rm{S}}}$ \citep[Equation 2]{cappellari2013effect}:
\begin{equation} \label{eq:2MASSmass}
\textrm{log}_{10} M_*^{\textrm{2MASS}} \approx 10.58 - 0.44 \times (\textrm{M}_{\textrm{K}_{\textrm{S}}} + 23),
\end{equation}
where the stellar mass $M_*^{\rm{2MASS}}$ is approximately equal to the dynamical mass $M_{\rm{JAM}}$. As noted by \cite{cappellari2013effect}, \autoref{eq:2MASSmass} is for ETGs only. However, it is not thought to vary significantly for spirals/S0s as the mass-to-light ratios in $\rm{K}_{\rm{S}}$ band differ by less than $\sim50\%$ for spirals and ETGs \citep{williams2009kinematic}. We calculate $\textrm{M}_{\rm{K}_{\rm{S}}}$ from $m_{\rm{K}_{\rm{S}}}$ using the luminosity distance estimated from the NSA heliocentric redshift.\par

\begin{figure}
\centering
\includegraphics[width=0.49\textwidth]{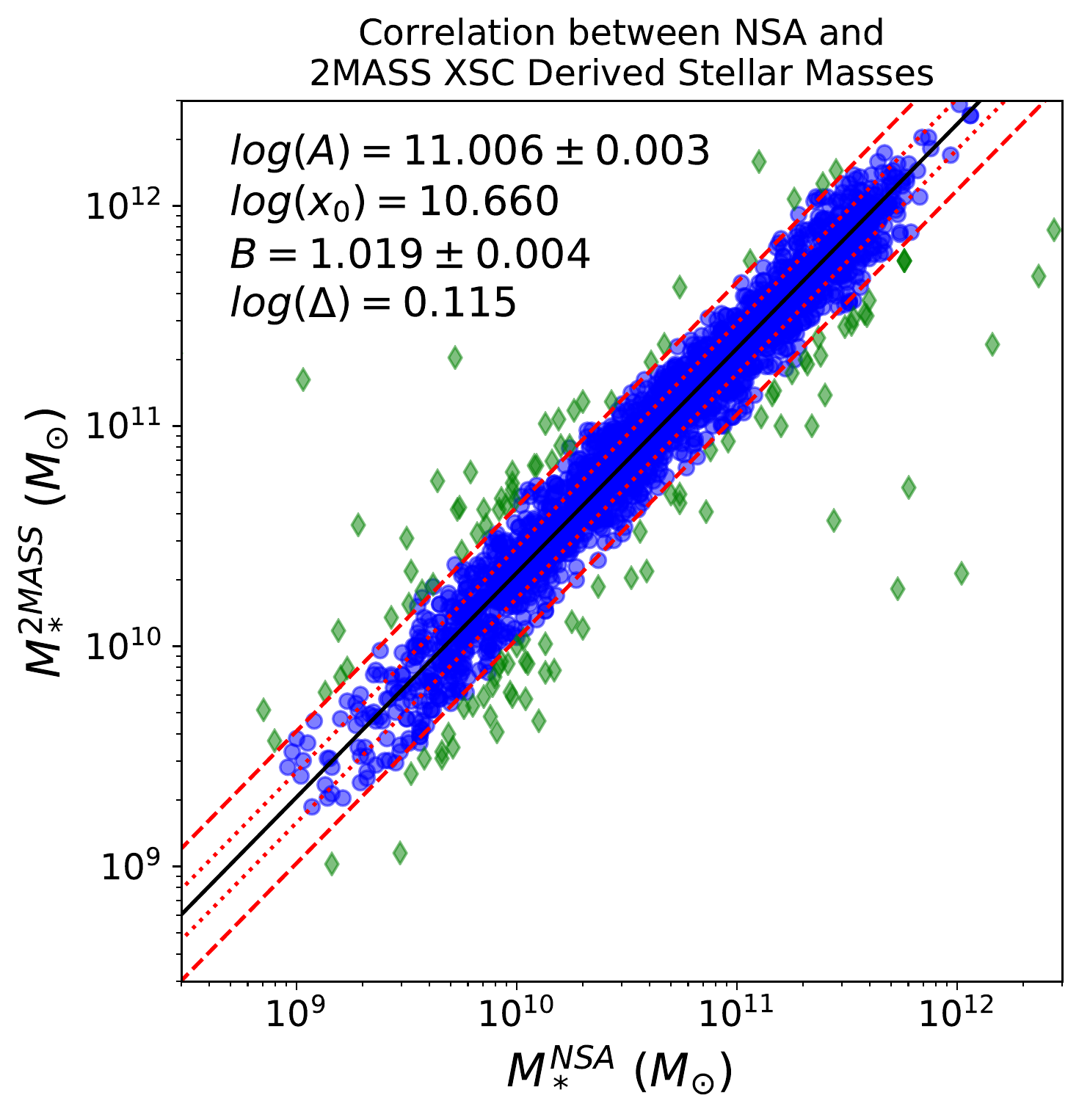}
\caption{Correlation between the stellar mass estimated from apparent $\rm{K}_{\rm{S}}$ band magnitudes from the 2MASS XSC and the stellar mass derived from the K-correction tabulated in the NSA. The red dashed lines indicate the $2.6\Delta$ region where $\Delta$ is the intrinsic scatter. Points which are outside the $2.6\Delta$ region are not included in the fit and are shown as green diamonds.}
\label{fig:masscorr}
\end{figure}

%
%
%
%
%
%
%
In order not to exclude the $\sim$17$\%$ of galaxies that are not in the 2MASS XSC, we estimate the value that would be obtained using \autoref{eq:2MASSmass} assuming the NSA value which we have for all galaxies. We find that there is a tight one-to-one correlation between the two catalogues with a best-fitting relation $M_*^{\rm{2MASS}} = 2.22{{M_*^{\rm{NSA}}}^{1.019}}$ (see \autoref{fig:masscorr}). There is a non-trivial scale factor of 2.22 which is due in part to different assumptions in the initial mass function (IMF). For both the NSA stellar masses and the ATLAS$^{\rm{3D}}$ dynamical modelling, the stellar mass is calculated as a product of the mass-to-light ratio ($M/L$) and the stellar luminosity. In calculating the stellar $M/L$ from the stellar population models, \cite{blanton2007k} assume a Chabrier IMF \citep{chabrier2003stellar}, whereas the masses obtained from dynamical modelling (which are assumed by \cite{cappellari2013effect} to represent the stellar mass) are found to be consistent with an IMF mass normalisation which varies systematically as a function of galaxy $\sigma$ from that of a Chabrier to heavier than a \cite{salpeter1955luminosity} IMF \citep{cappellari2012variation}. Updated trends are given by \cite{posacki2015stellar} and \cite{li2017variation}. For reference, \cite{bernardi2010galaxy} quote a factor ~1.78 between the stellar $M/L$ predicted by a Chabrier and a Salpeter IMF.\par

\subsection{Quality control}
\label{sec:qualcon}
For the \textit{whole sample} of 2722 galaxies, we can in principle calculate $\lambda_{R_e}$ for each one and place them on the ($\lambda_{R_e},\epsilon$) diagram. However, in practice, there are a number of reasons where this is not possible or sensible. In this section, we describe our methods for achieving a \textit{clean sample} of MaNGA galaxies to use in our final results. We list the six criteria which we then expand on in turn. We intentionally leave out the number of affected galaxies to simplify the discussion. Instead, we provide a summary table at the end of the subsection (see \autoref{tab:qualcon}).\par
We exclude galaxies that fall into the following criteria:\par

\begin{enumerate}
\item are identified to be a merger or part of a close pair,
\item are flagged as having bad or unclassifiable kinematics or morphology,
\item have $R_e^{\rm{maj}}$ small enough that $\sigma_{\rm{PSF}}/R_e^{\rm{maj}}>1$,
\item have more than 10\% of spaxels within the half-light ellipse flagged by the MaNGA DRP as DONOTUSE,
\item have been classified as a FR but have a misalignment between the kinematic and photometric major axes that is greater than 30$\degree$ (except for when $\epsilon<0.4$, see Appendix \ref{app:misalignFR}),
\item have more than 50\% of pixels within the half-light ellipse with $\sigma_{\ast}=0$ where $\sigma_{\ast}$ is the intrinsic velocity dispersion (except for disks, see Appendix \ref{app:zerodisp}).
\end{enumerate}

We exclude galaxies identified to be mergers or in close pairs as often the half-light ellipse is difficult to define photometrically. If the galaxies are interacting through a merger process and are close enough to lie within the field of view of the IFU, then the velocity and velocity dispersion maps are usually too chaotic to measure a meaningful value of $\lambda_{R_e}$. Apart from mergers or close pairs, there are some galaxies for which the kinematic structure does not conform to the categories outlined in \autoref{sec:kinprop}. These are not the border cases between the adopted kinematic classifications, but are kinematics which are affected by foreground stars for example. We also perform our own morphological classification using the SDSS true colour image for each galaxy in the MaNGA sample according to the Hubble scheme, i.e. ellipticals, S0s, spirals and irregular galaxies \citep{hubble1926extragalactic, hubble1936realm, sandage1961hubble}. We distinguish edge-on spirals from S0s as having a dust lane that extends across the disk. We also separate face-on S0s from ellipticals by considering the galaxy edge. If the edges are (well) defined, then we classify the galaxy as an S0. If the surface brightness decreases to the sky level, then we classify the galaxy as an elliptical. We exclude galaxies where the classification is ambiguous or not possible.\par
In \autoref{sec:beamsmearingsummary}, we present a correction to account for the distortion due to atmospheric seeing. The correction is largest for small galaxies with $R_e^{\rm{maj}} \approx \sigma_{\rm{PSF}}$ where $\sigma_{\rm{PSF}}$ is the width of the Gaussian PSF of the 2.5-metre Sloan telescope. The correction becomes inaccurate for galaxies small enough such that $R_e^{\rm{maj}} < \sigma_{\rm{PSF}}$. We also exclude galaxies where more than 10$\%$ of spaxels within the half-light ellipse have been flagged by the MaNGA DRP with a MANGA\_DRP3PIXMASK value of 1024, indicating that a particular spaxel is unsuitable for science (see Table B11 of \citealp{law2016data}). The MANGA\_DRP3PIXMASK also contains masks indicating dead fibres, lack of coverage and foreground stars, but these are not sufficient to render a spaxel unsuitable for science purposes.\par
Regarding criterion (v), FRs are known to be aligned to within 10$\degree$ whereas SRs are naturally misaligned \citep{emsellem2007sauron, krajnovic2011atlas3d}. Here, we define SRs to be galaxies that lie within the region where $\lambda_{R_e}<0.08+\epsilon/4$ and $\epsilon<0.4$ (C16, Equation 19). The requirement that SRs should be rounder than $\epsilon=0.4$ is useful to exclude counter-rotating disks that are physically related to FRs (section 3.4.3 of C16) but also have low values of $\lambda_{R_e}$. In Appendix \ref{app:misalignFR}, we show that for FRs rounder than $\epsilon = 0.4$, our measurement of $\lambda_{R_e}$ within a randomly orientated half-light ellipse is underestimated at most by 0.1 when compared to an ellipse that is aligned with the major axis. For this reason, we include misaligned FRs with $\epsilon<0.4$.\par
Finally, we exclude elliptical and irregular galaxies where the fraction of pixels within the half-light ellipse with $\sigma_{\ast}=0$ is greater than 50\%. We choose this cut based on the simple test described in Appendix \ref{app:zerodisp}, where we show that for a reasonable estimate of the noise in the instrumental dispersion, the intrinsic value of $\lambda_{R_e}$ can be recovered to within 0.1 provided that the maximum velocity and velocity dispersion are $\gtrsim 50$ and $\gtrsim 25$ km s$^{-1}$ respectively. As we assume the noise in the instrumental dispersion to be random with a Gaussian distribution, the maximum fraction of pixels with $\sigma_{\ast}=0$ is 50\%, and hence we exclude galaxies without disks where we observe a higher fraction.

Having applied these steps to the \textit{whole sample}, we arrive at a \textit{clean sample} containing all the surviving galaxies. These galaxies have complete photometric and kinematic data as well as morphological and kinematical classifications. We give the number of galaxies removed due to each criterion in \autoref{tab:qualcon}. For MPL-5, 2286 galaxies remain forming the $\textit{clean sample}$ which we use in all results unless specified otherwise.\par

\begin{table}
\centering
\caption{The number of galaxies not included in the \textit{clean sample} due to the reasons outlined in \autoref{sec:qualcon}. Column 1 lists each criteria according to the list found at the beginning of \autoref{sec:qualcon}, and column 2 list the total number of galaxies that have been flagged due to each criterion. Column 3 lists the number of galaxies not already excluded due to previous criteria and column 4 lists the cumulative number of galaxies excluded.}
\label{tab:qualcon}
\begin{tabular}{cccc}
\toprule
Criterion & \begin{tabular}[c]{@{}c@{}}Total No. of \\ Galaxies\end{tabular} & \begin{tabular}[c]{@{}c@{}}No. Not Already \\ Excluded\end{tabular} & \begin{tabular}[c]{@{}c@{}}Cumulative \\ No.\end{tabular} \\ \midrule
(i) & 171 & 0 & 171\\
(ii) & 106 & 96 & 267\\
(iii) & 4 & 3 & 270\\
(iv) & 109 & 82 & 352\\
(v) & 97 & 56 & 408\\
(vi) & 62 & 28 & 436\\
\bottomrule
\end{tabular}
\end{table}

\subsection{Volume correction and kernel smoothing}
\label{sec:volumecorrection}
The MaNGA sample is designed to have a roughly flat distribution in absolute i-band magnitude, M$_i$, as a proxy for stellar mass. However, as detailed in the survey design paper \citep{yan2016sdss}, the range in redshift at which a galaxy with a given M$_i$ can be observed varies systematically with M$_i$. Hence the survey is volume-limited at a given M$_i$. The minimum redshift is set by the angular size of the IFU bundles (the galaxy has to fit onto the detector), and the maximum redshift is set in order to maintain a constant number density of galaxies in each M$_i$ bin \citep{wake2017sdss}. As a result, the brightest, largest galaxies are sampled at a higher redshift than the dimmest, smallest galaxies. In order to correct the MaNGA sample to a volume-limited sample, we weight each galaxy based on the volume it occupies.\par
In order to visualise the effect of correcting the MaNGA sample to a volume-limited sample, we ``smooth" the galaxy distribution using a weighted kernel density estimation (KDE) routine\footnote{We use a modified version (\url{http://nbviewer.jupyter.org/f844bce2ec264c1c8cb5}) of the scipy implementation in Python (\url{http://docs.scipy.org/doc/scipy/reference/index.html}).} (\citealp{scott2015multivariate, silverman1986density}, see the right hand side of \autoref{fig:anisotropy_morph} for an example). Each data point is replaced by a 2D Gaussian, the amplitude of which can be weighted according to the volume correction. At any location on the diagram, the total density is a linear combination of each Gaussian. The result is a 2D probability density field. To eliminate regions of low density extending out to $\infty$, we restrict the smoothed regions within 80\% probability contours, unless otherwise stated.\par

\subsection{Analytic correction to account for atmospheric seeing}
\label{sec:beamsmearingsummary}
In order to measure $\lambda_{R_e}$ accurately, it is necessary to correct for the atmospheric seeing. At the spatial resolution of the PSF, the LOSVD is smeared out such that values measured for $V$ and $\sigma$ are converged towards the mean (zero and $\bar{\sigma}$ for $V$ and $\sigma$ respectively). The overall effect is that the observed value of $\lambda_{R_e}$ is \textit{lower} than the intrinsic value (i.e. without seeing). In this work, we present an approximate analytic correction to account for this effect that can be applied to any dataset. In Appendix \ref{app:beamsmearing}, we discuss in detail the derivation, properties, application and limitations of the correction. Here, we provide a brief summary with all the necessary information required to successfully implement the correction.\par
The correction is derived by convolving simple galaxy models with a Gaussian PSF using the JAM method described in \cite{cappellari2008measuring}, and measuring $\lambda_{R_e}$ for a range of PSF sizes. We find the best-fitting function that describes the observed behaviour with the fewest number of variables: 

\begin{equation} \label{eq:lambdamoffatsersic}
\lambda_{R_e}^{\rm{mod}} = \lambda_{R_e}^{\rm{true}} gM_2 \Bigg( \frac{\sigma_{\rm{PSF}}}{R_e^{\rm{maj}}} \Bigg) f_n \Bigg( \frac{\sigma_{\rm{PSF}}}{R_e^{\rm{maj}}} \Bigg),
\end{equation}
where
\begin{equation}
gM_2 \Bigg( \frac{\sigma_{\rm{PSF}}}{R_e^{\rm{maj}}} \Bigg) = \Bigg[ 1+\Bigg( \frac{\sigma_{\rm{PSF}}/R_e^{\rm{maj}}}{0.47} \Bigg)^{1.76} \Bigg]^{-0.84} \rm{ ,}
\end{equation}
and
\begin{equation}
f_n \Bigg( \frac{\sigma_{\rm{PSF}}}{R_e^{\rm{maj}}} \Bigg) = \Bigg[1 + (n-2) \Bigg(0.26 \frac{\sigma_{\rm{PSF}}}{R_e^{\rm{maj}}} \Bigg) \Bigg]^{-1}\rm{,}
\end{equation}
where $\lambda_{R_e}^{\rm{mod}}$ and $\lambda_{R_e}^{\rm{true}}$ are the model (``observed") and true (``intrinsic") values of $\lambda_{R_e}$, $\sigma_{\rm{PSF}} = \rm{FWHM}_{\rm{PSF}}/2.355$, $R_e^{\rm{maj}}$ is the semi-major axis and $n$ is the Sérsic index. The function $gM_2$ is a generalised form of the Moffat function \citep{moffat1969theoretical}, and $f_n$ is an empirical function required to model the dependence on the Sérsic index.\par

\begin{figure*}
\centering
\includegraphics[width=\textwidth]{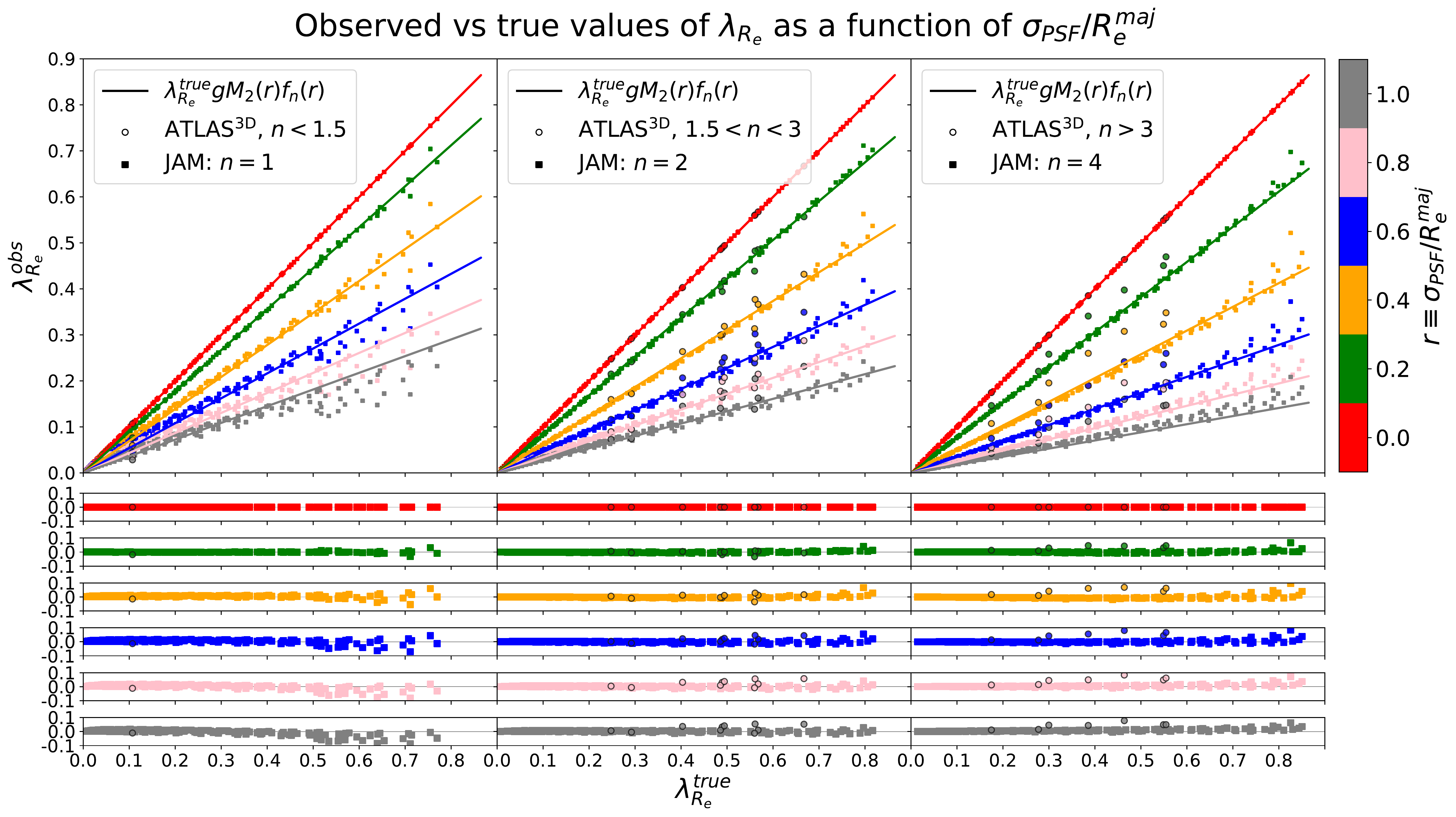}
\caption{The best-fitting generalised Moffat (gMoffat) function describing how $\lambda_{R_e}$ changes with increasing \ratio for Sérsic indices $n=1$, $n=2$ and $n=4$ (left to right). The gMoffat fit is shown as the solid lines for values of $\sigma_{\rm{PSF}}/R_e^{\rm{maj}}=0,0.2,0.4,0.6,0.8,1$ indicated by the colourbar. The points are the data for which the model is fit at the same ratios. ATLAS$^{\textrm{3D}}$ galaxies are shown by the large circles whereas the JAM models are shown by small squares. The six lower panels for each column indicate the residuals (data - model) for each ratio, where the model is shown as the grey horizontal line at $y=0$.}
\label{fig:moffat}
\end{figure*}

At constant \ratio and $n$, \autoref{eq:lambdamoffatsersic} is a linear function of $\lambda_{R_e}^{\rm{true}}$. Six linear slices for $n=1\textrm{, }2\textrm{ and }4$ are shown in \autoref{fig:moffat}. The residuals between the JAM data and the correction are largely due to inclination effects. We supplement the JAM data with a subsample of 18 galaxies from the ATLAS$^{\textrm{3D}}$ survey, which are assumed to be unaffected by seeing.\par
The correction is applicable only for regular rotators where $R_e^{\rm{maj}} \geq \sigma_{\rm{PSF}}$ (as described in \autoref{sec:qualcon}) and $0.5 \leq n \leq 6.5$. In \autoref{sec:results}, we apply our correction to observed values of $\lambda_{R_e}$. For a given galaxy, the intrinsic $\lambda_{R_e}$ corrected for seeing is given by solving \autoref{eq:lambdamoffatsersic} for $\lambda_{R_e}^{\rm{true}}$. The error in the value of $\lambda_{R_e}^{\rm{true}}$ is given by $[+0.03(\sigma_{\rm{PSF}}/R_e^{\rm{maj}}), -0.08n(\sigma_{\rm{PSF}}/R_e^{\rm{maj}})]$ (see \autoref{fig:profileresiduals}). In the case that $\lambda_{R_e}^{\rm{true}}-0.08n(\sigma_{\rm{PSF}}/R_e^{\rm{maj}})<\lambda_{R_e}^{\rm{obs}}$, which can happen for low $\lambda_{R_e}^{\rm{obs}}$, then the value of $\lambda_{R_e}^{\rm{obs}}$ itself provides the lower limit. We find that for regular rotators in the \textit{clean sample}, the mean error is $[0.005, -0.041]$ and the median error is $[0.004, -0.027]$.\par
To facilitate the application of our correction, we provide a \texttt{Python} code that calculates Equations \ref{eq:lambda_R} and \ref{eq:lambdamoffatsersic} as well as Equation \ref{eq:sigma} for stellar kinematic data. The code is available at this address: \url{https://github.com/marktgraham/lambdaR_e_calc}.


%
\section{Results and discussion}
\label{sec:results}
Here we present our results for the 2286 galaxies in the \textit{clean sample}. We present the $(\lambda_{R_e}, \epsilon)$ diagram both with and without the beam correction and classified by kinematic morphology, Hubble morphology, stellar mass and $\sigma_e$. We also show the kinematic misalignment as a function of $\epsilon$, including some galaxies that are not in the \textit{clean sample}. Finally, we present the mass-size relation for the \textit{clean sample}, classifying for Hubble and kinematic morphology as well as the fast/slow classification. All the quantities required to reproduce these plots are given in \autoref{tab:galaxydata}.

\begin{figure*}
\centering
\subfigure{\includegraphics[width=0.49\textwidth]{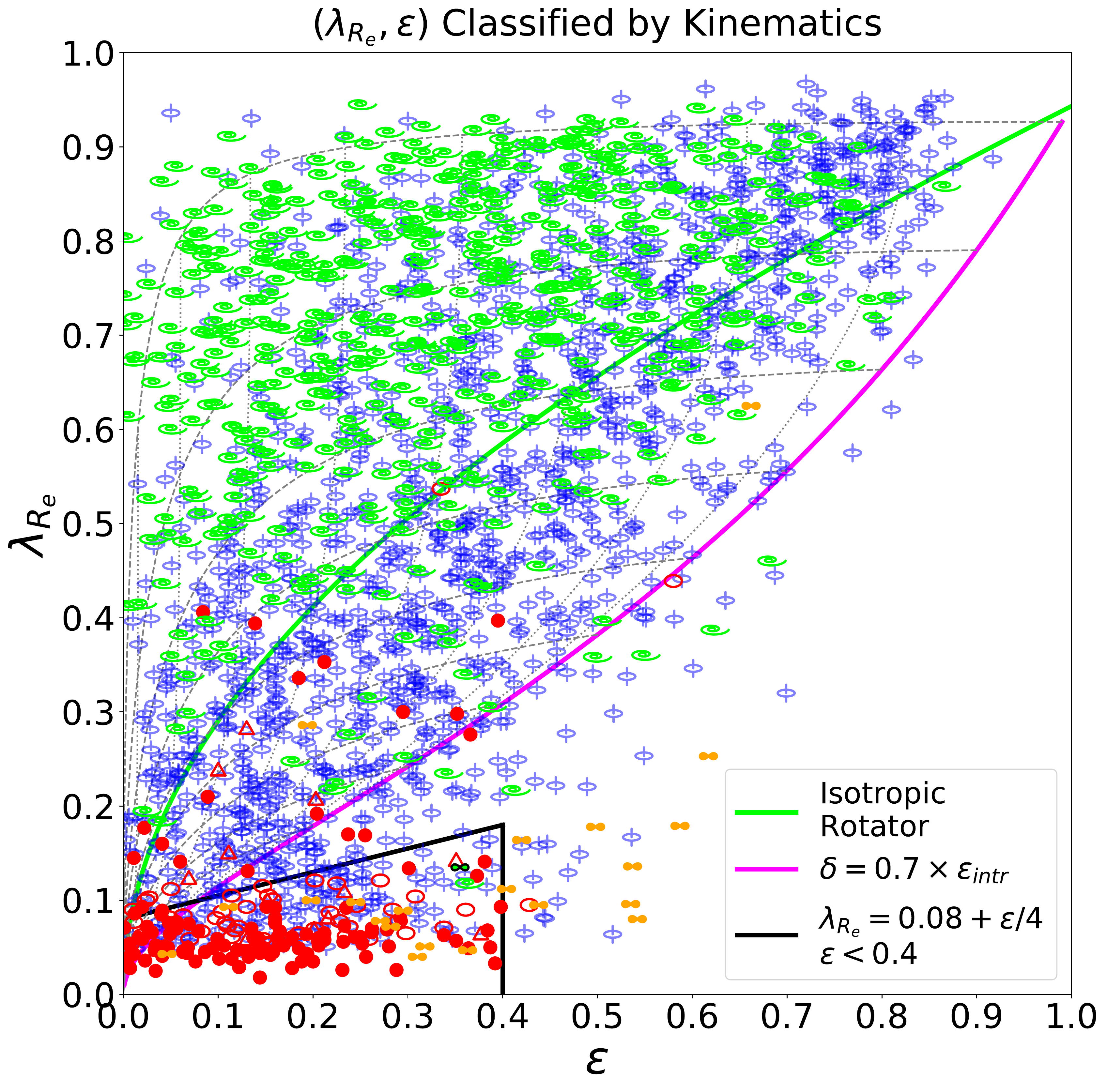}}\quad
\subfigure{\includegraphics[width=0.49\textwidth]{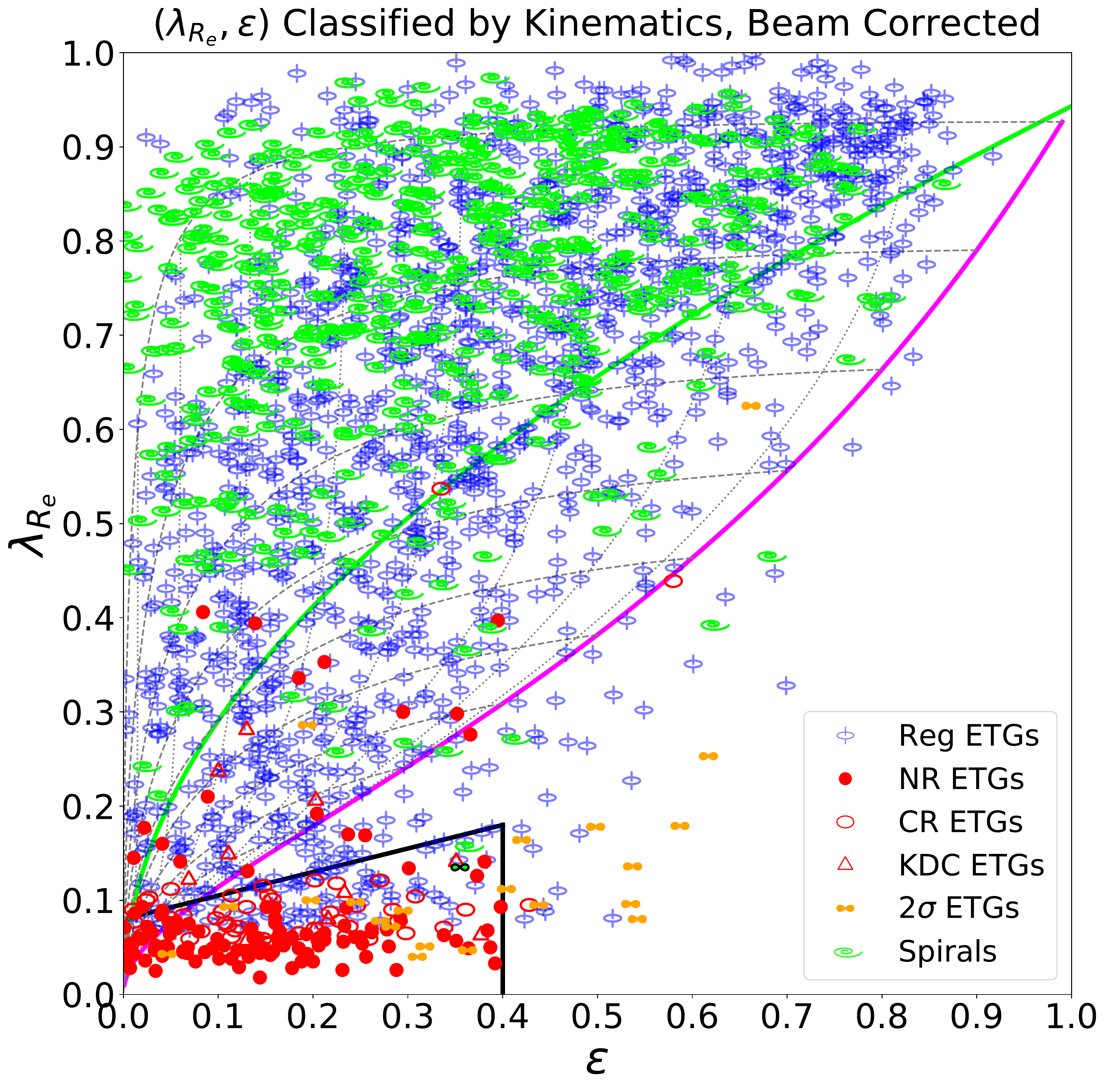}}
\caption{\textbf{Left:} The $(\lambda_{R_e}, \epsilon)$ diagram labelled by kinematic morphology. The points are the \textit{observed} $(\lambda_{R_e}, \epsilon)$ values for the 2286 galaxies that form the \textit{clean sample}. We label the ETGs according to the kinematic classes detailed in \autoref{sec:kinprop}, where ``Reg'' = regular rotator, ``NR'' = non-rotator, ``CR'' = complex rotator, ``KDC'' = kinematically decoupled core and ``2$\sigma$'' = counter-rotating disk. All spirals (apart from MaNGA-ID 1-236144, see text) are regular rotators and so they are not labelled kinematically. MaNGA-ID 1-236144 is shown as a lime-green 2$\sigma$ galaxy with black edges. The thick green line is the prediction for an edge-on ($i=90\degree$) isotropic rotator from \protect\cite{binney2005rotation} (see Equation 14 in C16), and the magenta line is the edge-on relation from \protect\cite{cappellari2007sauron} (see Equation 11 in C16). The thin dotted lines show how the magenta line changes at different inclinations ($\Delta i = 10\degree$), while the black dashed lines trace how galaxies with a particular value of $\epsilon_{\rm{intr}}$ at $i=90\degree$ move across the diagram with changing inclination. Finally, the black lines at the lower-left corner define the region occupied by SRs: $\lambda_{R_e}<0.08+\epsilon/4$, $\epsilon<0.4$ (see Equation 19 in C16). \textbf{Right:} The same as left except that the beam correction has been applied to regular ETGs and spirals using \autoref{eq:lambdamoffatsersic}. Out of 2286, 15 galaxies have a value of $\lambda_{R_e}^{\rm{true}}>1$ and hence 2271 galaxies remain in this version of the diagram.}
\label{fig:anisotropy_kine}
\end{figure*}

\subsection{Stellar angular momentum}
\label{sec:angmom}
The MaNGA survey is designed such that each internal data release (MPL-5 etc.) contains observations for a representative subset of the complete sample of $\sim$10,000 galaxies. This means that any initial results we draw from a subset of $\sim2300$ galaxies should be indicative of the full sample. \autoref{fig:anisotropy_kine} shows the $(\lambda_{R_e}, \epsilon)$ diagram classified by kinematic morphology. Each point is a single galaxy in the \textit{clean sample}. There are some galaxies for which we have more than one kinematic observation with different IFUs. For these cases, we plot the $\lambda_{R_e}$ value derived from the kinematic maps which have the lowest error in $\Psi_{\rm{kin}}$ (or in some cases the lowest misalignment if the error is the same).\par
We apply our beam correction given in \autoref{sec:beamsmearingsummary} to values of $\lambda_{R_e}$ for regular rotator ETGs and spirals, shown right in \autoref{fig:anisotropy_kine}. For the rest of this paper, we consider the beam-corrected version to be the ``true" diagram. We find the mean and median increase in $\lambda_{R_e}$ due to the beam correction for regular rotator ETGs and spirals to be 0.09 and 0.07 respectively.\par
There are 15 galaxies that, once beam-corrected, overshoot $\lambda_{R_e} = 1$ and therefore are not shown in the right-hand side of \autoref{fig:anisotropy_kine}. The beam corrected values are affected by errors in $\lambda_{R_e}^{\rm{obs}}$, $\sigma_{\rm{PSF}}$, $R_e^{\rm{maj}}$ (i.e. $R_e$ and $\epsilon$), and $n$. \cite{yan2016sdss} estimated the uncertainty on $\lambda_{R_e}^{\rm{obs}}$ by generating 100 random normal distributions for $V$ and $\sigma$ according to the measurement errors on both quantities, and computed $\lambda_{R_e}$ for each (see Section 8.4 and Figure 29). For $\lambda_{R_e} \sim 0.1$, the error is within 0.05 but can be as high as $\sim0.08$ for $\lambda_{R_e}>0.5$. The error in $\sigma_{\rm{PSF}}$ is less than 10\% \citep{law2016data} and it is possible that the Sérsic index can have systematic uncertainties of $\pm2$. Finally, as described in \autoref{sec:2mass} and shown in \autoref{fig:mgere}, effective radii are poorly defined and can easily have uncertainties of a few arcseconds. Hence, a combination of these errors can result in large errors in the beam-corrected $\lambda_{R_e}$ which are impossible to quantify. A value of $\lambda_{R_e} > 1$ can only be as a result of these errors and hence we remove these galaxies.\par
Before this work, there was much debate on whether there is a continuum of galaxy properties on the $(\lambda_{R_e}, \epsilon)$ plane, or if there is a dichotomy with a break between slow and fast rotators. A dichotomy was claimed using dynamical masses (Figure 11 of C16), but the same was not apparent in the $(\lambda_{R_e}, \epsilon)$ diagram itself. Thanks to the large number statistics provided by MaNGA, we are able for the first time to provide unambiguous evidence of a bimodality between fast and slow rotators, which can be seen in the right-hand side of \autoref{fig:anisotropy_kine}.\par
We compare the right-hand side of \autoref{fig:anisotropy_kine} with the equivalent diagram in Figure 13 of C16, which plots values for 340 ETGs from ATLAS$^{\rm{3D}}$ \citep{emsellem2011atlas3d} and the SAMI Pilot survey \citep{fogarty2015sami}. In both diagrams, the non-regular rotators cluster in the SR region (indicated by the black lines), while the majority of regular ETGs and, in our case nearly all spirals, occupy the FR regime, with almost all of the population lying above the empirical relation for edge-on galaxies (magenta line). In our work, there are a small number of regular ETGs in the SR region which are not seen in C16. Although these galaxies have ordered rotation, the maximum rotation velocity within the half-light ellipse is $\sim30-50$ km s$^{-1}$. Since $\sigma$ is high (see \autoref{fig:anisotropy_sigma_e}), these galaxies have a low value of $\lambda_{R_e}$. While most non-rotators lie within the SR regime, a small population lies in the FR regime. Many of these galaxies, especially above $\lambda_{R_e} \sim 0.3$, are either face on disks with no overall rotation, or irregular galaxies. In some of these cases, $\lambda_{R_e}$ has been inflated due to the low dispersion found in these galaxies (see Appendix \ref{app:zerodisp}). Another effect to consider is the standard $|V|$ effect in the summation over pixels when calculating $\lambda_{R_e}$ as illustrated in Figure B1 of \cite{emsellem2007sauron}. In theory, a perfectly noisy velocity map should result in $\lambda_{R_e}=0$ if the sign of $V$ were taken into account. However, the absolute $V$ in \autoref{eq:lambda_R} prevents $\lambda_{R_e}$ lying below about 0.025. It is likely that this effect contributes to the high $\lambda_{R_e}$ seen in these galaxies. The counter-rotating ETGs occupy similar regions in both diagrams with the exception of one at $\lambda_{R_e} \sim 0.6$. We note that these galaxies are more difficult to recognise in MaNGA than they were in ATLAS$^{\rm{3D}}$ due to the lower spatial resolution (both due to larger distances and to larger fibres compared to smaller lenslets), and this implies that some will be visually classified as regular rotators.\par
We note that our kinematic classification applies only to ETGs and not to spirals as, by nature, all spirals are regular rotators. However, we find one galaxy (MaNGA-ID 1-236144, Plate-IFU 8980-12703) that has a clear dust lane indicative of spiral structure, but also has a discernible counter-rotating core, suggesting that this galaxy is in fact a 2$\sigma$ galaxy. It is possible that the dust lane is in fact gas accreted after formation, and may itself be counter rotating. Furthermore, this galaxy appears in the SR region. However, as the galaxy is edge-on, it is impossible to rule out spiral structure.\par
\begin{figure*}
\centering
\includegraphics[width=0.686\textwidth]{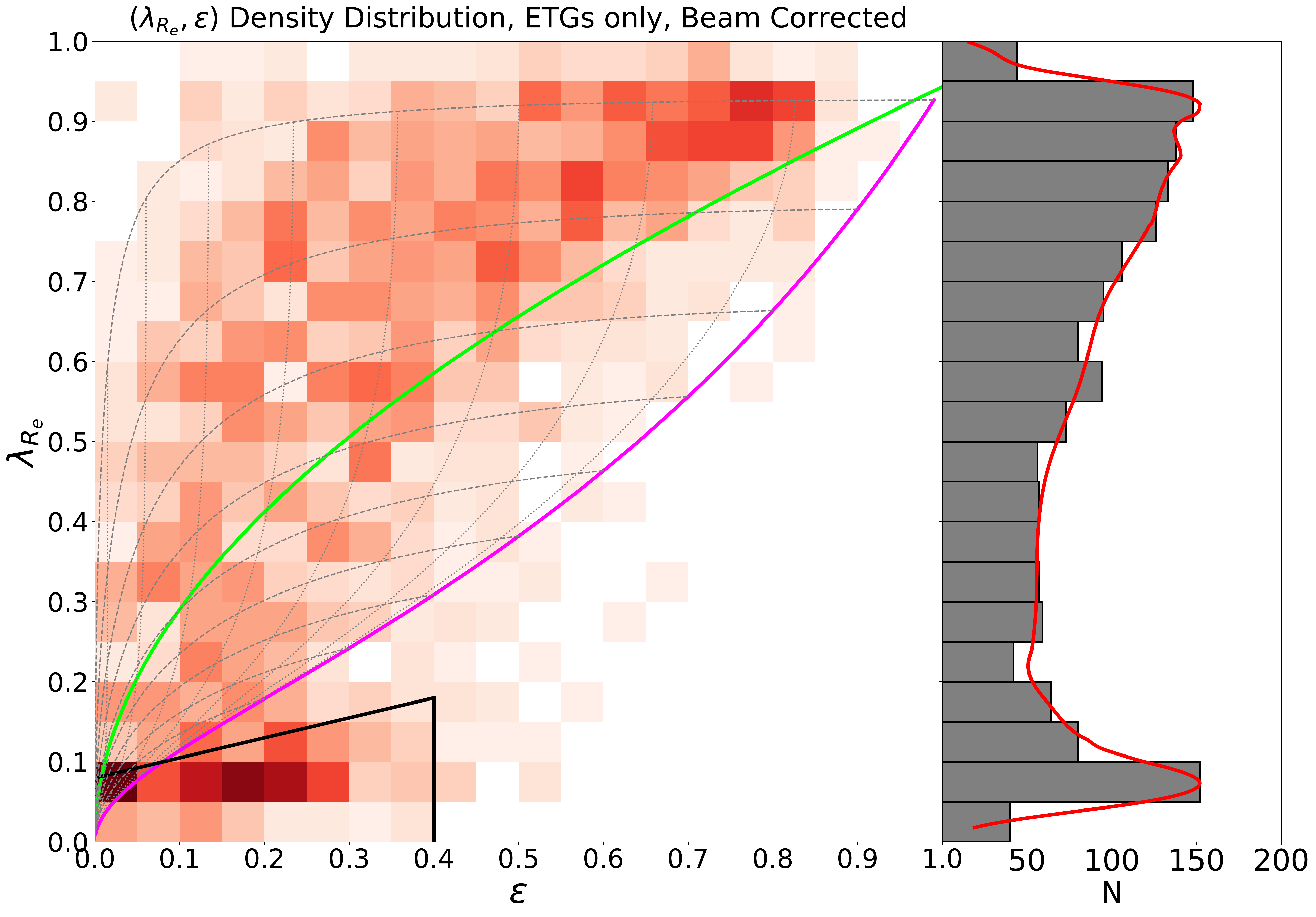}
\caption{\textbf{Left:} The beam-corrected $(\lambda_{R_e}, \epsilon)$ diagram for ETGs only colour coded by the number of points in a grid with cells of size 0.05$\times$0.05. Dark red indicates a high density of points while pale red indicates a low density of points. The black, magenta, green and dashed lines are the same as in \autoref{fig:anisotropy_kine}. \textbf{Right:} A histogram of $\lambda_{R_e}$ collapsed along the $x$-axis. The red curve is a one-dimensional smoothing of the individual data points using an adaptive KDE routine.}
\label{fig:anisotropy_density}
\end{figure*}
\begin{figure*}
\centering
\includegraphics[width=0.686\textwidth]{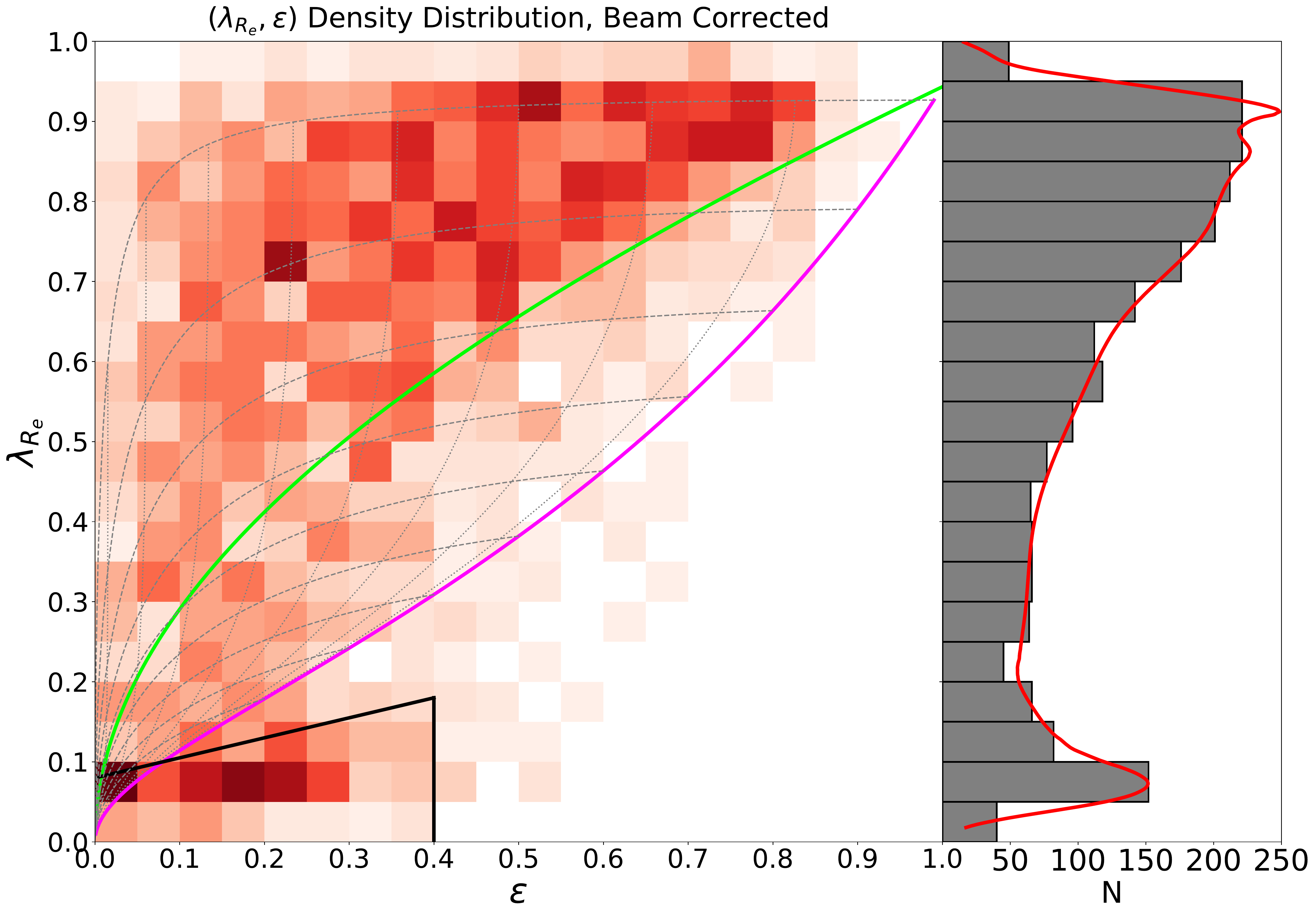}
\caption{The same as \autoref{fig:anisotropy_density} except that spirals are included.}
\label{fig:anisotropy_density_spirals}
\end{figure*}

While there are more outliers in \autoref{fig:anisotropy_kine} than can be seen in C16, their small number compared to the size of the sample allows us to treat these galaxies as noise. In order to highlight the bimodality without first classifying by any galaxy property, we split the $(\lambda_{R_e}, \epsilon)$ diagram into a grid with cells 0.05$\times$0.05 in size. We then colour-code each cell by the number of points within that cell as shown in \autoref{fig:anisotropy_density} for ETGs only. We only show ETGs at first as this is the class of galaxies that visually appear homogeneous, and hence require IFS observations in order to separate them in terms of angular momentum. We include a histogram of $\lambda_{R_e}$ with bin widths of 0.05. We also apply an adaptive KDE routine to the original unbinned values (i.e. not the histogram) which allows for local variations in the kernel size to give a smoothed 1D distribution. Using an adaptive kernel is advantageous in that it preserves density information on a range of scales. We use the Python implementation\footnote{\url{https://github.com/cooperlab/AdaptiveKDE}} of the method outlined in \cite{shimazaki2010kernel}. All three measures of density agree qualitatively with the conclusion that slow and fast rotators form distinct galaxy populations on the $(\lambda_{R_e}, \epsilon)$ diagram with local maxima within and outside the SR region respectively. However, there is a non-zero density of points at $\lambda_{R_e} \sim 0.3$ which constitutes a minimum between the two distributions. For this reason, the observed bimodality does not necessarily imply a dichotomy, which is however indicated by the dynamical models (see C16). The bimodality in $\lambda_{R_e}$ does correspond to a bimodality in the intrinsic $\epsilon_{\rm{intr}}$ (see \citealp[Figure 8]{weijmans2014atlas} and \citealp[Figure 5]{foster2017intrinsic}) which is unavailable to us without accurate inclinations. In \autoref{fig:anisotropy_density_spirals}, we include spirals and irregular galaxies which can be distinguished by visual morphology from ETGs.\par
We note in passing that there is a lower density of points in the region bound by the green and magenta lines for $\epsilon>0.4$ than has been seen in previous studies such at ATLAS$^{\rm{3D}}$ and CALIFA. It is possible that there could be some circularisation of the ellipticity due to seeing and lower spatial sampling, which in principle should be corrected for, but maybe not fully. As MaNGA galaxies are more distant than ATLAS$^{\rm{3D}}$ and CALIFA, this may explain what we observe. Even if we include all 2722 galaxies in the \textit{whole sample}, we do not see an increase in number density in this region.\par

\begin{figure*}
\centering
\subfigure{\includegraphics[width=0.49\textwidth]{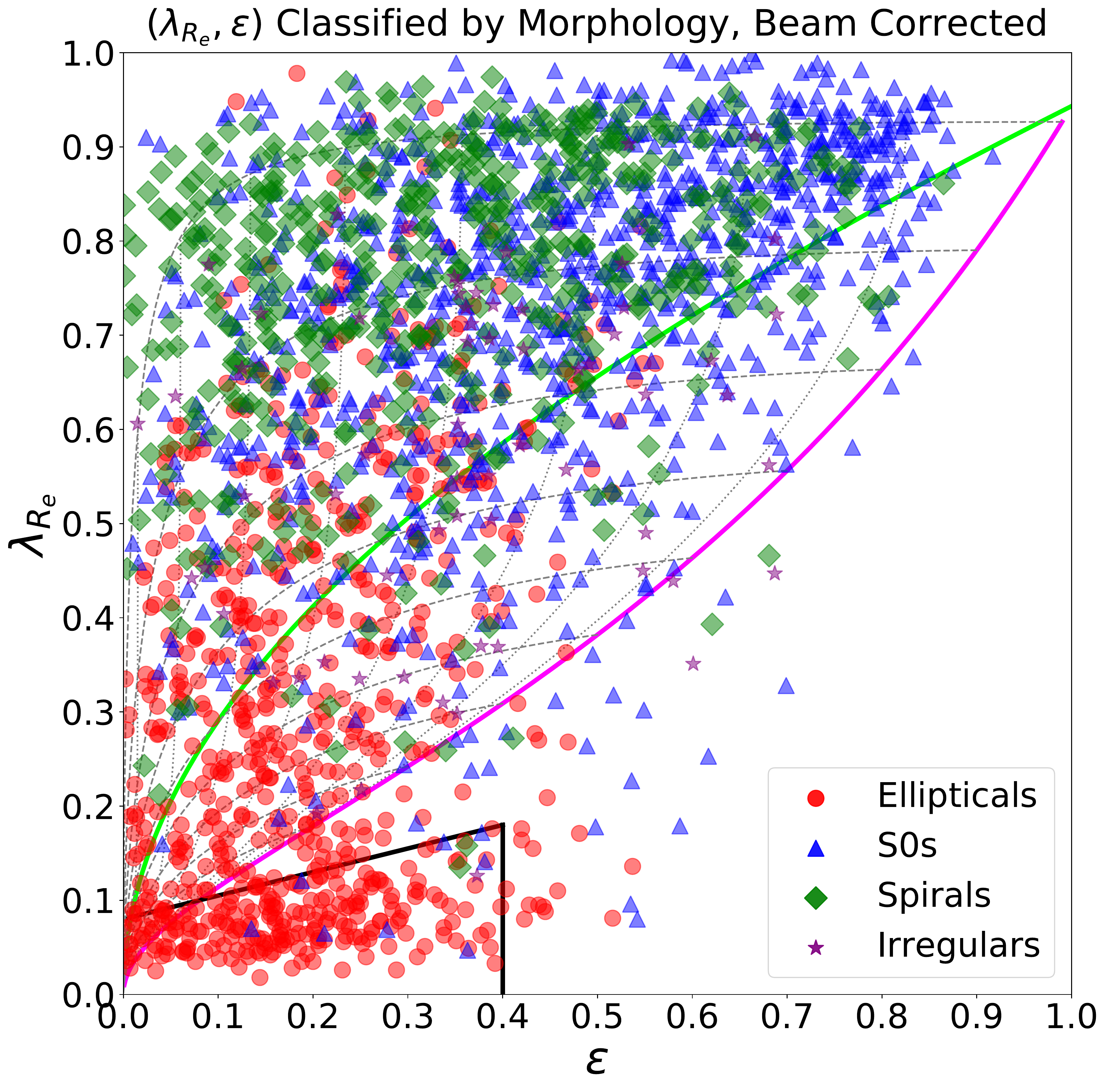}}\quad
\subfigure{\includegraphics[width=0.49\textwidth]{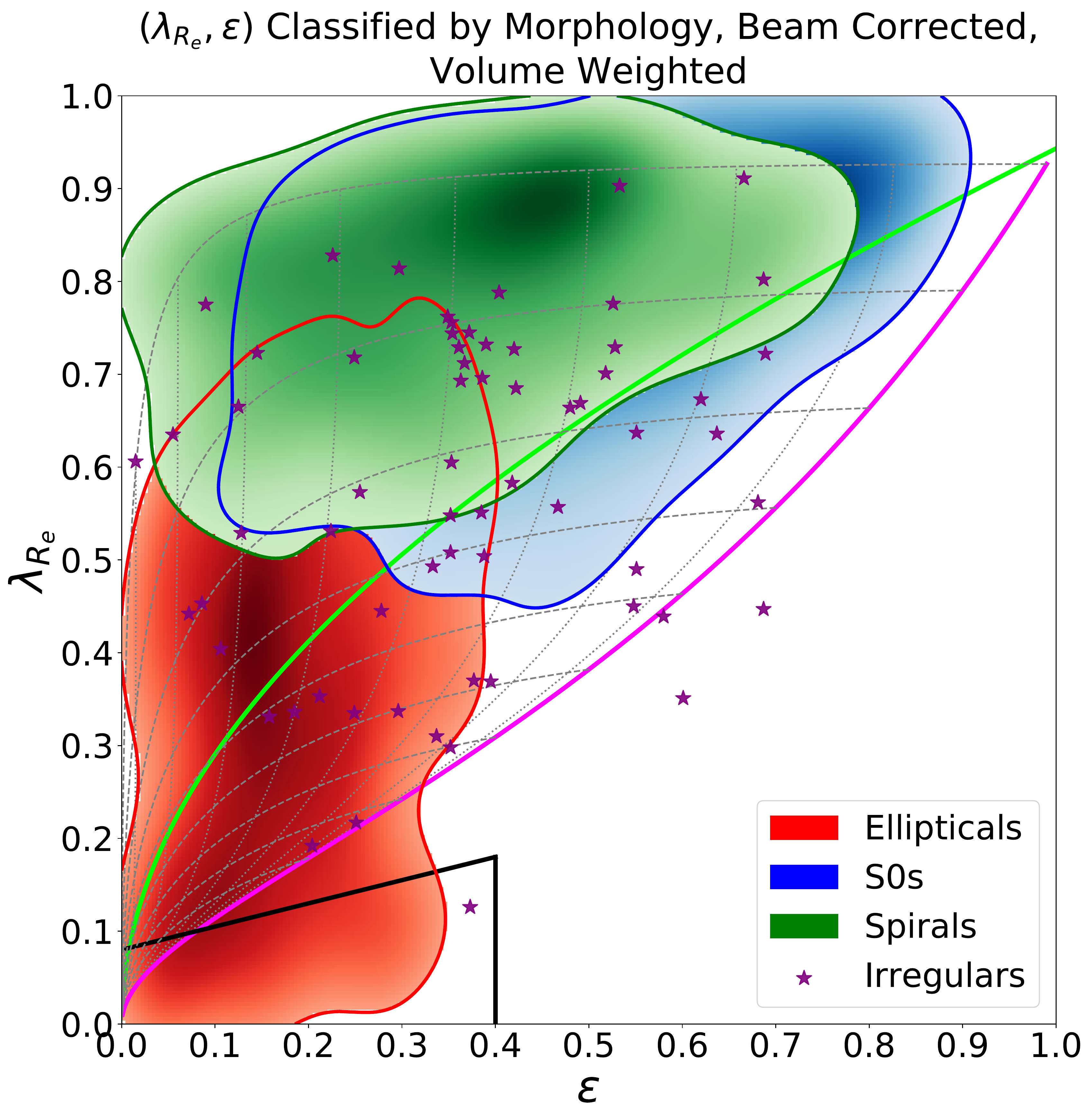}}
\caption{\textbf{Left:} The beam-corrected $(\lambda_{R_e}, \epsilon)$ diagram colour coded by Hubble morphology as indicated. The black, magenta, green and dashed lines are the same as in \autoref{fig:anisotropy_kine}. \textbf{Right:} The same as left except that ETGs and spirals (i.e. excluding irregular galaxies) are smoothed and volume weighted using the method described in \autoref{sec:volumecorrection}.}
\label{fig:anisotropy_morph}
\end{figure*}

\begin{figure*}
\centering
\subfigure{\includegraphics[width=0.462\textwidth]{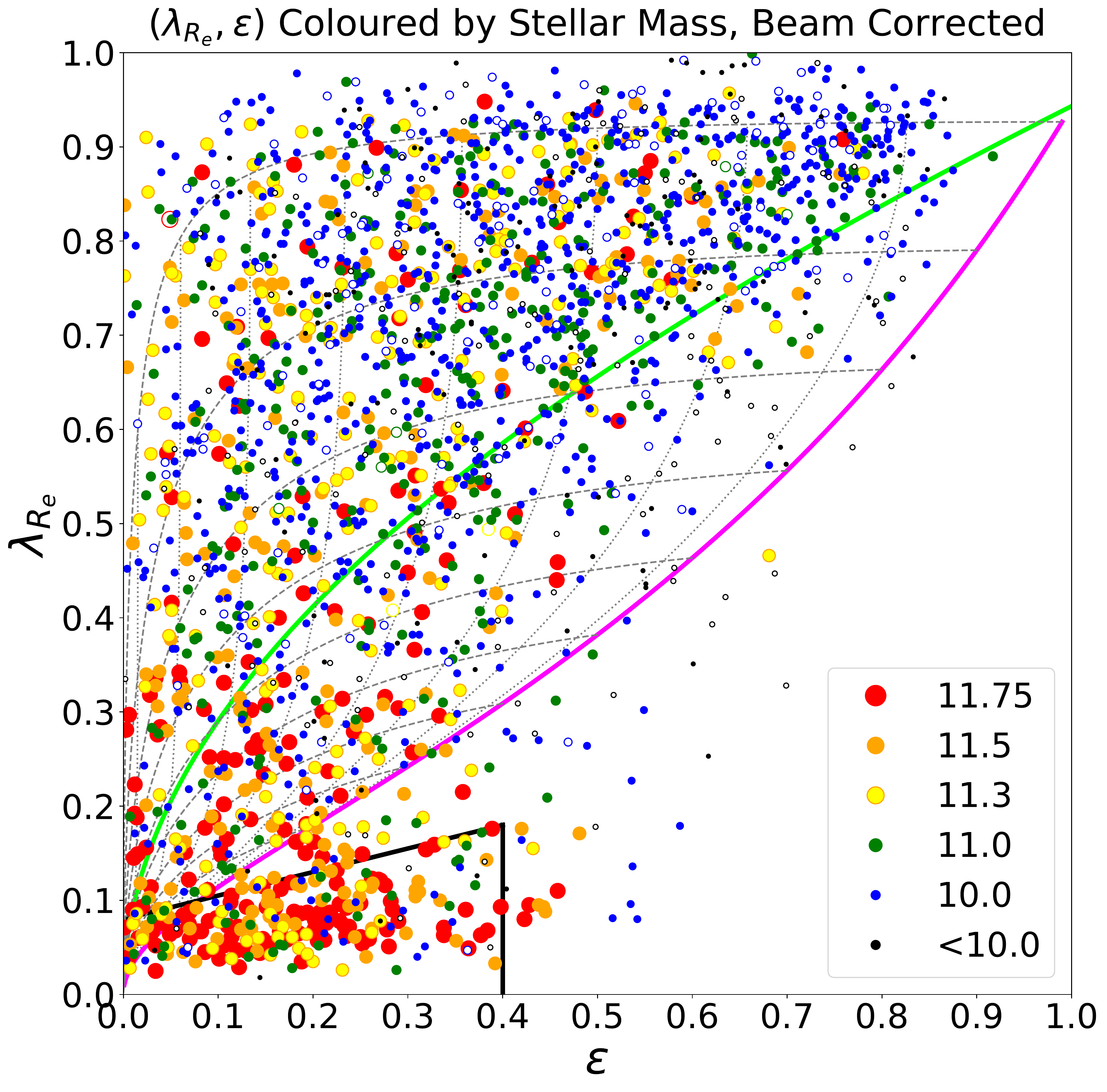}}\quad
\subfigure{\includegraphics[width=0.518\textwidth]{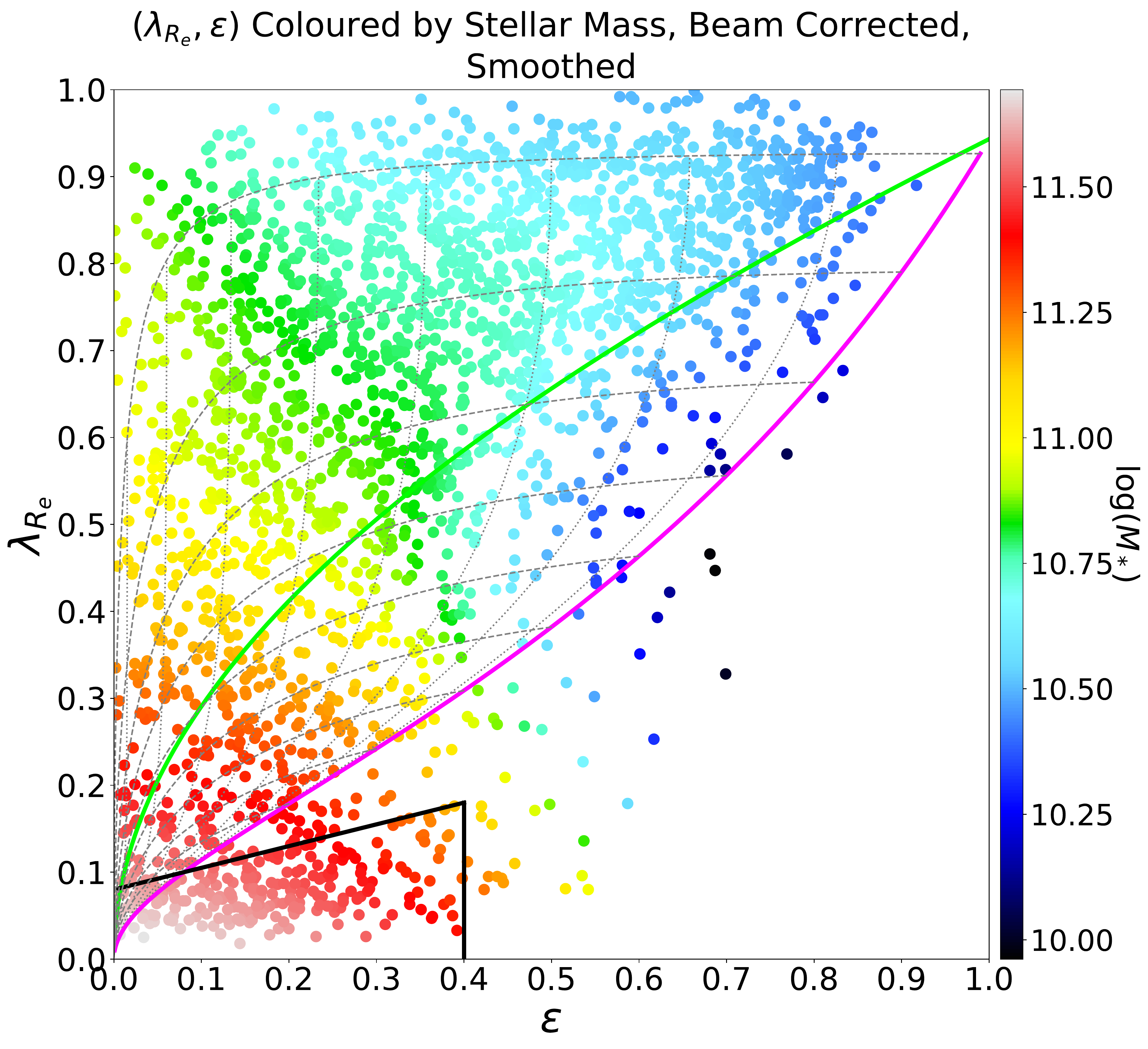}}
\caption{\textbf{Left:} The beam-corrected $(\lambda_{R_e}, \epsilon)$ diagram colour coded by $\log(M_\ast)$. The values for each colour are indicated in $\log(M_{\odot})$. $\log(M_{\ast})=11.3$ corresponds to the critical mass of $2\times10^{11}$ M$_{\odot}$. If a galaxy is found in the 2MASS XSC, the stellar mass is calculated using \autoref{eq:2MASSmass} and the point is shown as a filled circle. Otherwise, we use a best fit to calculate the 2MASS stellar mass from the NSA stellar mass (see \autoref{fig:masscorr}). These galaxies are indicated with a hollow circle and are almost exclusively found in the lowest three mass bins. The black, magenta, green and dashed lines are the same as in \autoref{fig:anisotropy_kine}. \textbf{Right:} The same as left except that all colours have been LOESS smoothed. The colours are indicated by the colourbar.}
\label{fig:anisotropy_mass}
\end{figure*}

\begin{figure*}
\centering
\subfigure{\includegraphics[width=0.467\textwidth]{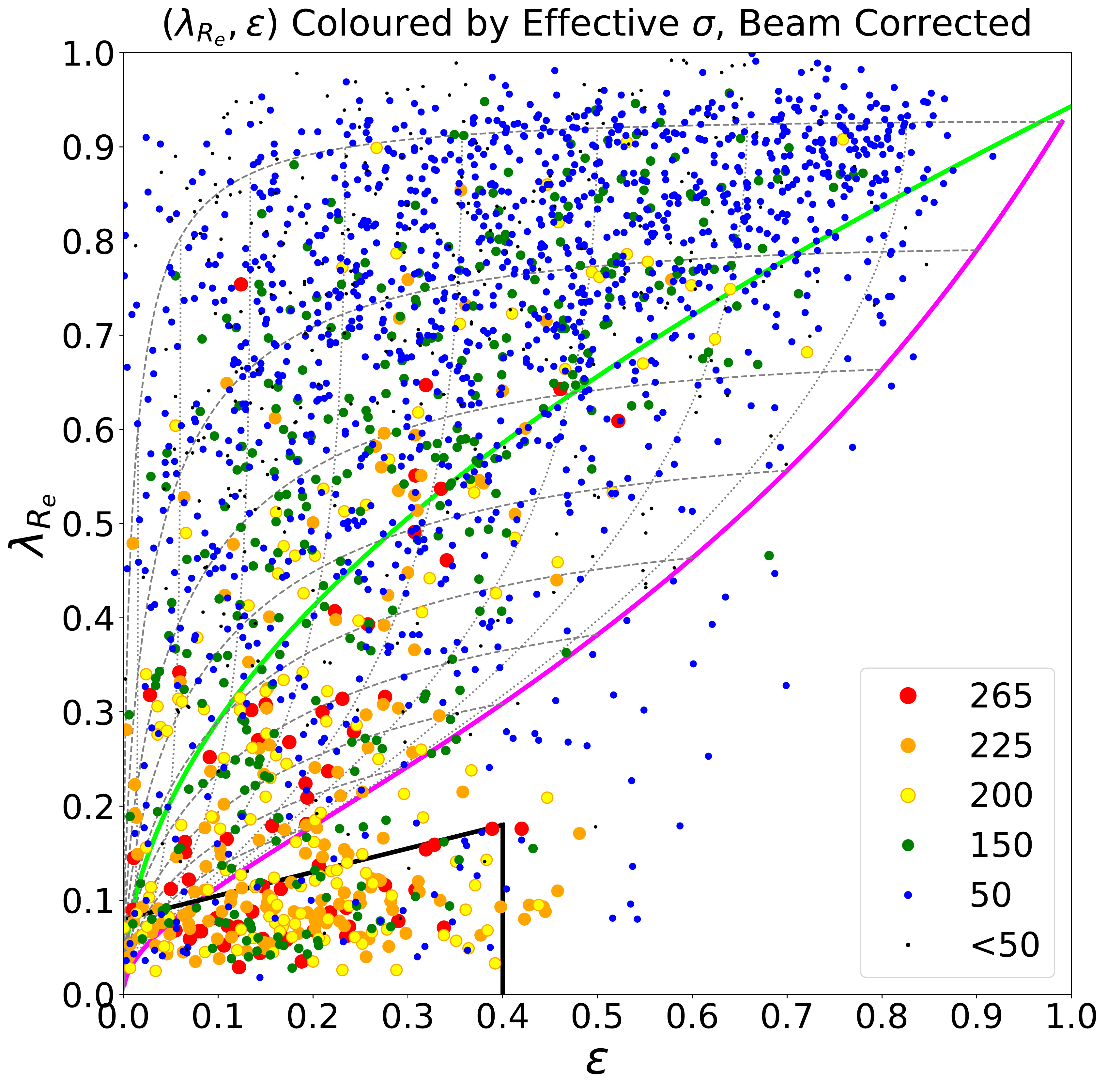}}\quad
\subfigure{\includegraphics[width=0.513\textwidth]{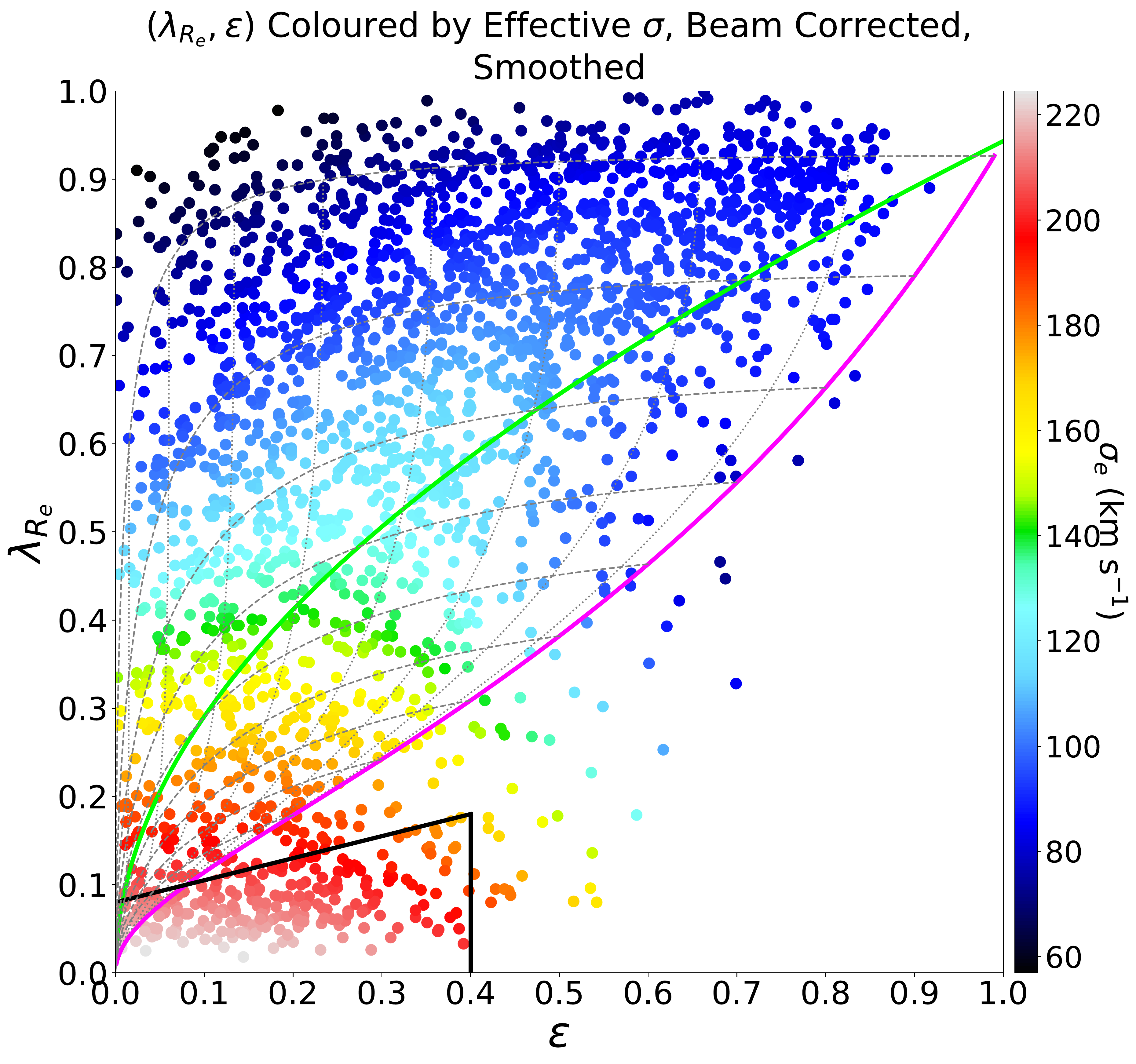}}
\caption{\textbf{Left:} The beam-corrected $(\lambda_{R_e}, \epsilon)$ diagram colour coded by $\sigma_e$ where each bin in $\sigma_e$ is indicated in km s$^{-1}$. The black, magenta, green and dashed lines are the same as in \autoref{fig:anisotropy_kine}. \textbf{Right:} The same as left except that all colours have been LOESS smoothed. The colours are indicated by the colourbar.}
\label{fig:anisotropy_sigma_e}
\end{figure*}


In \autoref{fig:anisotropy_morph}, we show the $(\lambda_{R_e}, \epsilon)$ diagram classified by morphology. In the right-hand side of \autoref{fig:anisotropy_morph}, we smooth the points using a KDE routine, applying the appropriate volume weighting to each data point. In doing this, we recover the distribution as would be obtained from a volume-limited sample. For ellipticals, there is a discernible difference between the position of the peak of the smoothed distribution and the highest density of points in the raw data. This is because the MaNGA selection function, which samples galaxies with a flat distribution in stellar mass, selects a higher fraction of massive SR galaxies than would be observed in a random, volume-limited sample. Hence, massive elliptical SRs are down-weighted. We show both the unweighted and weighted distributions as both are valid representations of the galaxies in the MaNGA sample.\par
We compare \autoref{fig:anisotropy_morph} with Figure 15 of C16 which shows 666 spirals and ETGs taken from different surveys: 260 ETGs from ATLAS$^{\rm{3D}}$ \citep{emsellem2011atlas3d}, 300 (mostly) spirals from CALIFA \citep{falcon2015angular} and $\sim$100 (mostly) ETGs from SAMI \citep{fogarty2015sami}. Elliptical galaxies occupy similar regions in Figure 15 of C16 and \autoref{fig:anisotropy_morph} with a maximum $\lambda_{R_e}$ of 0.5 and $\sim$0.8 respectively, and a maximum ellipticity of 0.4. The strongest concentration of ellipticals in \autoref{fig:anisotropy_morph} is at $(\lambda_{R_e}, \epsilon)=(0.125,0.4)$ which is higher than C16, who finds the peak to be at $(\lambda_{R_e}, \epsilon)=(0.1,0.15)$. The higher peak in our work is a direct effect of the volume weighting as the highest density of points is closer to the SR boundary.\par
While there is good agreement on the distribution of spiral galaxies, there are many spiral galaxies above $\lambda_{R_e}=0.9$ which are not seen in Figure 15 of C16. The lack of spirals with $\epsilon<0.2$ in Figure 15 of C16 is due to the sample selection of CALIFA which omits very round and very flat spirals from their sample \citep{walcher2014califa}. However, the lower and right-hand-side extent of both distributions agree to about 0.1. There are few spirals flatter than $\epsilon=0.75$ because for very flat discs ($\epsilon \sim 0.8$), it is difficult to verify the presence of spiral arms without a clear dust lane present. Therefore, it is possible that many edge-on spirals will be misclassified as S0s. ATLAS$^{\rm{3D}}$, CALIFA and SAMI are all at a lower redshift than MaNGA ($z<0.01$, $z<0.03$ and $z<0.095$ respectively) and so are less likely to suffer from the same bias. In our work, the peak(s) of the distribution are at slightly lower $\epsilon$ and higher $\lambda_{R_e}$. We find that many galaxies with a beam corrected $\lambda_{R_e}>0.9$ have a high fraction of low $\sigma$ pixels, with a mean of 32$\%$. However, as many of these intrinsically have low $\sigma$, we are confident in their $\lambda_{R_e}$. There is less agreement between the right-hand side of \autoref{fig:anisotropy_morph} and Figure 15 of C16 for S0s. In our work, we find S0s have about 0.3 higher $\lambda_{R_e}$ on average and are flatter by about 0.1 in $\epsilon$. Irregular galaxies are randomly distributed throughout the diagram, but all lie in the FR region. \autoref{fig:anisotropy_morph} agrees qualitatively with \cite{cortese2016sami} who found that $\lambda_{R_e}$ is on average lower for elliptical galaxies than spirals and S0s. \par


The $(\lambda_{R_e}, \epsilon)$ diagram coloured by intervals of stellar mass is shown in the left-hand side of \autoref{fig:anisotropy_mass}. While the majority of the most massive galaxies (M $>10^{11.75}$ M$_{\odot}$) are in the SR region, a significant number are in the FR region. Galaxies with M $>10^{11}$ M$_{\odot}$ are evenly distributed across the diagram. To highlight the overall trend, we also show a smoothed version of the same diagram using the \texttt{cap\_loess\_2d}\textsuperscript{\ref{Capwebpage}} routine of \cite{cappellari2013atlas3d}, which implements the multivariate, locally weighted regression (LOESS) algorithm of \cite{cleveland1988locally}. The technique is able to find the best-fitting two-dimensional surface on the $(x,y)$ plane that describes the mean values of a third variable (i.e. $z$). The smoothed plot reveals the underlying trend where the stellar mass decreases with increasing $\lambda_{R_e}$ and $\epsilon$. We note that the absolute range in the LOESS smoothed diagram is less than what is shown in the left-hand side of \autoref{fig:anisotropy_mass}.\par
We compare the left-hand side of \autoref{fig:anisotropy_mass} with the left-hand side of \autoref{fig:anisotropy_sigma_e} which is colour-coded by the luminosity-weighted effective velocity dispersion, $\sigma_e$, within the half-light ellipse. We find that $\sigma_e$ performs significantly better at differentiating between regions on the $(\lambda_{R_e}, \epsilon)$ diagram than stellar mass, with fewer outliers for a clearer transition between high/low $\sigma_e$ and slow/fast rotators. There is no clear trend (by eye) between the stellar mass and $\lambda_{R_e}$ in the FR region. There are a large number of galaxies more massive than the critical mass (yellow, orange and red) which are uniformly distributed throughout the FR region up to the highest $\lambda_{R_e}$. However, (almost) all high $\sigma$ galaxies are concentrated in the low $\lambda_{R_e}$ region of FRs and at high $\lambda_{R_e}$, (almost) all galaxies have low $\sigma$ (black, blue and green). $\sigma_e$ correlates positively with bulge mass fraction or central mass density (\citealp{cappellari2013atlas3d}; C16) such that a larger bulge fraction results in a higher ratio of random to ordered motion within the effective radius. Therefore, a high value of $\sigma_e$ corresponds to a lower value of $\lambda_{R_e}$. We also show the LOESS smoothed version (see the right-hand side of \autoref{fig:anisotropy_sigma_e}). $\sigma_e$ correlates strongly with $\lambda_{R_e}$ and, in contrast with the stellar mass, only has a weak dependence on $\epsilon$.\par
\begin{figure*}
\centering
\includegraphics[width=\textwidth]{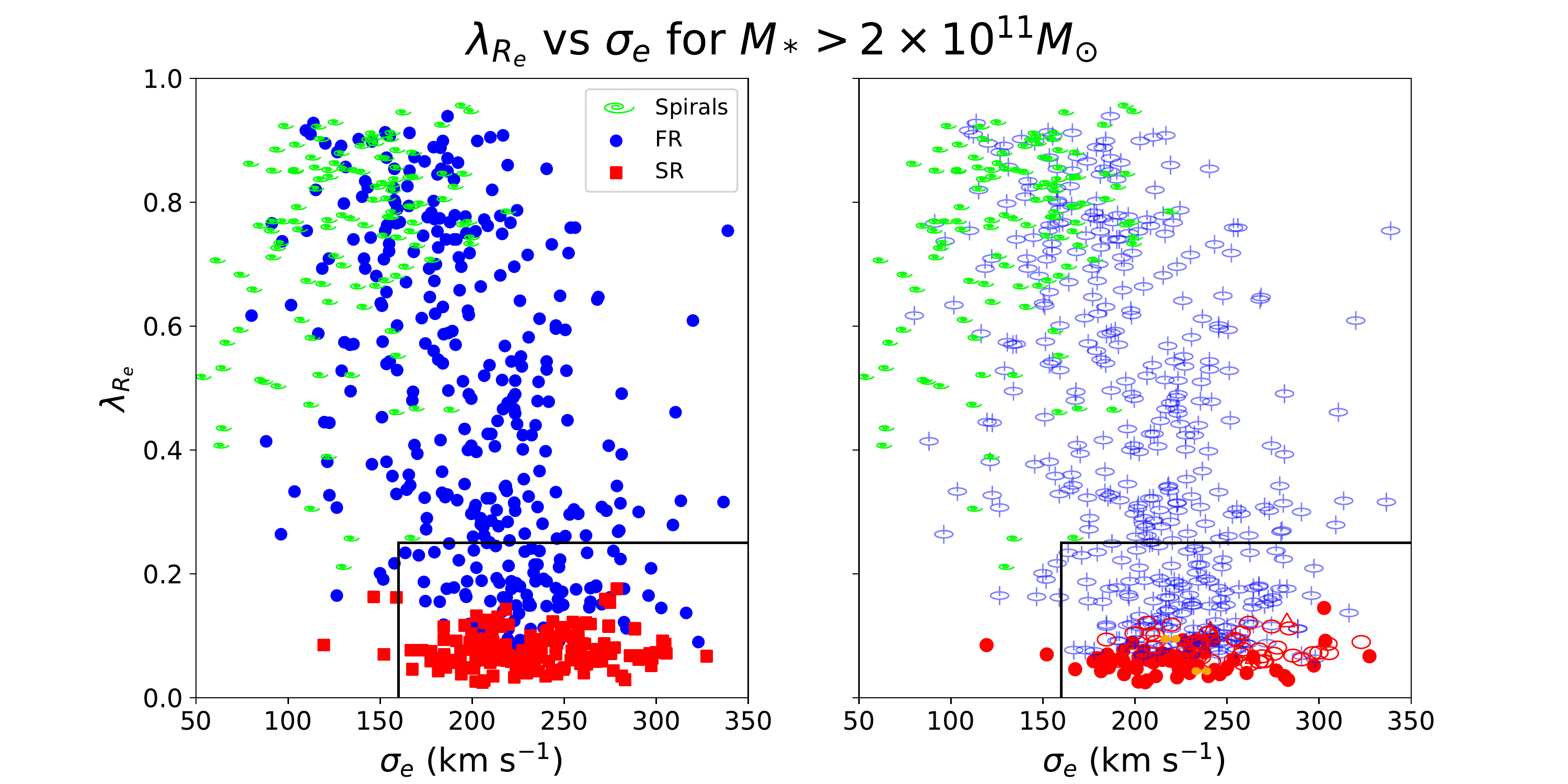}
\caption{Correlation between $\lambda_{R_e}$ and $\sigma_e$ for galaxies more massive than $2 \times 10^{11}$ M$_{\odot}$. The region enclosed by the black solid lines indicates where SRs with a core nuclear profile are likely to be found. \textbf{Left:} Galaxies are separated into fast and slow rotators. \textbf{Right:} Galaxies are classified by kinematic morphology as in \autoref{fig:anisotropy_kine}.}
\label{fig:lambda_sigma_e}
\end{figure*}

To illustrate the relationship between $\lambda_{R_e}$ and $\sigma_e$ more explicitly, we follow Figure 7 of \cite{krajnovic2013atlas3d}. \cite{krajnovic2013atlas3d} defined an area on the $(\lambda_{R_e}, \sigma_e)$ plane which cleanly separates core and core-less SRs. In \autoref{fig:lambda_sigma_e}, we plot the same quantities selecting only galaxies that lie above the critical mass of $2 \times 10^{11}$ M$_{\odot}$ (see \autoref{sec:mass-size}). We find that all but four SRs lie within the region corresponding to core galaxies (i.e. $\sigma_e > 160$ km s$^{-1}$, $\lambda_{R_e} < 0.25$), while the majority of FRs and nearly all spirals lie outside this region. Although we cannot say for certain without obtaining higher resolution photometry from HST for example, the best candidates for true dry merger relics are the high-mass SRs found inside the `core' region in \autoref{fig:lambda_sigma_e}. (In any case, the distances involved are too great to resolve the cores even if they do exist in these galaxies.)

%

\subsection{Kinematic misalignment}
\begin{figure*}
\begin{center}
\includegraphics[width=\textwidth]{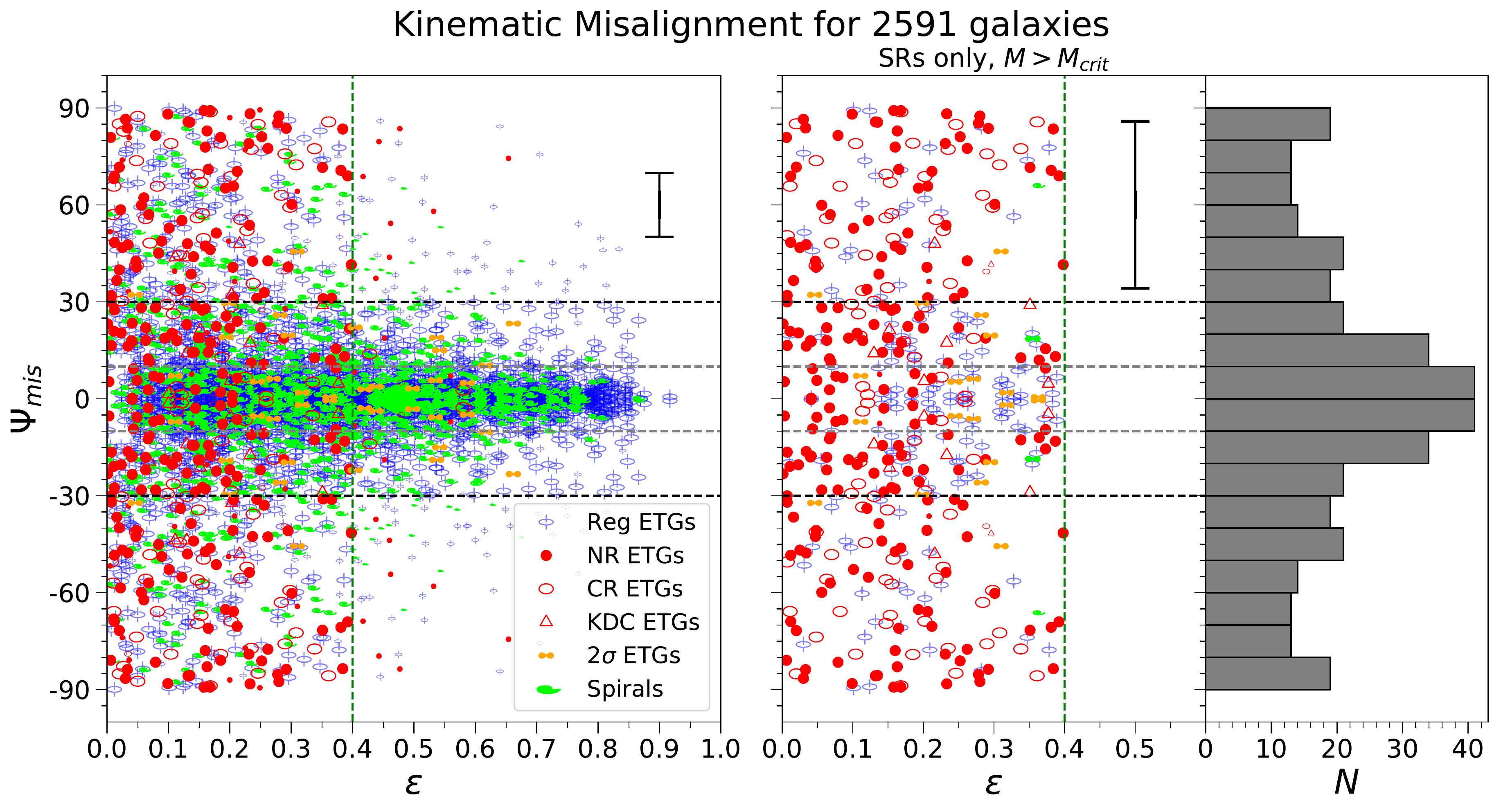}
\caption{\textbf{Left:} Misalignment $\Psi_{\rm{mis}}$ between the photometric axis $\Psi_{\rm{phot}}$ and the kinematic axis $\Psi_{\rm{kin}}$ as a function of $\epsilon$ for all galaxies that are not mergers/close pairs and otherwise have classifiable kinematics. Galaxies that are not in the \textit{clean sample} and do not appear in the other figures in this section are indicated here by a smaller symbol. Each galaxy has been symmetrised to appear above and below the zero line to remove any potential bias between positive and negative misalignments. The vertical dashed green line indicates the $\epsilon<0.4$ criterion for SRs. The black horizontal dashed lines indicate the cut-off at $|\Psi_{\rm{mis}} = 30\degree|$ above which we discard FRs that are flatter than $\epsilon=0.4$ (criterion (v) in \autoref{sec:qualcon}). The horizontal grey dashed lines indicate the more stringent limit of $|\Psi_{\rm{mis}} = 10\degree|$ \protect\citep{emsellem2007sauron}. The mean error for all galaxies in the \textit{clean sample} is indicated. \textbf{Right:} The same selection as left except that only SRs more massive than M$_{\rm{crit}}$ diagram are shown. A histogram of the distribution is shown to the right of the scatter plot.}
\label{fig:misalign}
\end{center}
\end{figure*}

\autoref{fig:misalign} illustrates the kinematic misalignment for galaxies in the \textit{clean sample}. We also include galaxies are not in the \textit{clean sample}, but have classifiable kinematics that fall into the categories described in \autoref{sec:kinprop} (i.e. are not part of mergers or close pairs, and do not have flagged kinematics). To remove any potential bias between positive and negative misalignments, we symmetrise the diagram about $\Psi_{\rm{mis}}=0$. The population of disks at high $|\Psi_{\rm{mis}}|$ and low $\epsilon$ are not necessarily intrinsically misaligned, as described in \autoref{sec:qualcon}, but are most likely due to uncertain or inaccurate $\Psi_{\rm{phot}}$.\par
There is a strong peak of disk galaxies centred around $\Psi_{\rm{mis}}=0$. We find that 83.7\% of regular ETGs and 84.5\% of spiral galaxies are aligned within 10$\degree$. This is in good agreement with \cite{krajnovic2011atlas3d} who suggest that 80\% of all regular ETGs lie within $|\Psi_{\rm{mis}}=10\degree|$. While we note that our accuracy is generally lower due to the larger distances, this is offset by the larger numbers providing a powerful statistical measurement. C16 measured the 1$\sigma$ rms scatter in misalignment to be 4$\degree$ for the combined ATLAS$^{\rm{3D}}$ and SAMI Pilot survey. If we consider only regular ETGs with $\epsilon \geq 0.4$ in the \textit{clean sample}, we find the 1$\sigma$ rms scatter to be 4.2$\degree$. The agreement between the two is impressive considering that we measure the photometric position angle at $R_e$, whereas C16 used the photometric position angle measured at 3$R_e$. If we consider spirals in the \textit{clean sample} with $\epsilon \geq 0.4$, then the 1$\sigma$ rms scatter is 6.0$\degree$.\par
We also show the same diagram but for SRs only, and we include a histogram of the distribution to highlight overdensities. While there is some uniformity in the distribution of SRs between $\Psi_{\rm{mis}}=\pm90\degree$, we find very interesting evidence for a peak in the distribution for a misalignment of $\pm90\degree$. This is the misalignment one would expect for prolate galaxies. Previous evidence of 90$\degree$ misalignment exists. For example, \cite{krajnovic2011atlas3d} found misalignments of $\sim \pm 90 \degree$ (see their Figure 8), but made no statement about mass. Simulations have found that the fraction of prolate galaxies increases with stellar mass \citep{li2017prolate, ebrova2017prolate}. While some claims exist that prolate galaxies may be more numerous at large mass \citep{emsellem2016prolate, tsatsi2017rotation}, those studies have very small statistics and cannot make any claims about the intrinsic shape distribution. Moreover, the fact that we find a peak at $\pm90\degree$ does not by itself imply any of the galaxies are necessarily prolate, as 90$\degree$ misalignment can be expected in triaxial galaxies too (e.g. \citealp{franx1991ordered}). However, the large sample of SRs in the MaNGA sample makes it possible to reveal the excess at $\pm90\degree$ for the first time. We will discuss and quantitatively interpret this interesting result in Li et al. in prep.\par

\subsection{Mass-size relation}
\label{sec:mass-size}
\begin{figure*}
\begin{center}
\includegraphics[width=\textwidth]{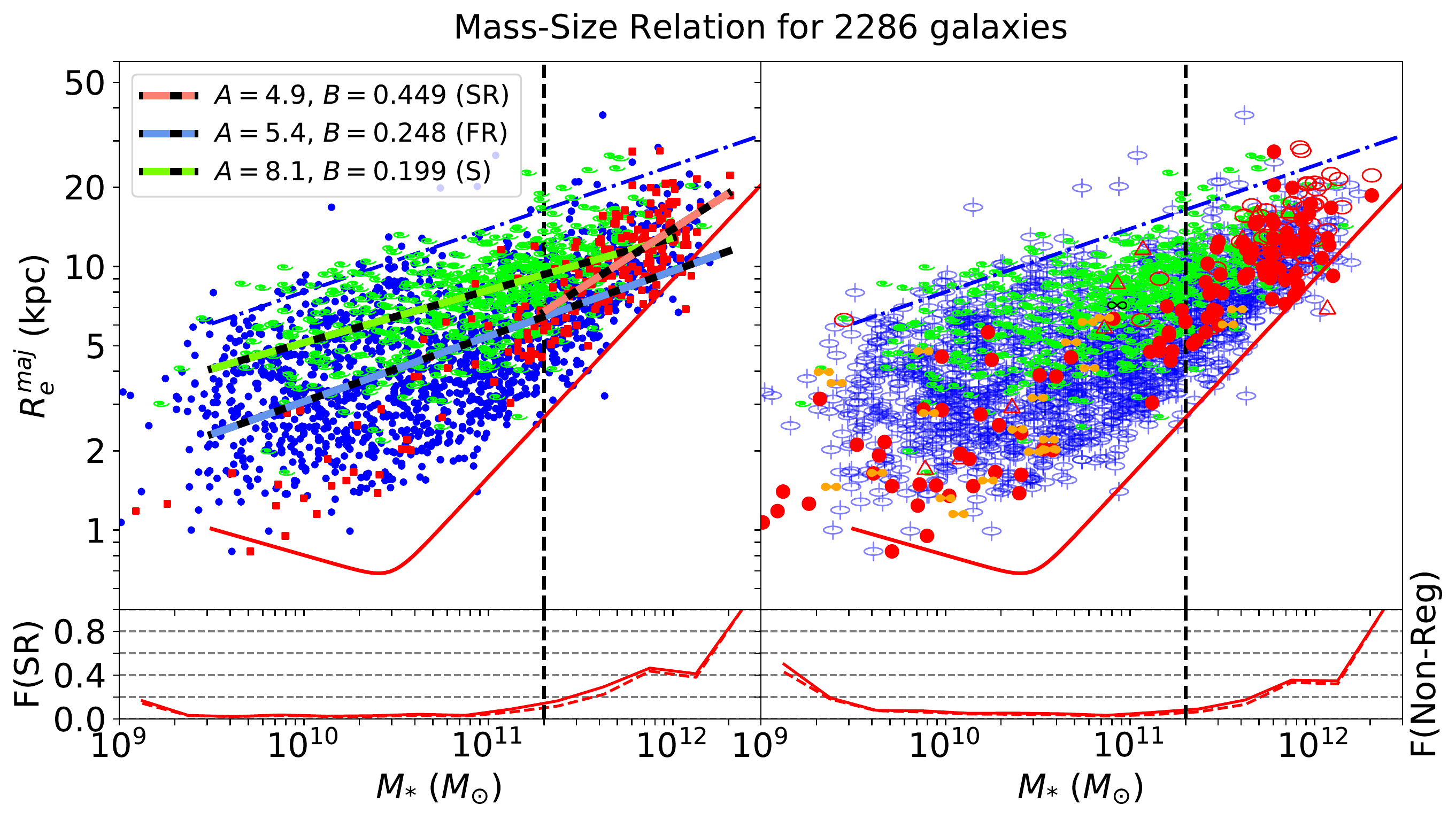}
\caption{\textbf{Top Left:} The mass-size relation for galaxies in the \textit{clean sample} classified by spirals and fast/slow ETGs where the symbols are as in \autoref{fig:lambda_sigma_e}. The galaxies are enclosed between two regimes defined by Equation 4 (red) and Figure 9 (dashed blue) of \protect\cite{cappellari2013atlas3d}. The effective radius in kpc is calculated from the effective radius in arcsec using the angular diameter distance \protect\citep{hogg1999distance}: $R_e\rm{ [kpc]} = R_e\rm{ [\arcsec]} \times D_A\rm{ [Mpc]}/206.265$. SRs are expected to lie above $2\times10^{11}$ M$_{\odot}$ (vertical dashed black line) if they possess a `core' surface brightness profile. Overplotted are fits to FR ellipticals and S0s (dashed blue), FR spirals (dashed green) and SRs with $M \geq M_{\rm{crit}}$ (dashed red). Each relation is shown in the legend where $A$ is the intercept at $M_\ast = 10^{11}$ M$_{\odot}$ and $B$ is the slope. The errors in $A$ are all $\pm 0.1$ and the errors in $B$ are $\pm 0.01$ apart from the red SR relation where the error is $\pm 0.04$. \textbf{Top Right:} The same relation classified by visual kinematic morphology. The symbols are as in \autoref{fig:anisotropy_kine}. \textbf{Bottom Left:} The fraction of slow rotators compared to the total i.e. $\rm{F(SR)}=N_{\rm{SR}}/[N_{\rm{SR}}+N_{\rm{FR}}]$ where $N_{\rm{SR}}$ is the number of SRs, as a function of stellar mass. The fraction is calculated in bins of width $\textrm{log}(M_\ast)=0.25$. The solid red line does not include spirals whereas the dashed red line does include spirals. \textbf{Bottom Right:} The fraction of non-regular rotators compared to the total i.e. $\rm{F(Non-Reg)}=N_{\rm{Non-Reg}}/[N_{\rm{Non-Reg}}+N_{\rm{Reg}}]$, as a function of stellar mass. As bottom left, the solid red line does not include spirals whereas the dashed red line does include spirals.}
\label{fig:mass-size-kine}
\end{center}
\end{figure*}

\autoref{fig:mass-size-kine} shows the kinematic mass-size relation for galaxies in the \textit{clean sample}, using the same kinematic classification as in \autoref{fig:anisotropy_kine}. We use $R_e^{\rm{maj}}$ as it is far less dependent on inclination than $R_e$ \citep{cappellari2013atlas3db}. This is particularly important for a sample like ours, which includes a large population of disk galaxies (spirals and FRs). We compare the left-hand side of \autoref{fig:mass-size-kine} with Figure 21 of C16 which shows the mass-size diagram separately for galaxies in the field and two large clusters. In both the field and the Virgo and Coma Clusters, there exists a critical mass $M_{\rm{crit}} \approx 2 \times 10^{11}$ M$_{\odot}$ above which core SRs dominate while FRs and spiral galaxies are essentially absent. In \autoref{fig:mass-size-kine}, we confirm this characteristic mass where 73.1\% of SRs lie above $M_{\rm{crit}}$ (however some SRs below $M_{\rm{crit}}$ are unlikely to be genuine core SRs, see discussion below), while 77.0\% of FRs and 76.2\% of spirals lie below $M_{\rm{crit}}$. Of the ETGs that lie above $M_{\rm{crit}}$, 31.7\% are SRs (25.8\% including spirals). We also show the fraction of SRs (F(SR)) and non-regular rotators (F(Non-Reg)) in bins of width $\textrm{log}(M_\ast)=0.25$. We find that below M$_{\rm{crit}}$, both fractions are less than about 0.1, while above M$_{\rm{crit}}$, both fractions are above 0.1. (The spikes at low mass are due to low number statistics). The result is unchanged if spirals are included.\par
There are exceptions on both sides. A greater fraction of FRs lie above $M_{\rm{crit}}$ in \autoref{fig:mass-size-kine} than in Figure 21 of C16. This may still be consistent with C16 when one considers that the MaNGA sample is dominated by the field environment, and the separation in mass between fast and slow rotator is cleaner in a cluster environment (see \citealp{cappellari2011atlas3db} for Virgo, \citealp{houghton2013densest} for Coma, \citealp{cappellari2013effect} for Virgo and Coma, \citealp{d2013fast} for Abell 1689, \citealp{scott2014distribution} for Fornax and \citealp{brough2017kinematic} for eight clusters observed with SAMI). Another source of the massive FRs could be major mergers in circular orbits. It has been shown in simulations of binary major mergers \citep{bois2011binary} and cosmological simulations \citep{naab2014atlas3d, li2017prolate} that gas-rich major mergers in circular orbits can preserve or even spin-up the merger remnants \citep{lagos2017mergers}. Hence, it is possible that these massive galaxies could be in a cluster environment, having built up their mass through mergers. Therefore, the left panel of \autoref{fig:mass-size-kine} is qualitatively consistent with the left panel of Figure 21 in C16, but we have vastly superior number statistics.\par
Recent studies have argued that there is no significant dependence of $\lambda_{R_e}$ on environment at fixed stellar mass \citep{brough2017kinematic, greene2017kinematic, oliva2017central, veale2017angular}. These results may be superficially interpreted as implying that environment is not important for the formation of slow rotators. And in this way they may appear in contrast with clear observations, from better resolved IFS data, that the massive slow rotators are only found near the peak density of clusters or groups (\citealp{cappellari2011atlas3db, cappellari2013effect, d2013fast, houghton2013densest, scott2014distribution}; see a review of this topic in Section 5 of C16). But actually, even Figure 11 from \cite{brough2017kinematic} of SAMI observations, which likely represent the current state-of-the art in terms of sample size and environment sampled, shows that slow rotators are rare and they are essentially {\em all} massive and {\em all} lie near the peak overdensities, in agreement with previous studies. It is currently unclear whether the SAMI results of \cite{brough2017kinematic} are in tension with the MaNGA results of \cite{greene2017kinematic}, or whether the differences are due to the larger environmental ranges sampled by SAMI. But overall, all these observations seems consistent with a picture in which the formation of slow rotators is driven by mass growth via dry mergers in groups (as reviewed in Section 7 of C16). We plan to look at this interesting aspect in more detail in a future paper.\par

\begin{figure*}
\begin{center}
\includegraphics[width=\textwidth]{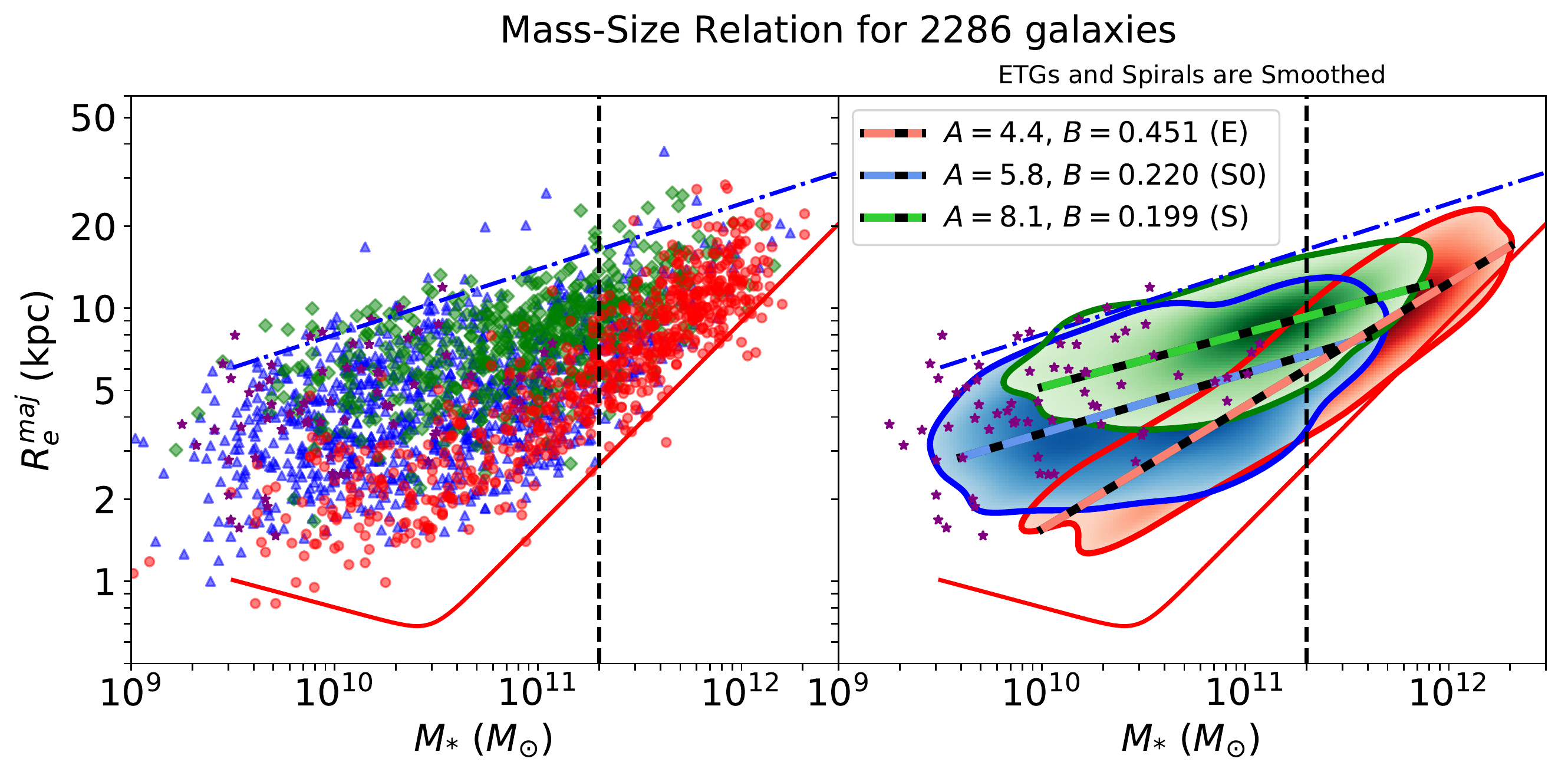}
\caption{\textbf{Left: }The mass-size relation for galaxies in the \textit{clean sample} classified by visual morphology. The symbols are the same as in \autoref{fig:anisotropy_morph}. \textbf{Right: } The same as left except ETGs and spirals are smoothed. Overplotted are fits for each major galaxy population (i.e. not irregulars) with each relation given in the legend. The errors in $A$ are all $\pm 0.1$ and the errors in $B$ are all $\pm 0.01$.} \label{fig:mass-size-morph}
\end{center}
\end{figure*}

There are also a few non-regular rotators at the low mass end (see the right-hand side of \autoref{fig:mass-size-kine}). Of this population, those which lie inside the SR region are smaller in size than those which lie outside. This population of low mass SRs also existed in ATLAS$^{\rm{3D}}$ (see Figure 8 of \citealp{cappellari2013atlas3d}), but none of those galaxies have a core in the surface brightness profile (see Appendix A in \citealp{krajnovic2013atlas3d}), suggesting that those galaxies in ATLAS$^{\rm{3D}}$ are not genuine dry merger relics. It is likely these low mass SRs shown in \autoref{fig:mass-size-kine} would disappear if we showed only SRs with cores as in C16, but as mentioned above, we cannot observe all of them with the resolution of HST. Some galaxies visually identified as KDCs are likely to be inclined 2$\sigma$ (counter-rotating) galaxies. Compared with true KDCs which show ordered rotation within the core but no overall rotation on a global scale, these galaxies have rising velocity profiles out to the borders, indicating a counter-rotating disc. However, these galaxies show no obvious peaks in velocity dispersion and so cannot be classified as 2$\sigma$ galaxies. Aside from these examples, some of the low-mass non-rotators are face-on discs with low $\sigma_e$. These galaxies appear as the FR non-rotators in \autoref{fig:anisotropy_kine}. We find that the fraction of galaxies with masses less than $5\times10^{10}$ M$_{\odot}$ that are either non-rotators, complex rotators or KDCs (i.e. not regular rotators or counter-rotating discs) is $\sim1\%$ which can be expected with a large sample of galaxies.\par
In \autoref{fig:mass-size-kine}, we provide linear relations using \texttt{lts\_linefit} for SRs with $M \geq M_{\rm{crit}}$ (dashed red), FR ETGs (dashed blue) and FR spirals (dashed green). We also compare relations (not shown) from the GAMA survey for $z=0$ disks and spheroids (\citealp{lange2016relations}; L16) with similar populations in our work. We fit FR ellipticals and FR disks (i.e. spirals and S0s) separately below $M_{\rm{crit}}$, and SRs above $M_{\rm{crit}}$ using Equation 2 from L16,
\begin{equation}
R_e=A \Bigg (\frac{M_\ast}{10^{10} M_{\odot}} \Bigg)^B,
\end{equation}
where $R_e$ is the effective radius in kpc and $M_\ast$ is the stellar mass. We find that the best-fit relation for SRs with $M \geq M_{\rm{crit}}$ has a slope of $0.449 \pm 0.039$ which is steeper than the L16 $z=0$ spheroid relation ($B=0.263$) but not as steep as the L16 elliptical, $M>10^{10}$ M$_{\odot}$ relation ($B=0.643$). However, the slope is very close to that found by \cite{lange2015relations} for ETGs ($B=0.46$). The FR spheroids have a slope $B=0.293 \pm 0.018$ which is similar to that of the L16 $z=0$ spheroids ($B=0.263$), while the FR disks have a slightly shallower slope ($B=0.201 \pm 0.011$) than the L16 $z=0$ disks ($B=0.274$).\par 
Finally, we show the mass-size relation coloured by morphology (\autoref{fig:mass-size-morph}). We include relations for ellipticals, S0s and spirals of all masses. Ellipticals lie almost parallel to the solid-red line (Equation 4 of \citealp{cappellari2013atlas3d}) while spirals and S0s lie almost parallel with the dashed blue line (Figure 9 of \citealp{cappellari2013atlas3d}). All irregular galaxies lie below $10^{11}$ M$_{\odot}$ with the majority at lower masses still. Here we compare relations (not shown) for ellipticals with $M>10^{10}$ M$_{\odot}$ and for disks, i.e. spirals and S0s, at all masses with the analogous relations from L16. We find that ellipticals in our work have a similar slope ($B=0.485 \pm 0.009$) to that of SRs given above and hence is less steep than the $M>10^{10}$ M$_{\odot}$ elliptical relation from L16. The slope of our disk relation, $B=0.249 \pm 0.008$, agrees well with the L16 $z=0$ disk relation. However, our effective radii are generally measured from MGE fitting, while L16 used bulge-disk decompositions to derive $R_e$, and so our results are unlikely to agree in all cases.\par

\begin{table*}
\centering
\caption{Table containing all the quantities and properties required to plot all the figures in \autoref{sec:results}. Column (1) is the plate-IFU combination associated with a MaNGA observation. As most galaxies in MaNGA only have a single observation, this is usually a single galaxy. Column (2) indicates whether the galaxy associated with the plate-IFU combination given in Column (1) is in the \textit{clean sample} as decided by the criteria given at the beginning of \autoref{sec:qualcon}. Column (3) lists the Hubble classification determined visually by MTG and MC (see \autoref{sec:qualcon}, U = unclassified, M/CP = merger or close pair) and Column (4) lists the kinematic classification (F = flagged) described in \autoref{sec:kinprop}. Column (5) lists the circular effective radius given in log units which is taken from either the MGE fit (\autoref{sec:mgemethod}), the 2MASS XSC (\autoref{sec:2mass}) or the NSA catalogue (\autoref{sec:NSA}) as described in \autoref{sec:effectiveradius}. Column (6) lists the semi-major axis given in physical log units and calculated using the angular diameter distance (see \autoref{fig:mass-size-kine}) estimated from the redshift given in Column (15). These two radii are given in log units as they have constant relative errors. Column (7) gives the FWHM of the PSF in arcsec taken from \texttt{drpall-v2\_1\_2} and Column (8) lists $\lambda_{R_e}$ corrected for the PSF given in Column (7) using \autoref{eq:lambdamoffatsersic}. Columns (9) and (10) list the final $\epsilon$ and photometric position angle $\Psi_{\rm{phot}}$, defined as East of North, taken from either the MGE fit, the photometric fit from \texttt{find\_galaxy} or the NSA catalogue (see \autoref{fig:epscorr} and \autoref{sec:ellipticity}). For 143 galaxies, the measured $\Psi_{\rm{phot}}$ is replaced by a more accurate value supplied by HL (see Appendix \ref{app:misalignFR}). Column (11) lists the kinematic position angle $\Psi_{\rm{kin}}$ East of North and column (12) lists the error in $\Psi_{\rm{kin}}$. Column (13) lists $\textrm{log}(M_\ast)$ either calculated from the absolute $\rm{K}_{\rm{S}}$ using \autoref{eq:2MASSmass}, or taken from the NSA catalogue and scaled to 2MASS (see \autoref{fig:masscorr}). Column (14) lists the effective velocity dispersion in log units calculated using \autoref{eq:sigma}. Finally, columns (15) and (16) list the redshift and Sérsic index taken from the NSA catalogue. A complete table will be available from the journal website.}
\label{tab:galaxydata}
\resizebox{\textwidth}{!}{%
\begin{tabular}{@{}ccccccccccccccccc@{}}
\toprule
\begin{tabular}[c]{@{}c@{}}Plate-IFU\\ \\(1)\end{tabular} &
\begin{tabular}[c]{@{}c@{}}Clean\\Sample?\\(2)\end{tabular} &
\begin{tabular}[c]{@{}c@{}}Hubble\\Group\\(3)\end{tabular} &
\begin{tabular}[c]{@{}c@{}}Kinematic\\Classification\\(4)\end{tabular} &
\begin{tabular}[c]{@{}c@{}}$\textrm{log}(R_e^{\rm{circ}})$\\(\arcsec)\\(5)\end{tabular} &
\begin{tabular}[c]{@{}c@{}}$\textrm{log}(R_e^{\rm{maj}})$\\(kpc)\\(6)\end{tabular} &
\begin{tabular}[c]{@{}c@{}}PSF\\(\arcsec)\\(7)\end{tabular} &
\begin{tabular}[c]{@{}c@{}}$\lambda_{R_e}^{\rm{true}}$\\ \\(8)\end{tabular} &
\begin{tabular}[c]{@{}c@{}}$\epsilon$\\ \\(9)\end{tabular} &
\begin{tabular}[c]{@{}c@{}}$\Psi_{\rm{phot}}$\\(\degree)\\(10)\end{tabular} &
\begin{tabular}[c]{@{}c@{}}$\Psi_{\rm{kin}}$\\(\degree)\\(11)\end{tabular} &
\begin{tabular}[c]{@{}c@{}}Error in $\Psi_{\rm{kin}}$\\(\degree)\\(12)\end{tabular} &
\begin{tabular}[c]{@{}c@{}}$\textrm{log}(M_\ast)$\\ \\(13)\end{tabular} &
\begin{tabular}[c]{@{}c@{}}$\textrm{log}(\sigma_e)$\\(km s$^{-1}$)\\(14)\end{tabular} &
\begin{tabular}[c]{@{}c@{}}$z$\\ \\(15)\end{tabular} & 
\begin{tabular}[c]{@{}c@{}}$n$\\ \\(16)\end{tabular} & \\ \midrule
7443-12701 & Y & E & R & 0.699 & 0.417 & 2.6 & 0.793 & 0.342 & 154.8 & 153.0 & 5.5 & 10.38 & 1.747 & 0.0209 & 4.4\\
7443-12702 & N & U & R & 0.934 & 0.996 & 2.6 & 0.788 & 0.054 & 99.5 & 110.0 & 8.0 & 11.11 & 1.691 & 0.0579 & 1.7\\
7443-12703 & N & M/CP & R & 0.831 & 0.891 & 2.6 & 0.400 & 0.510 & 153.3 & 4.5 & 3.2 & 11.34 & 2.102 & 0.0406 & 2.2\\
7443-12704 & Y & S0 & R & 1.080 & 0.941 & 2.5 & 0.801 & 0.709 & 125.1 & 128.0 & 3.2 & 10.44 & 1.850 & 0.0193 & 0.9\\
7443-12705 & Y & S0 & R & 0.831 & 1.121 & 2.6 & 0.926 & 0.592 & 34.7 & 41.5 & 2.8 & 11.21 & 2.068 & 0.0648 & 1.0\\
7443-1901 & Y & S0 & R & 0.650 & 0.301 & 2.6 & 0.627 & 0.240 & 75.8 & 64.5 & 14.0 & 9.81 & 1.621 & 0.0193 & 2.2\\
7443-1902 & Y & S0 & R & 0.548 & 0.314 & 2.6 & 0.745 & 0.546 & 10.8 & 5.0 & 9.8 & 10.03 & 1.605 & 0.0194 & 2.8\\
7443-3701 & N & I & NR & 0.684 & 0.384 & 2.6 & 0.332 & 0.438 & 137.7 & 175.0 & 48.2 & 9.49 & 1.543 & 0.0185 & 1.3\\
7443-3702 & Y & E & NR & 0.545 & 0.858 & 2.5 & 0.025 & 0.034 & 141.8 & 125.5 & 81.2 & 11.84 & 2.313 & 0.1108 & 6.0\\
7443-3703 & N & E & F & 0.348 & 0.134 & 2.6 & 0.800 & 0.095 & 158.1 & 137.0 & 0.5 & 9.30 & 2.501 & 0.0290 & 0.9\\
7443-3704 & Y & E & R & 0.535 & 0.326 & 2.6 & 0.277 & 0.434 & 61.6 & 52.5 & 33.0 & 9.49 & 1.724 & 0.0230 & 1.6\\
7443-6101 & Y & S0 & R & 0.630 & 0.508 & 2.6 & 0.866 & 0.315 & 105.1 & 107.0 & 21.8 & 9.53 & 1.719 & 0.0313 & 1.1\\
7443-6102 & Y & S & R & 0.829 & 0.682 & 2.6 & 0.683 & 0.362 & 131.6 & 137.5 & 3.5 & 10.99 & 1.972 & 0.0284 & 1.5\\
7443-6103 & Y & S0 & R & 0.766 & 0.530 & 2.6 & 0.612 & 0.563 & 18.4 & 28.5 & 6.0 & 10.01 & 1.810 & 0.0189 & 0.8\\
7443-6104 & Y & E & R & 0.793 & 1.006 & 2.6 & 0.743 & 0.229 & 100.0 & 107.0 & 2.5 & 11.72 & 2.161 & 0.0755 & 3.1\\
7443-9101 & Y & S & R & 0.833 & 0.787 & 2.6 & 0.799 & 0.196 & 89.4 & 68.5 & 1.8 & 10.93 & 1.716 & 0.0408 & 2.4\\
7443-9102 & Y & E & R & 0.751 & 1.067 & 2.5 & 0.154 & 0.319 & 50.9 & 46.5 & 1.5 & 11.84 & 2.439 & 0.0920 & 6.0\\
\bottomrule
\end{tabular}}
\end{table*}

\section{Conclusions}
\label{sec:conclusions}
We have measured the stellar angular momentum parameterised by $\lambda_{R_e}$ from the kinematics of 2286 galaxies in the MaNGA sample, the largest sample with IFS observations to date. We have also derived an approximate analytic correction to the measurement of $\lambda_{R_e}$ that can be applied to any IFS dataset. For the first time, we see a clear bimodality in galaxy properties between slow and fast rotators which may result from fundamental differences in their formation history. The bimodality we see in the $(\lambda_{R_e}, \epsilon)$ diagram is necessarily a strict lower limit to the intrinsic bimodality we would observe in that same diagram if all galaxies were edge-on. This would be true even in noiseless galaxies due to the effect of inclination. However, in our sample, noise does plays a role in weakening the observed bimodality. This means that if we could deproject all galaxies and remove all noise, we would dramatically sharpen the separation between the fast and slow rotators. In this case, the histogram in \autoref{fig:anisotropy_density} would most likely look more similar to Figure 11 in C16, but with much better statistics.\par
The majority of regular rotators and spirals occupy the fast rotator region on the $(\lambda_{R_e}, \epsilon)$ diagram. There is a large concentration of non-rotating galaxies in the slow rotator region and all non-rotators are rounder than E4. If we consider only galaxies with a mass greater than $2 \times 10^{11} M_{\odot}$, there is a sharp cutoff in effective velocity dispersion and $\lambda_{R_e}$ associated with a core surface brightness profile. These galaxies are the best candidates for the genuine dry merger relics.\par
The strongest concentration of elliptical galaxies is on the upper boundary of the SR region. However, when weighted by volume, these galaxies are down-weighted as they are rare in a random volume of the Universe. There is a strong concentration of spirals and S0s at $\lambda_{R_e} \sim 0.9$ which is higher than seen in previous samples. This is caused by a combination of effects due to the instrumental resolution (i.e. unresolved $\sigma$) and the spatial resolution (i.e. the MaNGA PSF). As shown in Appendix \ref{app:zerodisp}, $\lambda_{R_e}$ is accurate to within $\sim$0.1 when $\sigma_{\rm{max}} \gtrsim 25\textrm{ km s}^{-1}$ for low instrumental noise. We find that the effective velocity dispersion $\sigma_e$ and kinematic morphology are the galaxy properties that give the cleanest separation between slow and fast rotators, as opposed to visual morphology and stellar mass. We find many more high-mass fast rotators than in previous studies (e.g. \citealp{cappellari2013effect}). This is likely due to MaNGA sampling a larger and more representative sample of the Universe than previous studies, which had a disproportionately high number of cluster satellite galaxies. However, it could also be due to gas-rich major mergers on circular orbits that preserve the spin of the merger remnant.\par
We also measure the misalignment between the photometric major axis and the kinematic axis for 2547 galaxies with kinematics that can be visually classified using the scheme described in \autoref{sec:kinprop}. We find that $\sim 80\%$ of regular ETGs and spiral galaxies are aligned within 10$\degree$. There is a large range of misalignments for slow rotators as expected from a triaxial distribution. Thanks to the large number of galaxies in the MaNGA sample, we are able for the first time to resolve a peak at $\Psi_{\rm{mis}}= \pm 90 \degree$, indicating minor axis rotation. This interesting result will be discussed further in Li et al. in prep.\par
Finally, we plot the mass-size relation for the MaNGA sample. For ETGs, we find a tight agreement between the \textit{quantitative} measurement of slow/fast rotator and the \textit{qualitative} classification of the kinematic morphology (regular/non-regular rotator), further highlighting the bimodality in galaxy properties. In a follow-up paper, we will look at the dependence that galaxy mass and size has on environment, using the large number statistics that the MaNGA sample provides.



\section*{Acknowledgements}
We thank the anonymous referee for their detailed and constructive comments which led to the significant improvement of this manuscript. We acknowledge helpful comments from Eric Emsellem that helped to improve this manuscript. MTG acknowledges support by STFC. MC acknowledges support from a Royal Society University Research Fellowship. AW acknowledges support of a Leverhulme Trust Early Career Fellowship.\par
Funding for the Sloan Digital Sky Survey IV has been provided by the Alfred P. Sloan Foundation, the U.S. Department of Energy Office of Science, and the Participating Institutions. SDSS acknowledges support and resources from the Center for High-Performance Computing at the University of Utah. The SDSS website is www.sdss.org.\par
SDSS is managed by the Astrophysical Research Consortium for the Participating Institutions of the SDSS Collaboration including the Brazilian Participation Group, the Carnegie Institution for Science, Carnegie Mellon University, the Chilean Participation Group, the French Participation Group, Harvard-Smithsonian Center for Astrophysics, Instituto de Astrofísica de Canarias, The Johns Hopkins University, Kavli Institute for the Physics and Mathematics of the Universe (IPMU) / University of Tokyo, Lawrence Berkeley National Laboratory, Leibniz Institut für Astrophysik Potsdam (AIP), Max-Planck-Institut für Astronomie (MPIA Heidelberg), Max-Planck-Institut für Astrophysik (MPA Garching), Max-Planck-Institut für Extraterrestrische Physik (MPE), National Astronomical Observatories of China, New Mexico State University, New York University, University of Notre Dame, Observatório Nacional / MCTI, The Ohio State University, Pennsylvania State University, Shanghai Astronomical Observatory, United Kingdom Participation Group, Universidad Nacional Autónoma de México, University of Arizona, University of Colorado Boulder, University of Oxford, University of Portsmouth, University of Utah, University of Virginia, University of Washington, University of Wisconsin, Vanderbilt University, and Yale University.\par
This publication makes use of data products from the Two Micron All Sky Survey, which is a joint project of the University of Massachusetts and the Infrared Processing and Analysis Center/California Institute of Technology, funded by the National Aeronautics and Space Administration and the National Science Foundation.





\bibliographystyle{mnras}
\bibliography{MasterBibliography} 



\appendix

\section{Misaligned Fast Rotators}
\label{app:misalignFR}
In our sample, we find a large number of FRs with misalignments greater than 30$\degree$. Many of these are barred spirals or S0s. For cases where the bar is misaligned with the photometric major axis of the galaxy (and $R_e^{\rm{circ}}$ is a similar scale to the size of the bar), \texttt{find\_galaxy} will fit to the bar. This implies that both the ellipticity and position angle are not representative of the galaxy as a whole.\par
Another case where our measurement of the misalignment might not reflect the intrinsic misalignment is for near face-on axisymmetric galaxies with low (apparent) $\epsilon$. As described by \cite{krajnovic2011atlas3d}, the position angle is essentially degenerate when $\epsilon$ is low, and so the orientation of the ellipse is not well defined meaning that $\lambda_{R_e}$ only has a weak dependency on $\Psi_{\rm{phot}}$. We check whether the NSA measurement of the position angle is more robust against the effect of bars and find that for barred galaxies with $\Psi_{\rm{mis}} > 30\degree$, NSA is comparable to \texttt{find\_galaxy}.\par
If we vary by eye the \texttt{FRACTION} parameter in \texttt{find\_galaxy} for 143 galaxies, as was carried out by HL, we are able to find a more accurate position angle. To determine whether we should include the galaxies that are still affected after performing this step, we take 181 barred spiral galaxies and S0s (all regular rotators) with $\Psi_{\rm{mis}} > 30\degree$ and measure $\lambda_{R_e}$ rotating the half-light ellipse with position angles between 0$\degree$ and 175$\degree$ in steps of 5$\degree$. As expected, we find that the maximum change in $\lambda_{R_e}$ increases with $\epsilon$. For galaxies with $\epsilon < 0.4$, more than 90$\%$ decrease by less than $\lambda_{R_e}=0.1$ when measuring $\lambda_{R_e}$ within an ellipse with minimal misalignment compared with maximal misalignment. Hence, the maximum expected decrease in $\lambda_{R_e}$ due to an incorrect position angle for galaxies with $\epsilon < 0.4$ is 0.1. For this reason, we include misaligned fast rotators with $\epsilon<0.4$.\par
As detailed in \autoref{tab:qualcon}, we exclude 56 galaxies based on this criterion. About half of these lie below the magenta line in \autoref{fig:anisotropy_kine} and so by performing this step, we are reducing the contamination from galaxies where the measured $\lambda_{R_e}$ is too low. However, we are including 220 FRs with $\epsilon<0.4$ which may bias us towards more face-on galaxies. Certainly for $\epsilon < 0.2$, the degeneracy in $\Psi_{\rm{phot}}$ means that we cannot exclude these galaxies based on misalignment. A quick test shows that by \textit{not} including the 220 FRs mentioned here, our conclusions remain the same, albeit with slightly different statistics.

\section{$\lambda_{R_e}$ in the low $\sigma_\ast$ regime}
\label{app:zerodisp}
The observed stellar velocity dispersion $\sigma_{\rm{obs}}$ must be corrected for the instrumental resolution $\sigma_{\rm{inst}}$ of the spectrograph. The correction is applied in quadrature as
\begin{equation} \label{eq:sigmacorrection} 
\sigma_{\ast} = \sqrt{\sigma_{\rm{obs}}^2 - \sigma_{\rm{inst}}^2}
\end{equation}
where $\sigma_{\ast}$ is the corrected, intrinsic velocity dispersion. While this is a necessary step to recover the true velocity dispersion, a caveat arises from the fact that it is entirely possible that at a S/N of 10, the error in $\sigma_{\rm{obs}}$ can be as much as 20$\%$ \citep{penny2016sdss}, leading to a situation where $\sigma_{\rm{obs}} < \sigma_{\rm{inst}}$. Numerically, this leads to an unphysical value of $\sigma_{\ast}$ in \autoref{eq:sigmacorrection}. We stress that these unphysical values do not indicate any problem with the data or any failure of the kinematic extraction. They simply imply that the dispersion at those locations is quite low with respect to $\sigma_{\rm{inst}}$ (typically $\sigma_\ast \lesssim \sigma_{\rm{inst}}/2$), but a more accurate $\sigma_\ast$ value cannot be determined, due to noise and uncertainties in $\sigma_{\rm{inst}}$. One option is simply to set $\sigma_{\ast}$ of the affected pixels to be zero. This may not be entirely satisfactory as it is clear that $\sigma_{\ast}$ must be very low at these pixels, but not necessarily equal to zero. Moreover, in the summation over all pixels within 1 $R_e$, the contribution of these pixels to the summation is the same at the denominator and numerator. The result is an inflated value of the intrinsic $\lambda_{R_e}^{\rm{true}}$.

\begin{figure} 
\begin{center}
\includegraphics[width=0.49\textwidth]{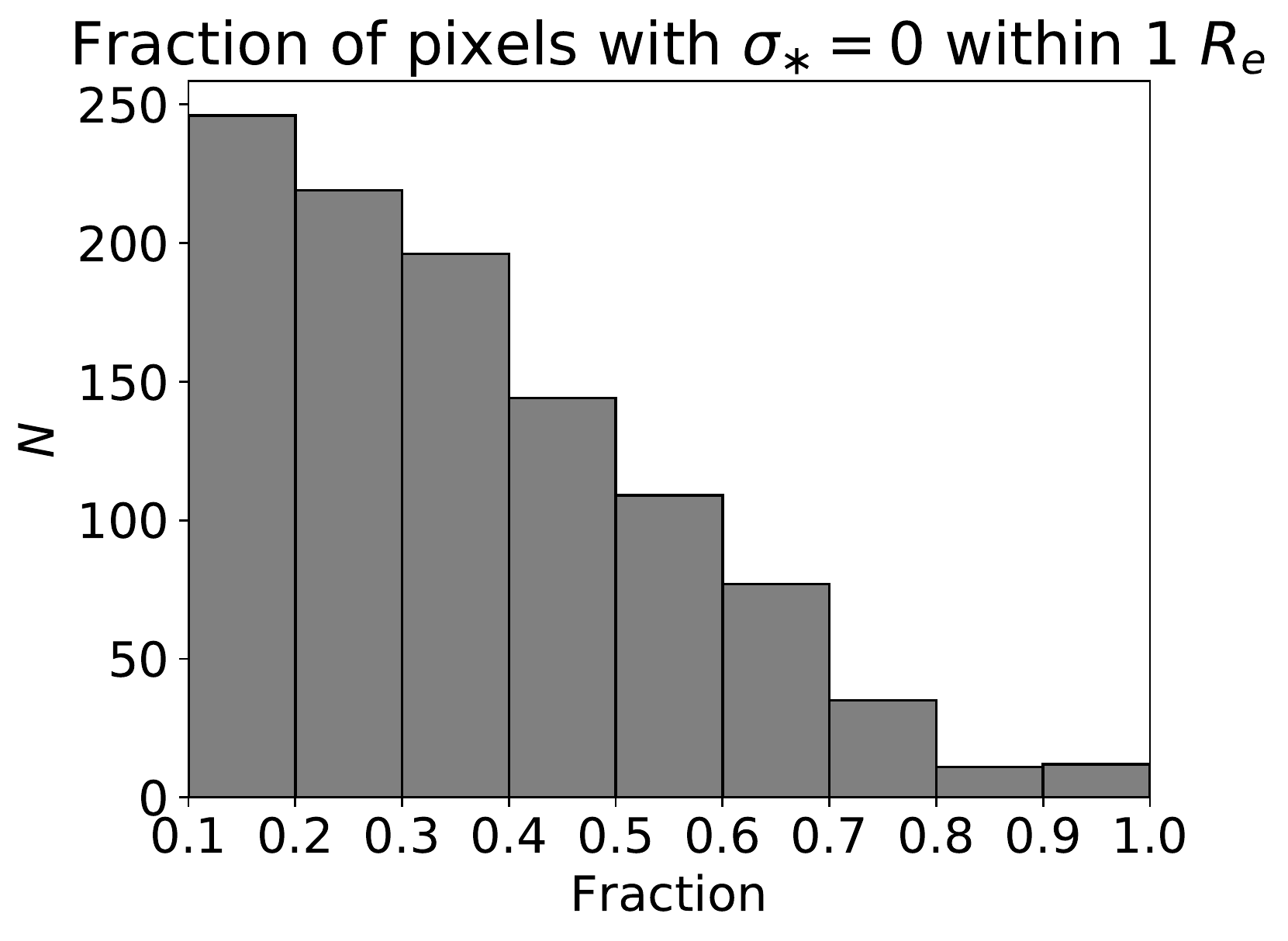}
\caption{Histogram indicating the number of galaxies $N$ in MPL-5 with at least 10$\%$ of pixels within 1 $R_e$ with $\sigma_{\ast}=0$.}
\label{fig:fraczero}
\end{center}
\end{figure}

\begin{figure*}
\begin{center}
\includegraphics[width=0.9\textwidth]{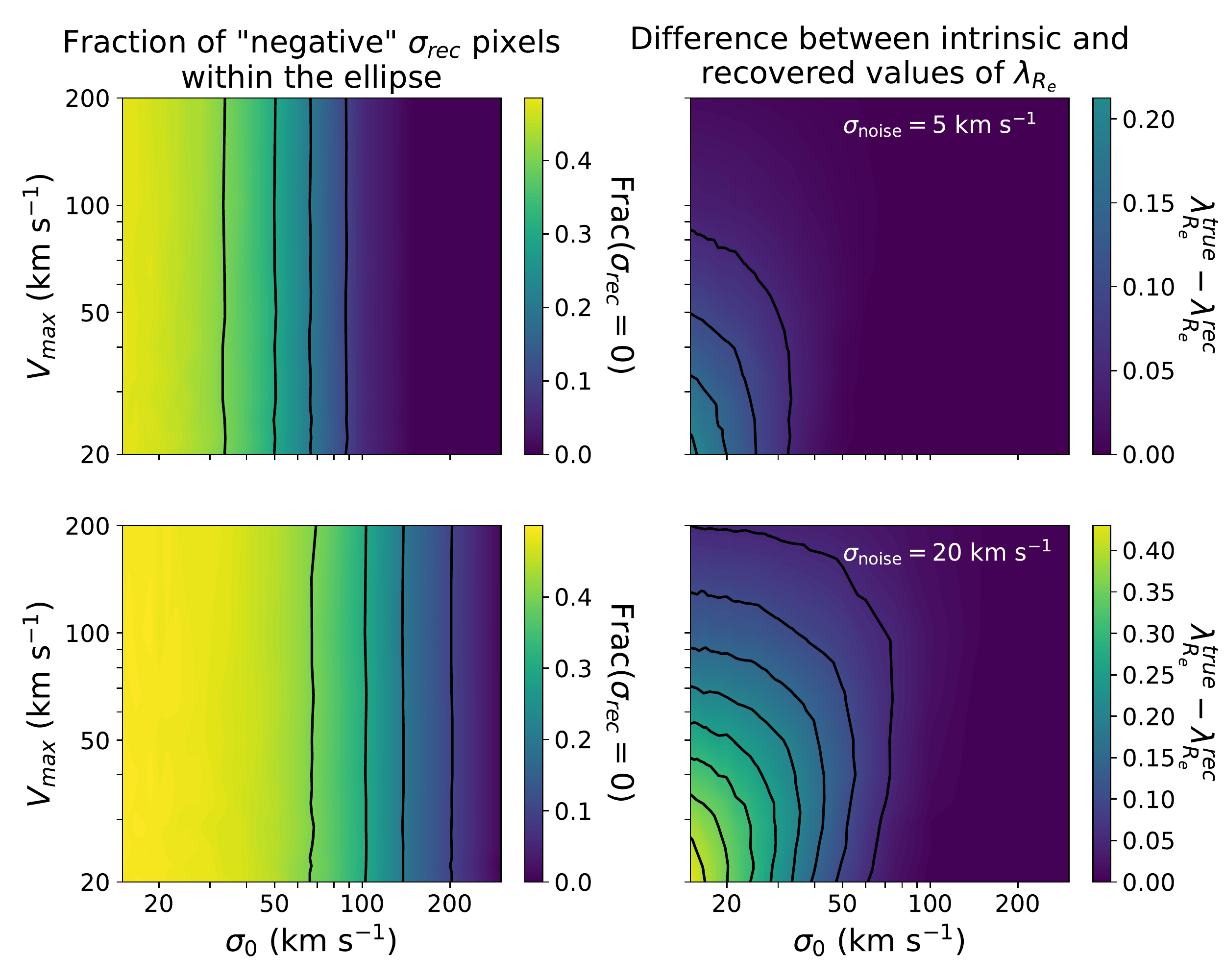}
\caption{\textbf{Top row:} On the left, we plot the fraction of pixels within the half-light ellipse with $\sigma_{\rm{rec}}=0$ km s$^{-1}$, indicated by the colour map, for the 200 models in the grid. When performing the $\sigma$ correction, we assume that the instrumental resolution is described by a Gaussian with a width of $\sigma=5 \textrm{ km s}^{-1}$. The black vertical lines are contours of constant fraction at values of 0.1, 0.2, 0.3 and 0.4. On the right, we plot on the same axes the difference between the \textit{intrinsic} $\lambda_{R_e}^{\rm{true}}$ and the \textit{recovered} $\lambda_{R_e}^{\rm{rec}}$. The contours now trace constant values of $\lambda_{R_e}^{\rm{true}}-\lambda_{R_e}^{\rm{rec}}=$ 0.05, 0.1, 0.15 and 0.2. For both panels in the top row, the colours are scaled to the equivalent panel in the bottom row. \textbf{Bottom row:} The same except we increase the width of the instrumental noise to $\sigma=20\textrm{ km s}^{-1}$. The contours on the left are at values of 0.1, 0.2, 0.3, and 0.4. The contours on the right are at values of 0.05, 0.1, 0.15, 0.2, 0.25, 0.3, 0.35 and 0.4.}
\label{fig:zerosigmacontours}
\end{center}
\end{figure*}

For MPL-5, there are approximately 1000 galaxies which have more than $10 \%$ of pixels with $\sigma_{\ast} = 0$ within 1 $R_e$ (see \autoref{fig:fraczero}). In order to assess whether to trust $\lambda_{R_e}^{\rm{true}}$ for the affected galaxies, we take a simple galaxy model and add noise to simulate the measurement errors and the uncertainties in the instrumental resolution before attempting to recover the intrinsic $\lambda_{R_e}^{\rm{true}}$ (i.e. without noise). For our galaxy model, we assume the circular velocity to be Eq. 16 from \cite{hernquist1990analytical}, reproduced here:
\begin{equation}
V_c=\frac{\sqrt{GMr}}{r+s}
\end{equation}
where $M$ is the total mass, $r$ is the radius and $s$ is the scale length. $V_c$ is equal to $V_{\rm{max}}$ when $r=s$. We note that the assumed velocity profile applies to regular rotators only. However, the effect we are studying is not likely to affect non-regular rotators which typically have high $\sigma_\ast$, and so our assumption is valid. We assume the intrinsic velocity dispersion takes the form $\sigma_{\ast}=\sigma_0/(r+s)$ where $\sigma_0$ is the velocity dispersion at the galaxy centre. To estimate the surface brightness, we use the example MGE given in the \texttt{test\_jam\_axi\_rms} function which can be found in the \texttt{jam\_axi\_rms} routine, part of the JAM package\textsuperscript{\ref{Capwebpage}}. The flux is given by the \texttt{jam\_axi\_rms} function, for which we assume an inclination of 60$\degree$. Finally, we assume $R_e=1.8153s$ \citep[Eq. 38]{hernquist1990analytical} where $s=40\arcsec$ (corresponding to 3.2 kpc at the assumed distance of the Virgo Cluster i.e. 16.5 Mpc) and $\epsilon=0.4$.\par
We take a grid of 200 models parameterised by $(V_{\rm{max}})_i=200/i\textrm{ km s}^{-1}$ where $1 \leq i \leq 10$ and $(\sigma_{0})_j=300/j\textrm{ km s}^{-1}$ where $1 \leq j \leq 20$. For each model in the grid, we calculate the intrinsic $\lambda_{R_e}^{\rm{true}}$ using intrinsic values of the velocity dispersion $\sigma_{\ast}$. We apply the correction in reverse to calculate the observed velocity dispersion, $\sigma_{\rm{obs}}$. We then add random Gaussian noise $N(\mu, \sigma_{\rm{noise}})$ where $\mu=0$ km s$^{-1}$ and $\sigma_{\rm{noise}}=5$ and 20 km s$^{-1}$ to simulate the uncertainty in the instrumental resolution:
\begin{equation}
\sigma_{\rm{obs}} = \sqrt{\sigma_{\ast}^2 + \sigma_{\rm{inst}}^2} + N(\mu, \sigma_{\rm{noise}}),
\end{equation}
where $\sigma_{\rm{inst}}$ is the instrumental resolution. For MaNGA, we take $\sigma_{\rm{inst}}$ to be the median value of 72 km s$^{-1}$ \citep{law2016data}. We note that $\sigma_{\rm{inst}}$ does have a dependence on wavelength (see Figure 18 of \citealp{law2016data}) which we ignore for this approximate test. To recover the intrinsic velocity dispersion, $\sigma_{\rm{rec}}$, from the observed velocity dispersion (as is done for MaNGA using \autoref{eq:sigmacorrection}), we apply the instrumental correction to the noisy data:

\begin{equation}
\sigma_{\rm{rec}} = \sqrt{\sigma_{\rm{obs}}^2 - \sigma_{\rm{inst}}^2}.
\end{equation}

We calculate the recovered $\lambda_{R_e}^{\rm{rec}}$ using $\sigma_{\rm{rec}}$ and compare with $\lambda_{R_e}^{\rm{true}}$ calculated with $\sigma_{\ast}$. Our results are shown in \autoref{fig:zerosigmacontours}. We find that for all values of $\sigma_0$ and $V_{\rm{max}}$, $\lambda_{R_e}^{\rm{rec}}$ \textit{underestimates} $\lambda_{R_e}^{\rm{true}}$. This seems at first to be counter-intuitive, since $\sigma_{\rm{rec}}$ can take a value of zero, whereas $\sigma_\ast$ can never equal zero (in the models). Hence, we would expect that $\lambda_{R_e}^{\rm{rec}}>\lambda_{R_e}^{\rm{true}}$ if this was the only effect. However, we find that the random \textit{positive} noise added at $r\approx0$ where $\sigma \approx \sigma_0$ gives a larger contribution to \textit{lowering} $\lambda_{R_e}^{\rm{rec}}$ than the pixels at large $r$ with $\sigma_\ast=0$ which serve to \textit{inflate} $\lambda_{R_e}^{\rm{rec}}$. We find that the regime where $\lambda_{R_e}^{\rm{true}}-\lambda_{R_e}^{\rm{rec}}>0.1$ for low instrumental noise of $\sigma_{\rm{noise}}=5\textrm{ km s}^{-1}$ is $\sigma_0 \lesssim 25\textrm{ km s}^{-1}$ and $V_{\rm{max}} \lesssim 50\textrm{ km s}^{-1}$. For higher noise where $\sigma_{\rm{noise}}=20\textrm{ km s}^{-1}$, this region expands to $\sigma_0 \lesssim 50 \textrm{ km s}^{-1}$ and $V_{\rm{max}} \lesssim 100 \textrm{ km s}^{-1}$. However, this level of noise is extreme and is expected to be much larger than any realistic instrumental noise.\par
As the noise we are adding is randomly positive and negative about the mean, the maximum percentage of pixels in the simulations where $\sigma_{\rm{rec}}=0$ is 50$\%$. Therefore, we exclude MaNGA galaxies where the fraction of pixels with $\sigma_\ast=0$ is greater than 50$\%$. The exception to this cut are disk galaxies (i.e. galaxies determined in \autoref{sec:qualcon} to be spirals or S0s) which intrinsically can have low velocity dispersion but also high circular velocities meaning that they occupy the top region of \autoref{fig:zerosigmacontours}. Hence, we are confident in the accuracy of $\lambda_{R_e}$ for disk galaxies regardless of the fraction of pixels with $\sigma_\ast=0$.\par
\section{Derivation of the beam smearing correction}
\label{app:beamsmearing}

\begin{figure*}
\centering
\includegraphics[width=\textwidth]{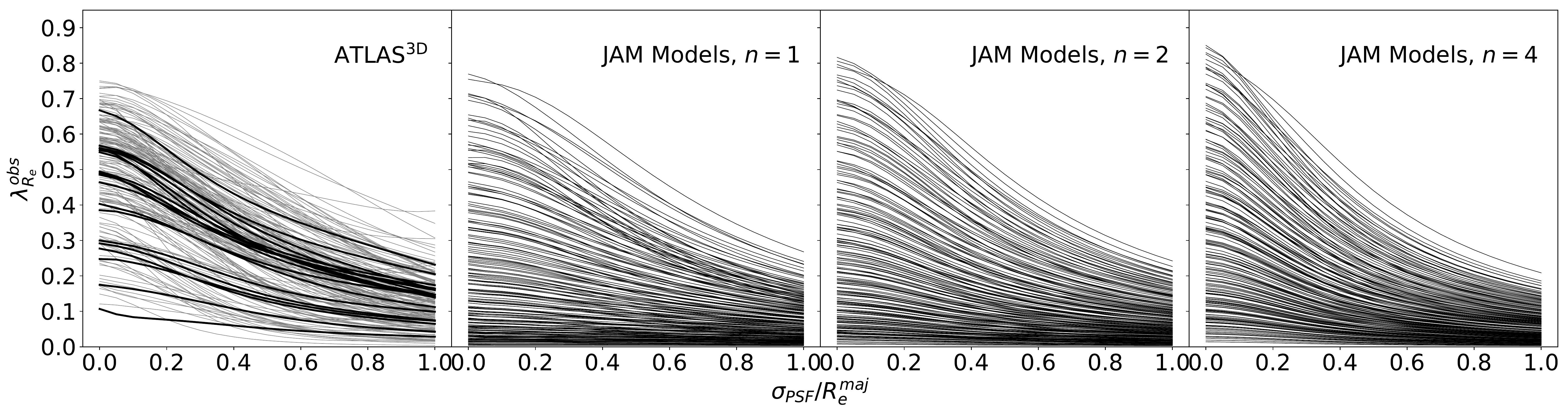}
\caption{The first panel shows profiles of $\lambda_{R_e}^{\rm{obs}}$ as a function of \ratio for 18 galaxies in the \texttt{small\_a3d} subsample (black) and 165 galaxies in the \texttt{large\_a3d} subsample (grey). The remaining panels show the same for 540 JAM Models with Sérsic indices of $n=1$, 2 and 4 shown as indicated. Each value of $\lambda_{R_e}^{\rm{obs}}$ is measured at intervals of 0.05 in \ratio. Constant errors are assumed in $\lambda_{R_e}^{\rm{obs}}$.}
\label{fig:profiles}
\end{figure*}

The two key quantities of interest are the size of the PSF, here quantified by $\sigma_{\rm{PSF}} = \rm{FWHM}_{\rm{PSF}}/2.355$, and the size of the galaxy, quantified here by the semi-major axis, $R_e^{\rm{maj}}$. Since the effect we are interested in is significant for large $\sigma_{\rm{PSF}}$ and small $R_e^{\rm{maj}}$, we choose our correction to be a function of \ratio. We choose the semi-major axis rather than the circular effective radius for the simple reason that the correction correlates with $R_e^{\rm{maj}}$ much better than with $R_e$. This is due to the fact that for very elliptical galaxies or models, the circular radius can be as much as three times smaller than the semi-major axis. By defining the ratio with respect to the semi-major axis, the correction depends only on the angular size and not inclination.\par
There is also a secondary dependence on the Sérsic index in that the exact nature of the ``blurring" effect depends on the concentration of the galaxy. To account for this, we derive a secondary correction as a function of Sérsic index and \ratio.\par

\subsection{Velocity moment convolution}
\label{sec:velocityconvolution}
In order to measure how $\lambda_{R_e}$ varies due to the size of the PSF, we use general formulas to convolve the first and second velocity moments and the surface brightness $\Sigma$ of all galaxies and models with the same circular Gaussian PSF (e.g. Equations 51-53 of \citealp{emsellem1994multi}):

\begin{equation}\label{eq:surfobs}
\Sigma_{\rm{\rm{obs}}} = \Sigma \otimes \rm{PSF},
\end{equation}

\begin{equation}\label{eq:velocityobs}
[\overline{v_{\rm{los}}}]_{\rm{\rm{obs}}} = \frac{(\Sigma \overline{v_{\rm{los}}}) \otimes \rm{PSF}}{\Sigma_{\rm{\rm{obs}}}},
\end{equation}

\begin{equation}\label{eq:sigmaobs}
[\overline{v_{\rm{los}}^2}]_{\rm{\rm{obs}}} = \frac{(\Sigma \overline{v_{\rm{los}}^2}) \otimes \rm{PSF}}{\Sigma_{\rm{\rm{obs}}}},
\end{equation}
where $\otimes$ represents convolution. We measure $\lambda_{R_e}$ for the convolved quantities using \autoref{eq:lambda_R} substituting $F$, $V$ and $\sigma$ for these three quantities.\par
There is a separate effect whereby the ``blurring'' of the visual image causes the observed flattening to be circularised. As a result, $\epsilon$ is also lowered due to beam smearing. However, we do not correct $\epsilon$ for this effect as doing so would require an \textit{intrinsic} measurement of the effective radius. From \autoref{eq:surfobs}, it is clear that the surface brightness becomes more extended with an increasing PSF, and hence the effective radius is also a function of the PSF. Hence, it is impossible to correct $\epsilon$ before correcting $R_e$. This argument nullifies the need for an $\epsilon$ correction as $\epsilon$ can then be measured within the intrinsic $R_e$. Furthermore, in the practical sense, $\epsilon$ is generally measured directly from photometry which is likely either already deconvolved (as in our case when using MGEs or the NSA which uses Sérsic models), or has a small enough PSF that this effect is small. In the case of $\lambda_{R_e}$, this circular argument does not apply as $\epsilon$ does not depend on $\lambda_{R_e}$. In any case, the effect due to the circularisation of the ellipse is likely to be much smaller than the uncertainties in our approximate correction.\par

\subsection{JAM and ATLAS$^{\rm{3D}}$ kinematics}
We model the first and second velocity moments, $V$ and $V_{\rm{RMS}} \equiv \sqrt{V^2+\sigma^2}$ respectively, by using the Jeans Anisotropic Modelling (JAM) method \citep{cappellari2008measuring}. The use of this method is justified by the fact that the resulting models were shown to provide quite realistic descriptions of the kinematics of real galaxies as a function of a single physical parameter (e.g. Figure 10 of C16). Given the surface brightness (i.e. a MGE) and two parameters (the inclination and the anisotropy $\beta_z \equiv 1-(\sigma_z/\sigma_R)^2$ where $\sigma_z$ is the velocity dispersion in the $z$ direction and $\sigma_R$ is the velocity dispersion in the $R$ direction), the method provides a solution to the Jeans equations \citep{jeans1922motions} assuming axisymmetry. Using simulated kinematics from JAM models brings a significant advantage in that there are no border effects.\par
We generate 1080 models with a range of intrinsic ellipticities ($0.1 \leq \epsilon_{\rm{intr}} \leq 0.94$), inclinations ($10\degree \leq i \leq 90\degree$ where 90$\degree$ is edge-on) and Sérsic indices ($1 \leq n \leq 6$). This range of parameters should cover ``normal'' late-type disks. We use a Sérsic profile to describe the stellar mass surface density as a function of radius. We fit the Sérsic profile using the MGE fitting method and software \texttt{mge\_fit\_1d}\textsuperscript{\ref{Capwebpage}} by \cite{cappellari2002efficient} to find the surface density and the dispersion $\sigma_j$ of each Gaussian in the unconvolved MGE. We calculate Equations C1-C3 assuming a self-consistent model whereby the same Gaussians are used to describe the surface brightness and the potential. For all models, we assume a semi-major axis $R_e^{\rm{maj}}$ of 10\arcsec  and an axisymmetric galaxy potential. We assume a distance of 41.2 Mpc (approximate maximum distance in ATLAS$^{\rm{3D}}$) and an anisotropy parameter $\beta_z=0.7 \times \epsilon_{\rm{intr}}$ \citep{cappellari2007sauron}.\par
Although we derive our correction purely from JAM models, we repeat the same method using kinematics from the ATLAS$^{\rm{3D}}$ survey. The targets of the survey are much closer than the MaNGA galaxies ($\gtrsim100$ Mpc), and so can be considered to be unaffected by seeing. We measure $\lambda_{R_e}^{\rm{true}}$ (i.e. without any convolution) within the half-light ellipse using the published values for $\epsilon$ (Table B1 from \citealp{emsellem2011atlas3d}), photometric PA (Table D1 from \citealp{krajnovic2011atlas3d}) and effective radius (Table 1 from \citealp{cappellari2013atlas3db}) for all the ATLAS$^{\rm{3D}}$ galaxies. We are able to reproduce the published values of $\lambda_{R_e}$ from \cite{emsellem2011atlas3d} to within about 0.1 for 247/257 of the ATLAS$^{\rm{3D}}$ galaxies. We therefore exclude 10 galaxies: five due to foreground star(s) (\textit{IC0676}, \textit{NGC2679}, \textit{NGC4478}, \textit{NGC4684}, \textit{NGC5770}), two due to border effects (\textit{UGC6014}, \textit{UGC03960}), one due to the presence of a foreground star and a companion galaxy (\textit{NGC5846}), one due to high velocity and velocity dispersion at the bulge (\textit{NGC4486A}) and one due to an unknown reason (\textit{NGC3032}).\par
We then remove 50/247 galaxies which are not regular rotators, as the correction is calibrated using JAM models which only describe regular rotators. Finally, we remove 14/197 galaxies which lie outside the range $0.5 \leq n \leq 6.5$, where $n$ are the values for the Sérsic index obtained from single fits to 1D light profiles and are found in Table C1 of \cite{krajnovic2013atlas3db}. We split the remaining 183 galaxies into two samples, the \texttt{small\_a3d} sample and the \texttt{large\_a3d} sample. The \texttt{small\_a3d} sample contains 18 galaxies which have 100\% coverage\footnote{We define the coverage to be the fraction of the half-light ellipse covered by spaxels, and is calculated as $N_{\rm{pix}} \Delta x^2 / (\pi R_e^2)$ where $N_{\rm{pix}}$ is the number of pixels within the half-light ellipse, $\Delta x$ is the pixel size in arcsec, and $R_e$ is the circularised effective radius.} within the half-light ellipse and contain no jumps in $V$ or $\sigma$ within the half-light ellipse that may affect the convolution. These galaxies are: \textit{NGC0502}, \textit{NGC2592}, \textit{NGC2778}, \textit{NGC3457}, \textit{NGC3458}, \textit{NGC3648}, \textit{NGC3674}, \textit{NGC4255}, \textit{NGC4262}, \textit{NGC4283}, \textit{NGC4377}, \textit{NGC4434}, \textit{NGC4660}, \textit{NGC5507}, \textit{NGC5845}, \textit{NGC6278}, \textit{UGC04551} and \textit{UGC06062}. The \texttt{large\_a3d} sample contains the other 165 surviving galaxies. For both subsamples, we convolve the galaxy MGEs, surface brightness and the first two velocity moments with Gaussian PSFs ranging from 0 to $R_e^{\rm{maj}}$ as described above.\par

\subsection{Sérsic-dependent generalised Moffat function}
\label{sec:psfcorrection}

For each JAM model and ATLAS$^{\rm{3D}}$ galaxy, we obtain profiles for ``observed'' values of $\lambda_{R_e}$, i.e. $\lambda_{R_e}^{\rm{obs}}$, as a function of \ratio (see \autoref{fig:profiles}). To describe the JAM profiles analytically, we adopt a generalised form of the Moffat function, which is widely used to provide an analytic description of the PSF\footnote{The Moffat function is similar to a Gaussian except with larger tails.}:
\begin{equation} \label{eq:moffat}
gM_n(\sigma_{\rm{PSF}}/R_e^{\rm{maj}})= \Bigg[ 1+\Bigg( \frac{\sigma_{\rm{PSF}}/R_e^{\rm{maj}}}{a_n} \Bigg)^{c_n} \Bigg]^{-b_n},
\end{equation}
where our generalised Moffat (gMoffat) reduces to the Moffat for $c_n=2$. Our choice for the gMoffat is arbitrary in that it provides a simple function which approximates the observed variations with few free parameters.\par



\begin{figure}
\centering
\includegraphics[width=0.49\textwidth]{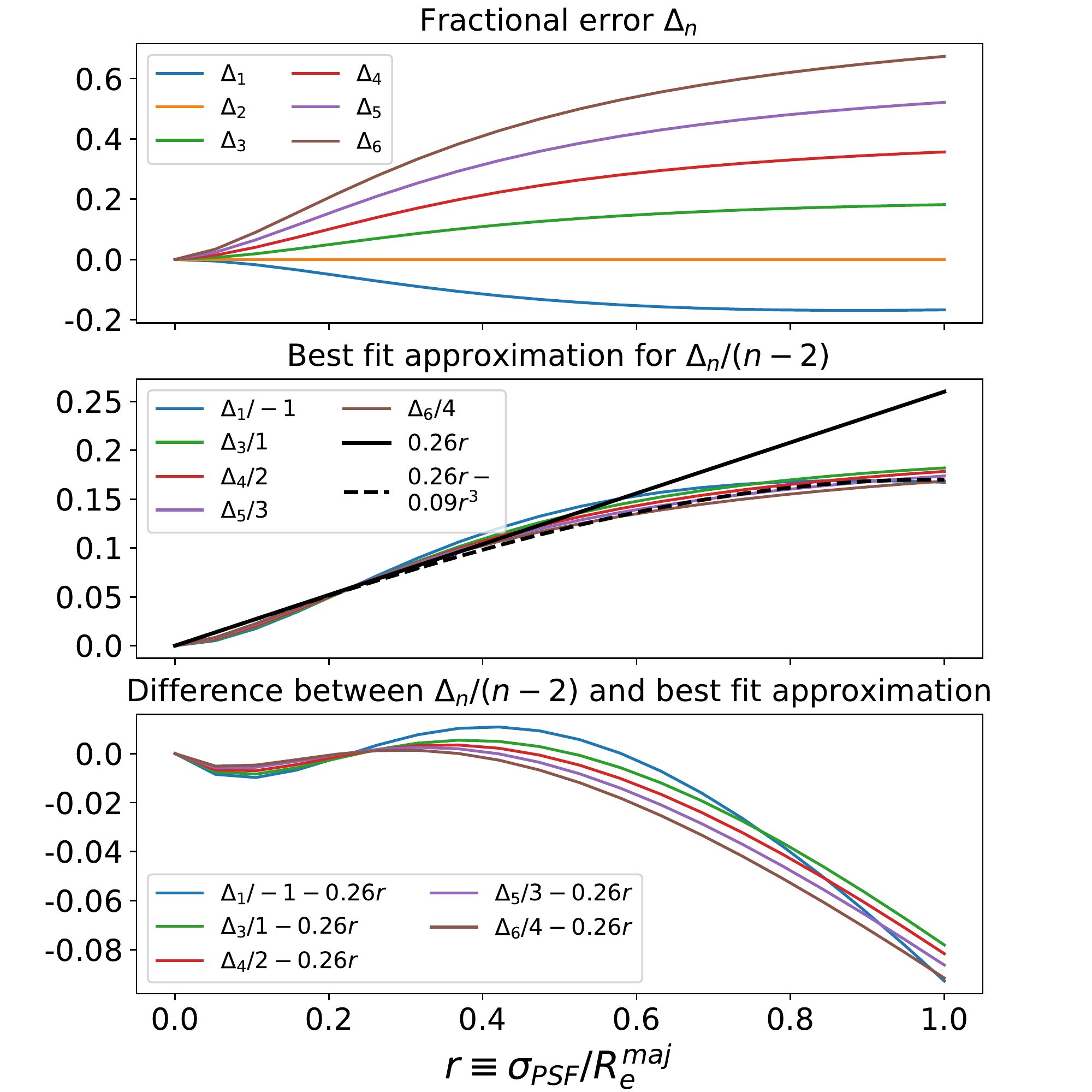}
\caption{\textbf{Top:} For a constant value of $\lambda_{R_e}^{\rm{true}}$, we plot $\Delta_n$, i.e. \autoref{eq:ratiominusone}, for each Sérsic-\textit{specific} gMoffat function $gM_n$ as a function of $r$ where $r \equiv $\ratio. \textbf{Middle:} We divide each $\Delta_n$ (apart from $\Delta_2$) in the top panel by $(n-2)$ to bring each curve to the same scale (i.e. $\Delta_3$). Two fits are shown: a cubic fit shown as the black dashed line, and a linear fit, shown as the black solid line. \textbf{Bottom:} We plot the difference between each $\Delta_n/(n-2)$ in the middle panel and the linear fit.}
\label{fig:sersic_derivation}
\end{figure}

We find the coefficients for our gMoffat function by minimising the least square of the residuals between the data and the function. The coefficients $a_n$, $b_n$ and $c_n$ are specific to the Sérsic index $n$. This fact is clearly illustrated by the different profile shapes in \autoref{fig:profiles}. However, rather than tabulating best-fitting coefficients for each Sérsic index which would limit the usability and accuracy of the correction, we find that we can approximate the correction at all $n$ between 1 and 6 by multiplying the gMoffat derived from $n=2$ JAM models by a factor $f_n(\sigma_{\rm{PSF}}/R_e^{\rm{maj}})$. While there is nothing particularly special about $n=2$, it is used as it gives the best results in the following analysis.\par
In order to find the functional form of $f_n$, we first find the coefficients $a_n$, $b_n$ and $c_n$ for each Sérsic index by fitting \autoref{eq:moffat} using the JAM models with the corresponding Sérsic index from $n=1$ to $n=6$. We then compare each of the six Sérsic-\textit{specific} gMoffat functions using fractional errors, i.e. 
\begin{equation} \label{eq:ratiominusone}
\Delta_n = \frac{gM_2(\sigma_{\rm{PSF}}/R_e^{\rm{maj}})}{gM_n(\sigma_{\rm{PSF}}/R_e^{\rm{maj}})}-1,
\end{equation}
as shown in the top panel of \autoref{fig:sersic_derivation}. We find that $\Delta_n$ takes the same form as a function of \ratio for $n=1$ to $n=6$ (apart from $n=2$ where $\Delta_2=0$). This non-trivial fact allows us to scale $\Delta_n$ by $1/(n-2)$ to match $\Delta_3$ (see the middle panel of \autoref{fig:sersic_derivation}).\par
While there does not exist a simple mathematical function that accurately describes $\Delta_n/(n-2)$, we find that we can approximate the scaled curves by fitting a straight line using \texttt{lts\_linefit} (see the solid black line in the middle panel of \autoref{fig:sersic_derivation}). Clearly, this simple linear fit becomes inaccurate at \ratio$ \gtrsim 0.5$ (see the bottom panel of \autoref{fig:sersic_derivation}) for all curves shown. While the addition of a cubic term does more accurately describe each curve (see the dashed black line in the middle panel of \autoref{fig:sersic_derivation}), we find that in practice, adopting the simple linear fit does not significantly affect the accuracy of the correction overall, considering that this effect is well within the errors of the approximate correction itself. Moreover, the fraction of galaxies in the \textit{clean sample} small enough to have \ratio $\gtrsim$ 0.5 is 0.6\%.\par
In summary, the Sérsic-\textit{dependent} gMoffat function (see \autoref{eq:lambdamoffatsersic}) provides an approximate but accurate description of the behaviour observed in \autoref{fig:profiles} for a range of JAM models with Sérsic indices from $n=1$ to $n=6$.

\begin{figure*}
\centering
\includegraphics[width=\textwidth]{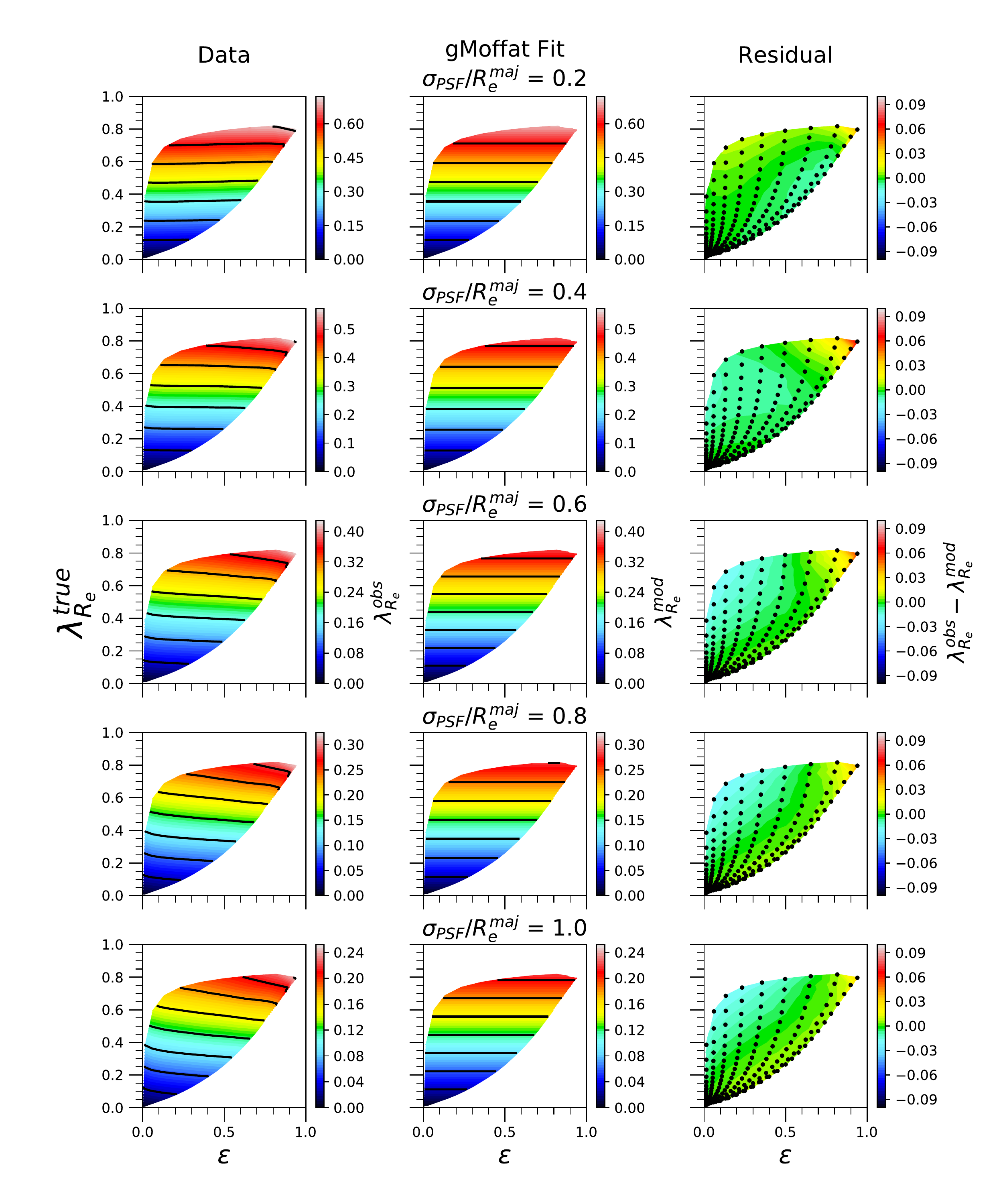}
\caption{Plots showing the observed values of $\lambda_{R_e}$ from the JAM models, the model given by \autoref{eq:lambdamoffatsersic} and the residuals on the $(\lambda_{R_e}^{\rm{true}}, \epsilon)$ plane for $n=2$. \textbf{Left:} Values of $\lambda_{R_e}^{\rm{obs}}$, indicated by the colourbar, shown for five values of \ratio, indicated for each row. The points with positions $(x, y)=(\epsilon, \lambda_{R_e}^{\rm{true}})$ are hidden in this column, but are shown in the right-hand column. \textbf{Middle:} The best-fitting gMoffat function where values are calculated using \autoref{eq:lambdamoffatsersic}. The colourbar ranges are the same as in the left column to allow for direct comparison. \textbf{Right:} Residuals, i.e the difference between the data and the model, for each value of \ratio. The black points are the $n=2$ JAM models and the colourbar is fixed for all rows between -0.1 and 0.1.}
\label{fig:lambdacontours}
\end{figure*}

\addtocounter{footnote}{-1}

\begin{figure*}
\centering
\includegraphics[width=\textwidth]{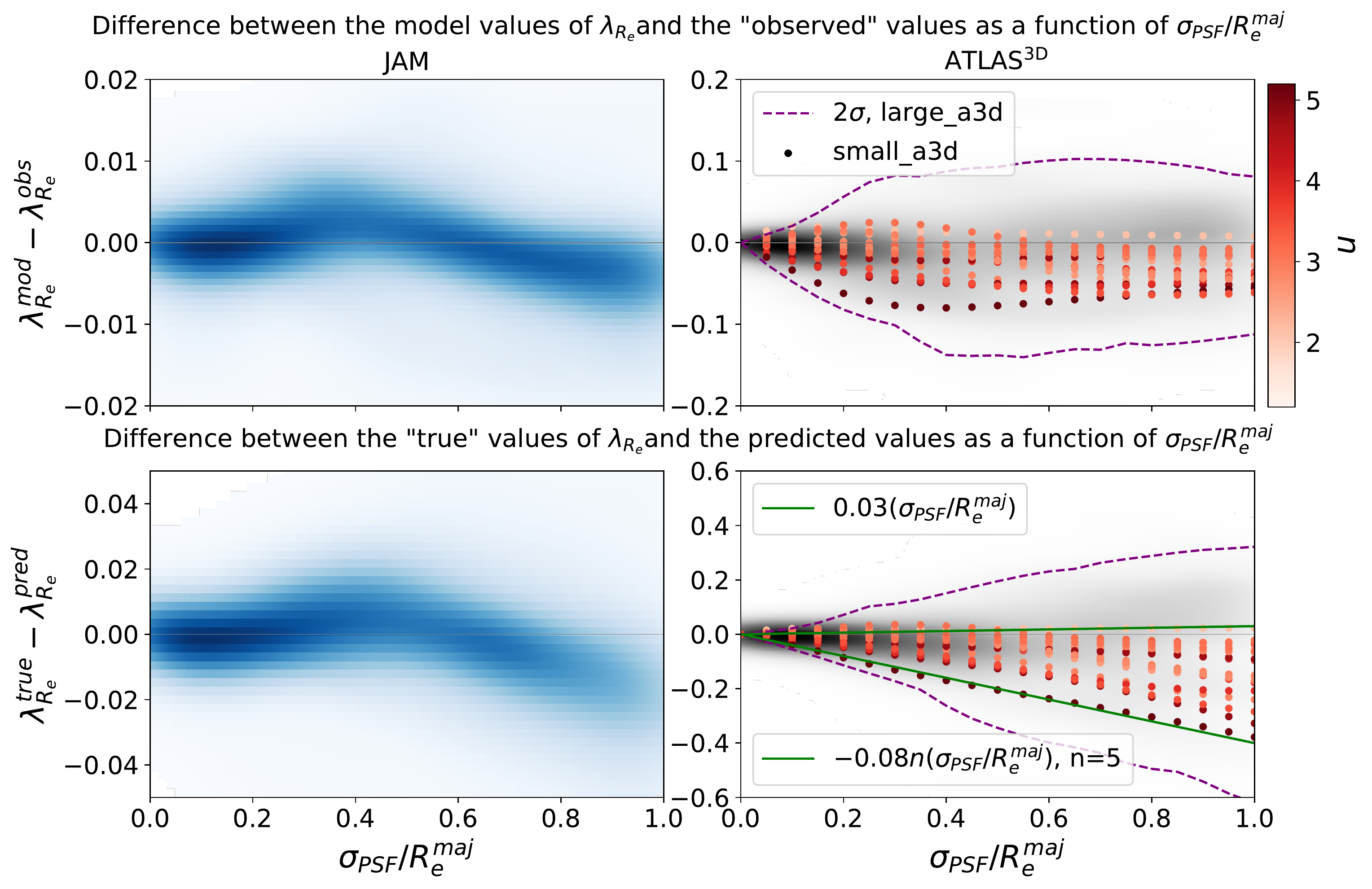}
\caption{\textbf{Top Left:} Difference between the model $\lambda_{R_e}$ ($\lambda_{R_e}^{\rm{mod}}$) for JAM predicted by \autoref{eq:lambdamoffatsersic} and the ``observed" (i.e. convolved) $\lambda_{R_e}$ ($\lambda_{R_e}^{\rm{obs}}$) as a function of \ratio. The 22680 points are smoothed using the KDE method described in \autoref{sec:volumecorrection}\protect\footnotemark. The model is indicated by the horizontal grey line where $\lambda_{R_e}^{\rm{mod}}-\lambda_{R_e}^{\rm{obs}}=0$. \textbf{Top Right:} The same as top left except for the \texttt{small\_a3d} subsample, shown as circles coloured according to $n$, and the \texttt{large\_a3d} subsample, shown as a smoothed grey density field. The purple dashed lines indicate the inner 95\% ($\sim 2 \sigma$) of the \texttt{large\_a3d} subsample. \textbf{Bottom Left:} The difference between the true values of $\lambda_{R_e}$ ($\lambda_{R_e}^{\rm{true}}$) and the values predicted by inverting \autoref{eq:lambdamoffatsersic} ($\lambda_{R_e}^{\rm{pred}}$) as a function of \ratio. The grey horizontal line indicates where the model accurately predicts the true (corrected) values of $\lambda_{R_e}$. \textbf{Bottom Right:} The same as bottom left except for the \texttt{small\_a3d} and \texttt{large\_a3d} subsamples. The purple dashed lines are as top right. The green lines are approximate linear fits to the outermost tracks of the \texttt{small\_a3d} subsample. The lower limit has an approximate linear dependence on $n$, shown here for $n=5$. The equations are shown in the legend, and can be used to predict the error in $\lambda_{R_e}^{\rm{pred}}$ as a function of \ratio.}
\label{fig:profileresiduals}
\end{figure*}

\footnotetext{22680 points $=$ 1080 models $\times$ 21 values of \ratio. To save computing time, we select a random sample of 3000 points from which to estimate the density field.}

\subsection{Application and error analysis}
\autoref{eq:lambdamoffatsersic} satisfies three physical boundary conditions: 

\begin{enumerate}
\item $\lim_{\sigma_{\rm{PSF}} \to 0} \lambda_{R_e}^{\rm{mod}} = \lambda_{R_e}^{\rm{true}}$,
\item $\lim_{\sigma_{\rm{PSF}} \to \infty} \lambda_{R_e}^{\rm{mod}} = 0$,
\item $\lim_{\lambda_{R_e}^{\rm{true}} \to 0} \lambda_{R_e}^{\rm{mod}} = 0$.
\end{enumerate}

For $0 \leq \sigma_{\rm{PSF}}/R_e^{\rm{maj}} \leq 1$, \autoref{eq:lambdamoffatsersic} is well determined, whereas for \ratio$> 1$, the function plateaus such that 

\begin{equation}
\lambda_{R_e}^{\rm{mod}}(\sigma_{\rm{PSF}}/R_e^{\rm{maj}}=1) \approx \lambda_{R_e}^{\rm{mod}}(\sigma_{\rm{PSF}}/R_e^{\rm{maj}}=\infty)\rm{.}
\end{equation}

This imposes an upper limit on our correction above which the function is too uncertain. Hence, we only define (and apply) our correction within the range $0 \leq \sigma_{\rm{PSF}}/R_e^{\rm{maj}} \leq 1$.\par
There is a low level of intrinsic scatter in the JAM models that can be seen in the residuals between the data and the model (e.g. \autoref{fig:moffat}). This intrinsic scatter is due to inclination effects. For each JAM model with intrinsic ellipticity $\epsilon_{\rm{intr}}$, we incline the model between $i=90\degree$ and $i=10\degree$. The LOSVD is dependent on inclination, but the Gaussian used to describe the PSF is always circular, and hence the effects due to convolution are dependent on inclination. As a result, the correction is most accurate at intermediate inclinations of $i \sim 50\degree$ as can be seen in \autoref{fig:lambdacontours}. However, as the scatter is less than 10\% for $\lambda_{R_e}^{\rm{true}} \gtrsim 0.05$ (and $\sim$0 as $\lambda_{R_e}^{\rm{true}}$ approaches 0), the correction can be considered valid for all inclinations.\par
Since \autoref{eq:lambdamoffatsersic} does not include an $\epsilon$ term, we are assuming that there is no trend in $\epsilon$. We check that this is the case by plotting the data from the JAM models, the model prediction and the difference between the two on the $(\lambda_{R_e}, \epsilon)$ plane for $n=2$ (see \autoref{fig:lambdacontours}). We find that there is a slight trend in the data at \ratio$\gtrsim 0.5$ whereby for a given $\lambda_{R_e}^{\rm{true}}$, $\lambda_{R_e}^{\rm{obs}}$ is higher for higher $\epsilon$. This trend is observed for all $n$ in our range. The consequence is that the model prediction $\lambda_{R_e}^{\rm{mod}}$ is slightly higher than $\lambda_{R_e}^{\rm{obs}}$ for lower $\epsilon$, and vice versa (see the right column of \autoref{fig:lambdacontours}). However, the magnitude of this effect is low considering that the disagreement between the data and the model is $\sim 0.03$ over the $(\lambda_{R_e},\epsilon$) plane.\par
In order to assess the accuracy of \autoref{eq:lambdamoffatsersic} when predicting $\lambda_{R_e}^{\rm{true}}$ from observed values of $\lambda_{R_e}^{\rm{obs}}$, we directly compare the true, unconvolved values of $\lambda_{R_e}^{\rm{true}}$ with values predicted by \autoref{eq:lambdamoffatsersic} i.e. $\lambda_{R_e}^{\rm{pred}}$. We apply our correction at the same ratios of \ratio and the same sample of JAM models used to derive \autoref{eq:lambdamoffatsersic}, as well as the two ATLAS$^{\rm{3D}}$ subsamples. In the bottom part of \autoref{fig:profileresiduals}, we plot the difference between $\lambda_{R_e}^{\rm{true}}$ and $\lambda_{R_e}^{\rm{pred}}$ for the JAM models (left) and the two ATLAS$^{\rm{3D}}$ subsamples (right). We find that \autoref{eq:lambdamoffatsersic} is able to predict $\lambda_{R_e}^{\rm{true}}$ to within 0.02 for the vast majority of JAM models. The scatter is due to a combination of inclination effects described above, as well as the range of Sérsic indices used. The turnover at \ratio$\sim 0.5$ is due to the linear approximation we adopted to keep the correction simple. The accuracy remains better than $\sim 0.04$ for all ratios.\par
We fit the outermost tracks on the lower right of \autoref{fig:profileresiduals} for the \texttt{small\_a3d} subsample with an approximate linear fit to estimate the error as a function of \ratio when applying the correction to real galaxies. We find that the correction essentially gives an upper limit for the true value of $\lambda_{R_e}$. The lower limit has an approximate linear dependence on $n$. As the errors are not a function of $\lambda_{R_e}^{\rm{obs}}$, it is entirely possible that the lower limit is below $\lambda_{R_e}^{\rm{obs}}$, in which case $\lambda_{R_e}^{\rm{obs}}$ itself provides the lower limit.\par
In \autoref{sec:angmom}, we apply our beam correction to our sample of MaNGA galaxies (e.g. \autoref{fig:anisotropy_kine}). In doing so, we note that 15 galaxies overshoot i.e. $\lambda_{R_e}^{\rm{true}}>1$. Almost all of these galaxies have high $\lambda_{R_e}^{\rm{obs}}$ values greater than 0.8. We note that our JAM models shown in \autoref{fig:lambdacontours} reach a maximum $\lambda_{R_e}^{\rm{true}}=0.8$ and hence we do not have any data above this range. This is likely a coincidence, as each $\lambda_{R_e}^{\rm{obs}}$ profile is essentially a scaled version of every other profile for constant $n$, as can be seen in \autoref{fig:profiles}. This fact is what allows the correction to be as accurate as it is for the range of $\lambda_{R_e}^{\rm{true}}$. While there is no reason to expect that this should change significantly above $\lambda_{R_e}^{\rm{true}}>0.8$, it may be the case that there are some border effects at $\lambda_{R_e}^{\rm{true}}\sim1$ for which we do not have data for. However, as discussed in \autoref{sec:angmom}, there are other sources of error that need to be considered for these particular galaxies.\par
In practice, one will only have single values for \ratio and $\lambda_{R_e}^{\rm{obs}}$ for any given galaxy. The correction relies on the assumption that if one were to measure $\lambda_{R_e}^{\rm{obs}}$ using an arbitrary number of telescopes with IFS capabilities, each with different PSFs, the resulting profile will exactly follow \autoref{eq:lambdamoffatsersic} (assuming of course that there are no systematic differences between telescopes other than the size of the PSF). As mentioned above, we derive our correction from JAM models with 100\% coverage within the half-light ellipse. While it is likely the correction is less accurate for galaxies without 100\% coverage (for example the \texttt{large\_a3d} subsample), there is no reason to refrain from using the correction in this case because the accuracy of \autoref{eq:lambdamoffatsersic} does not depend in a well-defined way on the coverage. Moreover, the most likely reason for a galaxy to have incomplete coverage is when the effective radius is large compared to the beam size, in which case the change in $\lambda_{R_e}$ due to the correction is likely to be small.\par
However, the correction is likely to be inaccurate for galaxies outside of the range $0.5\leq n \leq 6.5$ as the correction is calibrated for about that range\footnote{We extend the applicable range from $1\leq n \leq 6$ to $0.5\leq n \leq 6.5$ for the simple reason that the NSA values for $n$ start at 0.5.}. Finally, as the JAM method can only describe regular rotators (ETGs and spirals), the correction is only valid for regular rotators. We find that the correction does not describe well the profiles of non-regular ATLAS$^{\rm{3D}}$ galaxies. Furthermore, non-regular rotators have intrinsically low $\lambda_{R_e} \sim 0.2$ and so the effect due to seeing is small. Hence, in all our results where we apply the beam correction, we only correct $\lambda_{R_e}$ values for regular rotator ETGs and spirals, leaving non-regular rotators unchanged.


\bsp	
\label{lastpage}
\end{document}